\documentclass[journal]{IEEEtran}
\usepackage{textcomp}
\usepackage{gensymb}
\usepackage{colortbl}
\usepackage{xcolor}
\usepackage{array}
\usepackage{subcaption}
\usepackage{graphicx}
\usepackage{epsfig}
\usepackage{epstopdf}
\usepackage{hyperref}
\usepackage{cite}
\usepackage{amsmath, amssymb, amsthm, amsxtra}
\usepackage{bbm}
\usepackage{algorithmic}
\usepackage{algorithm}
\usepackage{authblk}
\usepackage{url}
\usepackage{comment}
\usepackage{siunitx}
\usepackage{lipsum}
\usepackage{dblfloatfix}
\usepackage{tabularx}
\usepackage{float}
\usepackage{booktabs}
\usepackage{longtable}
\usepackage{multirow}
\usepackage{makecell}
\usepackage{adjustbox}
\usepackage{pifont} 
 \usepackage{acro}
\usepackage{tikz}
\usepackage[utf8]{inputenc}
\usepackage[T1]{fontenc}
\usetikzlibrary{
    decorations.pathmorphing,
    positioning,
    fit,
    arrows.meta,
    backgrounds,
    shadows,
    shapes,
    calc,
    shapes.multipart
}
\usepackage{hhline}

\definecolor{buildingA}{RGB}{84,92,99}
\definecolor{buildingB}{RGB}{106,116,124}
\definecolor{windowC}{RGB}{206,235,245}
\definecolor{roofC}{RGB}{50,55,61}
\definecolor{satBlue}{RGB}{52,123,190}
\definecolor{satBody}{RGB}{244,176,74}
\definecolor{directgreen}{RGB}{0,158,73}
\definecolor{blockedred}{RGB}{210,45,38}
\definecolor{reflectedblue}{RGB}{0,114,178}
\definecolor{receiverPole}{RGB}{95,50,39}
\definecolor{antennaCream}{RGB}{250,244,220}

\definecolor{panelbg}{RGB}{249,250,252}
\definecolor{panelrule}{RGB}{210,216,222}
\definecolor{titlebg}{RGB}{42,57,72}
\definecolor{legendbg}{RGB}{246,248,250}
\definecolor{legendrule}{RGB}{205,210,216}

\tikzset{
    direct/.style={directgreen, line width=1.55pt, line cap=round},
    blocked/.style={blockedred, line width=1.45pt, dotted, line cap=round},
    reflected/.style={reflectedblue, line width=1.55pt, line cap=round, line join=round},
    panel title/.style={
        font=\sffamily\bfseries\footnotesize,
        text=white,
        align=center
    },
    legend text/.style={
        font=\sffamily\fontsize{7.4}{8.2}\selectfont,
        anchor=west
    },
    scenelabel/.style={
        font=\sffamily\bfseries\small,
        text=white
    }
}

\newcommand{\TowerOne}[2]{%
\begin{scope}[shift={(#1,#2)}]
  \draw[fill=buildingA, draw=black, line width=0.60pt] (0,0) rectangle (1.55,3.15);
  \draw[fill=roofC, draw=black, line width=0.50pt] (0.18,3.15) -- (0.38,3.45) -- (1.17,3.45) -- (1.37,3.15) -- cycle;
  \foreach \y in {0.72,1.23,1.74,2.25,2.76}{%
    \foreach \x in {0.25,0.90}{%
      \draw[fill=windowC, draw=black, line width=0.32pt] (\x,\y) rectangle ++(0.35,0.28);} }
  \draw[fill=windowC, draw=black, line width=0.32pt] (0.60,0.08) rectangle ++(0.36,0.48);
\end{scope}}

\newcommand{\TowerTwo}[2]{%
\begin{scope}[shift={(#1,#2)}]
  \draw[fill=buildingB, draw=black, line width=0.60pt] (0,0) rectangle (1.65,2.95);
  \draw[fill=roofC, draw=black, line width=0.50pt] (-0.05,2.95) rectangle (1.70,3.12);
  \foreach \y in {0.68,1.15,1.62,2.09,2.56}{%
    \foreach \x in {0.22,0.72,1.22}{%
      \draw[fill=windowC, draw=black, line width=0.30pt] (\x,\y) rectangle ++(0.24,0.27);} }
  \draw[fill=windowC, draw=black, line width=0.32pt] (0.64,0.07) rectangle ++(0.38,0.50);
\end{scope}}

\newcommand{\ReceiverIcon}[2]{%
\begin{scope}[shift={(#1,#2)}]
  \draw[fill=roofC, draw=black, line width=0.35pt, rounded corners=0.7pt]
    (-0.38,0.00) rectangle (0.38,0.12);
  \draw[fill=receiverPole, draw=black, line width=0.30pt]
    (-0.25,0.12) rectangle (0.25,0.20);
  \draw[fill=antennaCream, draw=black, line width=0.45pt]
    (-0.42,0.34) arc[start angle=180,end angle=0,x radius=0.42,y radius=0.16]
    -- (0.42,0.23) -- (-0.42,0.23) -- cycle;
  \draw[fill=antennaCream, draw=black, line width=0.30pt] (-0.34,0.20) rectangle (0.34,0.25);
  \draw[fill=antennaCream, draw=black, line width=0.30pt] (-0.27,0.15) rectangle (0.27,0.20);
  \draw[fill=antennaCream, draw=black, line width=0.30pt] (-0.20,0.10) rectangle (0.20,0.15);
\end{scope}}

\newcommand{\SatelliteIcon}[3]{%
\begin{scope}[shift={(#1,#2)}, rotate=#3, scale=0.48]
  \draw[fill=satBody, draw=black, line width=0.52pt, rounded corners=1.0pt]
    (-0.32,-0.32) rectangle (0.32,0.32);
  \draw[fill=roofC, draw=black, line width=0.45pt]
    (-0.18,-0.54) -- (0.18,-0.54) -- (0,-0.88) -- cycle;
  \draw[black, line width=0.60pt] (0,-0.88) -- (0,-1.12);
  \draw[black, line width=0.60pt] (-0.28,-1.12) arc[start angle=200,end angle=340,radius=0.30];
  \draw[fill=satBlue, draw=black, line width=0.52pt, rounded corners=0.7pt]
    (-1.50,-0.45) rectangle (-0.48,0.45);
  \draw[fill=satBlue, draw=black, line width=0.52pt, rounded corners=0.7pt]
    (0.48,-0.45) rectangle (1.50,0.45);
  \foreach \x in {-1.16,-0.82,0.82,1.16}{\draw[white, line width=0.30pt] (\x,-0.39) -- (\x,0.39);}
  \foreach \y in {-0.15,0.15}{
    \draw[white, line width=0.30pt] (-1.44,\y) -- (-0.54,\y);
    \draw[white, line width=0.30pt] (0.54,\y) -- (1.44,\y);}
\end{scope}}

\newcommand{\GroundLine}[1]{\draw[gray!55, line width=0.40pt] (#1,0) -- ++(6.30,0);}
\newcommand{\CityPair}{\TowerOne{0.18}{0}\TowerTwo{4.36}{0}}
\newcommand{\ScenarioPanel}[4]{%
\begin{scope}[shift={(#1,#2)}]
  \draw[rounded corners=2.3pt, fill=panelbg, draw=panelrule, line width=0.55pt]
    (0,0) rectangle (6.65,5.95);

  \fill[rounded corners=2.3pt, titlebg] (0,5.22) rectangle (6.65,5.95);
  \node[panel title] at (3.325,5.57) {#3};

  \begin{scope}[shift={(0.18,0.62)}]
    #4
  \end{scope}
\end{scope}
}

\makeatletter
\def\subsubsection{\@startsection{subsubsection}{3}{\parindent}{0ex plus 0.1ex minus 0.1ex}{0ex}{\normalfont\normalsize\bfseries}}
\renewcommand{\@IEEEpubidpullup}{32pt}
\makeatother

\begin{document}

\title{\LARGE GNSS Anomaly Detection in Urban Canyons:\\ A Comprehensive Survey of Multimodal Visibility Inference}

\author{Abdelgabar Ahmed,~\IEEEmembership{Student Member,~IEEE,} 
        Tarig Ballal,~\IEEEmembership{Member,~IEEE,} 
        Nour Kouzayha,~\IEEEmembership{Member,~IEEE,} 
        and~Tareq Y. Al-Naffouri,~\IEEEmembership{Fellow,~IEEE}
\thanks{Abdelgabar Ahmed, Nour Kouzayha, and Tareq Y. Al-Naffouri are with the Electrical and Computer Engineering Program, Division of Computer, Electrical and Mathematical Sciences and Engineering (CEMSE), King Abdullah University of Science and Technology (KAUST), Thuwal 23955-6900, Kingdom of Saudi Arabia (abdelgabar.ahmed@kaust.edu.sa; nour.kouzayha@kaust.edu.sa; tareq.alnaffouri@kaust.edu.sa).} \thanks{Tarig Ballal is with the Aerospace Engineering Department, IRC for Aviation \& Space Exploration, King Fahd University of Petroleum \& Minerals (KFUPM), Dhahran 34464, Kingdom of Saudi Arabia (tarig.ahmed@kfupm.edu.sa).}
\thanks{Accepted for publication in \textit{IEEE Aerospace and Electronic Systems Magazine}.
DOI: 10.1109/MAES.2026.3746915.}
}

\IEEEpubid{%
  \parbox[b]{\textwidth}{%
    \normalfont
    \fontsize{7}{8.5}\selectfont
    \raggedright
    \setlength{\parindent}{0pt}%
    \setlength{\parskip}{0pt}%
    \copyright~2026 IEEE. Personal use of this material is permitted.
    Permission from IEEE must be obtained for all other uses, in any
    current or future media, including reprinting/republishing this
    material for advertising or promotional purposes, creating new
    collective works, for resale or redistribution to servers or lists,
    or reuse of any copyrighted component of this work in other works.
  }%
}

\maketitle

\begin{abstract}
Reliable GNSS positioning in dense urban environments depends not only on positioning algorithms, but also on deciding which satellite measurements should be trusted. Buildings, vehicles, vegetation, and other urban structures can block, reflect, or attenuate satellite signals, producing Non-Line-of-Sight (NLOS), multipath-contaminated, and degraded measurements. This survey reviews methodologies for identifying these reception states before corrupted observations enter the navigation solution. The review first examines receiver-side signal evidence, showing how physical, tracking-level, and observable-domain features reveal different signatures of NLOS and multipath propagation. It then synthesizes algorithmic detection methods that transform these features into reception-state decisions, ranging from interpretable rule-based approaches to modern learning-based classifiers. The survey next turns to environmental awareness, separating 3D mapping modalities from the map-based algorithms that use them for satellite-visibility determination. Finally, it reviews auxiliary sensors, including IMU, camera, and LiDAR, as complementary sources of kinematic, visual, and geometric evidence for online measurement screening. Across these methodological families, we emphasize practical design trade-offs involving accuracy, interpretability, hardware burden, calibration, map dependence, computational cost, and real-time deployability. The survey is intended for researchers and engineers working on robust urban GNSS and multimodal navigation, and spans receiver configurations ranging from smartphones and mass-market automotive systems to sensor-rich and multi-antenna platforms. By connecting signal processing, classification algorithms, 3D mapping, visibility prediction, and exteroceptive sensing under a unified measurement-screening perspective, this survey provides a systems-level guide to robust urban GNSS anomaly detection and highlights open challenges in datasets, domain generalization, uncertainty modeling, edge implementation, map freshness, and trustworthy learning.
\end{abstract}


\section{Introduction}

Global Navigation Satellite Systems (GNSS) based positioning has become indispensable across a wide range of modern applications, from everyday navigation services to safety-critical systems such as autonomous vehicles and aerial drones. While the technology has matured significantly in open-sky environments, its performance remains highly vulnerable in urban areas. In particular, the urban canyon effect caused by tall buildings and complex infrastructure poses a major obstacle to reliable signal reception. Among the various sources of GNSS error, multipath (MP) propagation and Non-Line-of-Sight (NLOS) receptions have emerged as dominant challenges that are difficult to predict and mitigate. This section delves into the fundamental impact of multipath and NLOS signals on GNSS performance, laying the groundwork for understanding recent advances in detection and mitigation techniques.

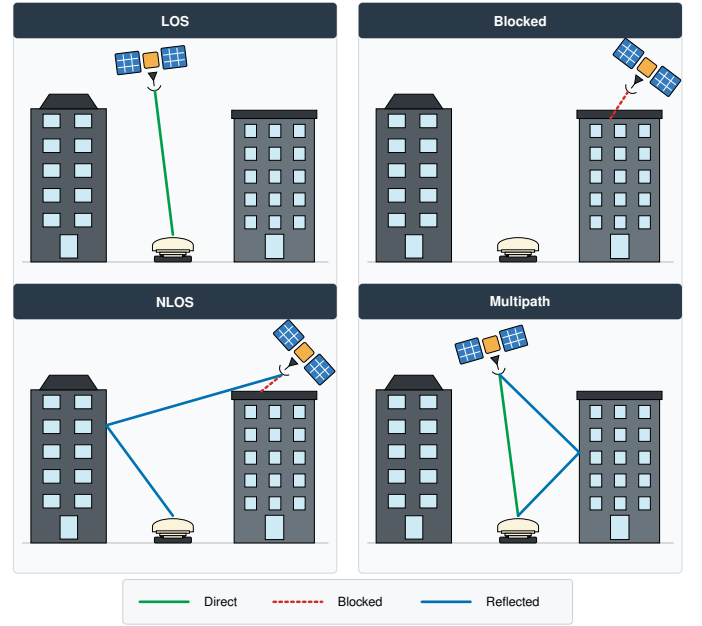
\begin{figure}[!t]
\centering
\resizebox{\columnwidth}{!}{%
\begin{tikzpicture}[x=1cm,y=1cm]

\ScenarioPanel{0.00}{5.78}{LOS}{
  \GroundLine{0}
  \CityPair
  \ReceiverIcon{3.10}{0}
  \SatelliteIcon{2.65}{4.15}{7}
  \draw[direct] (2.73,3.52) -- (3.10,0.56);
}

\ScenarioPanel{7.10}{5.78}{Blocked}{
  \GroundLine{0}
  \CityPair
  \ReceiverIcon{3.10}{0}
  \SatelliteIcon{5.75}{4.00}{-37}
  \draw[blocked] (5.36,3.50) -- (5.00,2.95);
}

\ScenarioPanel{0.00}{0.00}{NLOS}{
  \GroundLine{0}
  \CityPair
  \ReceiverIcon{3.10}{0}
  \SatelliteIcon{5.80}{3.92}{-48}
  \draw[blocked] (5.33,3.45) -- (4.92,3.12);
  \draw[reflected] (5.33,3.45) -- (1.73,2.42) -- (3.10,0.56);
}

\ScenarioPanel{7.10}{0.00}{Multipath}{
  \GroundLine{0}
  \CityPair
  \ReceiverIcon{3.10}{0}
  \SatelliteIcon{2.55}{4.08}{16}
  \draw[direct] (2.72,3.46) -- (3.10,0.56);
  \draw[reflected] (2.72,3.46) -- (4.36,1.86) -- (3.10,0.56);
}

\begin{scope}[shift={(2.25,-1.05)}]
  \draw[rounded corners=2pt, fill=legendbg, draw=legendrule, line width=0.50pt]
    (0,0) rectangle (9.25,0.92);

  \draw[direct] (0.35,0.46) -- (1.35,0.46);
  \node[legend text] at (1.55,0.46) {Direct};

  \draw[blocked] (3.10,0.46) -- (4.10,0.46);
  \node[legend text] at (4.30,0.46) {Blocked};

  \draw[reflected] (6.15,0.46) -- (7.15,0.46);
  \node[legend text] at (7.35,0.46) {Reflected};
\end{scope}

\end{tikzpicture}%
}
\caption{GNSS reception scenarios in urban environments.}
\label{fig_Scenarios}
\end{figure}

\IEEEpubidadjcol
This survey focuses on the identification of GNSS reception states---including line-of-sight, non-line-of-sight, and multipath-contaminated reception---in urban environments before the corresponding measurements are accepted, excluded, weighted, or corrected by the navigation solution. The reviewed methods span pedestrian and smartphone navigation, mass-market and autonomous vehicle platforms, and higher-complexity receiver configurations such as multi-antenna and sensor-rich systems. These platforms differ in their access to raw signal information, auxiliary sensors, computational resources, and calibrated hardware, and these receiver capabilities directly influence the applicability and practical trade-offs of the corresponding detection methods.

\subsection{Impact of Multipath and NLOS on GNSS Accuracy}
Tall buildings, reflective facades, vegetation, and dynamic objects in urban environments can block, reflect, diffract, or attenuate GNSS signals. These propagation effects directly corrupt the code pseudorange and carrier-phase observables used for positioning, making multipath and NLOS reception among the dominant error sources in urban canyons~\cite{misra2006global,braasch2017multipath}.
As illustrated in Fig.~\ref{fig_Scenarios}, GNSS reception can be grouped into three principal tracked-signal conditions. In the \emph{Line-of-Sight (LOS)} case, the direct satellite-to-receiver path is available and dominates the received signal. In the \emph{multipath} case, the direct path remains present, but one or more reflected components arrive together with it. In the \emph{NLOS} case, the direct path is blocked and the receiver tracks an indirect reflected or diffracted component. A geometrically blocked satellite may also be completely unavailable when no indirect component is strong enough to be acquired or tracked.
Taking the direct-path carrier phase as reference, these conditions can be represented by the complex received-signal model
\begin{equation}
r(t) =
g_0 s(t-\tau_0)
+
\sum_{i=1}^{N} g_i e^{j\theta_i} s(t-\tau_i)
+
n(t),
\label{eq:rx_model}
\end{equation}
where $g_0$ and $\tau_0$ denote the amplitude and delay of the direct component, $g_i$, $\tau_i$ and $\theta_i$ denote the amplitude, delay, and relative carrier phase of the $i$th reflected component, respectively, $N$ is the number of indirect paths, $s(t)$ is the transmitted signal waveform, and $n(t)$ is additive noise~\cite{braasch2017multipath,smyrnaios2013multipath}. Under nominal LOS reception, the first term dominates; under multipath, direct and indirect components jointly contribute to the tracked waveform; and under NLOS reception, the direct component is absent or strongly suppressed.

\begingroup
\definecolor{mpDirect}{RGB}{0,105,92}
\definecolor{mpReflected}{RGB}{0,92,175}
\definecolor{mpComposite}{RGB}{190,108,45}
\tikzset{
  mpAxis/.style={draw=black!48, line width=0.38pt, -{Latex[length=2.1pt]}},
  mpGrid/.style={draw=black!14, dash pattern=on 1.0pt off 1.8pt, line width=0.25pt},
  mpGuide/.style={draw=black!30, dashed, line width=0.32pt},
  mpLOS/.style={draw=mpDirect, line width=1.05pt},
  mpRefl/.style={draw=mpReflected, dashed, line width=0.72pt},
  mpComp/.style={draw=mpComposite, line width=1.35pt},
}
\def\mpAZero{1}
\def\mpAOne{0.75}
\def\mpHalfSpacing{0.25}
\def\mpReflectionDelay{0.40}
\def\mpRelativePhase{45}
\def\mpBiasRoot{0.1666432873823976}
\def\mpPhasorBaseLength{3.10}
\def\mpPhasorOriginX{1.35}
\def\mpPhasorOriginY{1.15}
\pgfmathsetmacro{\mpLOSTipX}{\mpPhasorOriginX+\mpPhasorBaseLength*\mpAZero}
\pgfmathsetmacro{\mpLOSTipY}{\mpPhasorOriginY}
\pgfmathsetmacro{\mpResultantTipX}{\mpLOSTipX+\mpPhasorBaseLength*\mpAOne*cos(\mpRelativePhase)}
\pgfmathsetmacro{\mpResultantTipY}{\mpLOSTipY+\mpPhasorBaseLength*\mpAOne*sin(\mpRelativePhase)}
\pgfmathsetmacro{\mpPromptPhase}{atan(\mpAOne*sin(\mpRelativePhase)/(\mpAZero+\mpAOne*cos(\mpRelativePhase)))}
\pgfmathdeclarefunction{mptri}{1}{%
  \pgfmathparse{max(1-abs(#1),0)}%
}
\pgfmathdeclarefunction{mplos}{1}{%
  \pgfmathparse{\mpAZero^2*(mptri(#1+\mpHalfSpacing)^2-mptri(#1-\mpHalfSpacing)^2)}%
}
\pgfmathdeclarefunction{mprefl}{1}{%
  \pgfmathparse{\mpAOne^2*(mptri(#1-\mpReflectionDelay+\mpHalfSpacing)^2-mptri(#1-\mpReflectionDelay-\mpHalfSpacing)^2)}%
}
\pgfmathdeclarefunction{mpcomp}{1}{%
  \pgfmathparse{mplos(#1)+mprefl(#1)+2*\mpAZero*\mpAOne*cos(\mpRelativePhase)*(mptri(#1+\mpHalfSpacing)*mptri(#1-\mpReflectionDelay+\mpHalfSpacing)-mptri(#1-\mpHalfSpacing)*mptri(#1-\mpReflectionDelay-\mpHalfSpacing))}%
}
\begin{figure}[t]
\centering
\begin{subfigure}[t]{\linewidth}
\centering
\begin{tikzpicture}[x=2.40cm,y=1.70cm,font=\sffamily\small]
  \path[use as bounding box] (-1.63,-2.12) rectangle (1.82,2.18);
  \begin{scope}
    \clip (-1.25,-1.46) rectangle (1.65,1.66);
    \path[fill=mpComposite,fill opacity=0.11] (0,0) rectangle (\mpBiasRoot,-1.46);
    \foreach \gridx in {-1.0,-0.5,0.5,1.0,1.5}
      \draw[mpGrid] (\gridx,-1.46) -- (\gridx,1.66);
    \foreach \gridy in {-1.25,-1.0,-0.75,-0.5,-0.25,0.25,0.5,0.75,1.0,1.25,1.5}
      \draw[mpGrid] (-1.25,\gridy) -- (1.65,\gridy);
    \draw[mpGuide] (0,0) -- (0,-1.46);
    \draw[mpGuide] (\mpBiasRoot,0) -- (\mpBiasRoot,-1.46);
    \draw[mpComposite,line width=0.42pt] (\mpBiasRoot,0) -- (0.39,0.60);
    \draw[mpDirect,line width=0.42pt] (0,0) -- (-0.10,-0.10);
  \end{scope}
  \draw[mpAxis] (-1.30,0) -- (1.72,0);
  \draw[mpAxis] (0,-1.54) -- (0,1.74);
  \foreach \tick/\ticklabel in {-1/-1.0,-0.5/-0.5,0/0,0.5/0.5,1/1.0,1.5/1.5} {
    \draw[draw=black!50,line width=0.32pt] (\tick,-0.027) -- (\tick,0.027);
    \node[font=\sffamily\tiny,anchor=north] at (\tick,-1.56) {\ticklabel};
  }
  \begin{scope}
    \clip (-1.25,-1.46) rectangle (1.65,1.66);
    \draw[mpLOS] plot[domain=-1.25:1.65,samples=117] (\x,{mplos(\x)});
    \draw[mpRefl] plot[domain=-1.25:1.65,samples=117] (\x,{mprefl(\x)});
    \draw[mpComp] plot[domain=-1.25:1.65,samples=117] (\x,{mpcomp(\x)});
  \end{scope}
  \fill[mpDirect] (0,0) circle (1.25pt);
  \fill[mpComposite] (\mpBiasRoot,0) circle (1.45pt);
  \node[mpDirect,anchor=north east,fill=white,fill opacity=0.92,text opacity=1,
        inner sep=1pt,font=\sffamily\tiny] at (-0.05,-0.08) {nominal LOS zero};
  \node[mpComposite,anchor=west,fill=white,fill opacity=0.94,text opacity=1,
        inner sep=1.2pt,align=left,font=\sffamily\tiny] at (0.45,0.72)
        {Composite zero crossing $\widehat{\epsilon}_{\rm MP}$};
  \draw[mpLOS] (-1.20,1.82) -- (-0.98,1.82);
  \node[mpDirect,anchor=west,font=\sffamily\tiny] at (-0.93,1.82) {Direct LOS};
  \draw[mpRefl] (-0.35,1.82) -- (-0.13,1.82);
  \node[mpReflected,anchor=west,font=\sffamily\tiny] at (-0.08,1.82) {Reflected only};
  \draw[mpComp] (0.70,1.82) -- (0.92,1.82);
  \node[mpComposite,anchor=west,font=\sffamily\tiny] at (0.97,1.82) {Composite (LOS + reflected)};
  \node[font=\sffamily\scriptsize] at (0.20,-1.98) {Code tracking error, $\epsilon$ [chips]};
  \node[font=\sffamily\scriptsize,rotate=90] at (-1.54,0.10) {EMLP discriminator output};
\end{tikzpicture}
\caption{}
\end{subfigure}\par\vspace{0.45ex}
\begin{subfigure}[t]{\linewidth}
\centering
\begin{tikzpicture}[x=1.22cm,y=1.22cm,font=\sffamily\small]
  \path[use as bounding box] (0.50,0.40) rectangle (6.98,3.48);
  \coordinate (mpO) at (\mpPhasorOriginX,\mpPhasorOriginY);
  \coordinate (mpD) at (\mpLOSTipX,\mpLOSTipY);
  \coordinate (mpR) at (\mpResultantTipX,\mpResultantTipY);
  \draw[mpAxis] (0.82,\mpPhasorOriginY) -- (6.75,\mpPhasorOriginY)
    node[font=\sffamily\scriptsize,anchor=west] {$I$};
  \draw[mpAxis] (\mpPhasorOriginX,0.56) -- (\mpPhasorOriginX,3.22)
    node[font=\sffamily\scriptsize,anchor=south] {$Q$};
  \draw[mpGuide] (mpR) -- (\mpResultantTipX,\mpPhasorOriginY);
  \draw[mpGuide] (mpR) -- (\mpPhasorOriginX,\mpResultantTipY);
  \draw[mpLOS,-{Latex[length=3.0pt]}] (mpO) --
    node[pos=0.48,below=4pt,mpDirect,font=\sffamily\scriptsize] {$Z_{\rm LOS}$} (mpD);
  \draw[mpRefl,-{Latex[length=3.0pt]}] (mpD) --
    node[pos=0.58,below right=2pt,mpReflected,font=\sffamily\scriptsize] {$Z_{\rm refl}$} (mpR);
  \draw[mpComp,-{Latex[length=3.2pt]}] (mpO) --
    node[pos=0.52,above left=2pt,mpComposite,font=\sffamily\scriptsize] {$Z_{\rm P}$} (mpR);
  \fill[black!55] (mpO) circle (0.75pt);
  \draw[mpComposite,line width=0.58pt] (\mpPhasorOriginX+0.62,\mpPhasorOriginY)
    arc[start angle=0,end angle=\mpPromptPhase,radius=0.62];
  \node[mpComposite,anchor=west,fill=white,fill opacity=0.92,text opacity=1,
        inner sep=1pt,align=left,font=\sffamily\scriptsize] at (2.20,1.30)
        {$\Delta\phi_{\rm MP}$};
  \node[font=\sffamily\tiny] at (4.00,3.00) {$Z_{\rm LOS}+Z_{\rm refl}=Z_{\rm P}$};
\end{tikzpicture}
\caption{}
\end{subfigure}
\caption{Receiver-level illustration of multipath effects on GNSS tracking: (a) code-domain DLL distortion and tracking-point shift, and (b) carrier-domain prompt-phasor rotation.}
\label{fig:code_carrier_mp_mechanism}
\end{figure}
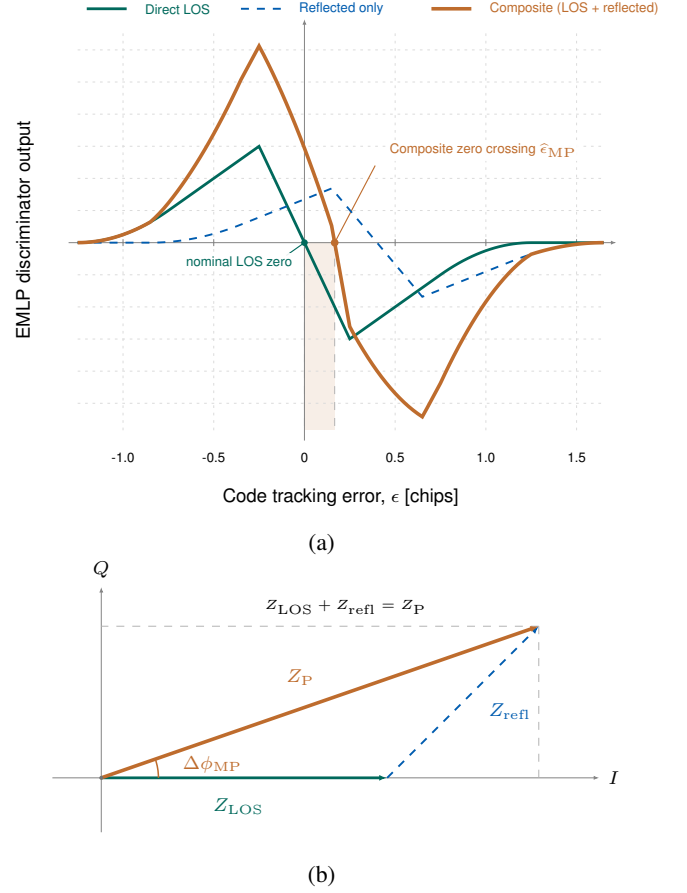
\endgroup

The resulting measurement error is determined not only by the excess propagation length, but also by how these components interact within the receiver tracking loops. Let $\Gamma_c(\cdot)$ denote the code autocorrelation function and define the excess delay of the $i$th reflected component in chips as 
\begin{equation} 
\Delta_i = \frac{\tau_i-\tau_0}{T_c}, 
\label{eq:delay_chip}
\end{equation} 
where $T_c$ is the code-chip period. Around the direct-path code delay, the complex correlator output can then be written as 
\begin{equation} 
Z(\epsilon) = g_0\Gamma_c(\epsilon) + \sum_{i=1}^{N} g_i e^{j\theta_i} \Gamma_c(\epsilon-\Delta_i), 
\label{eq:mp_correlator_model} 
\end{equation} 
where $\epsilon$ is the code-tracking error expressed in chips. The delayed and phase-shifted indirect components therefore deform the nominal correlation response observed by the receiver~\cite{braasch2017multipath}. Fig.~\ref{fig:code_carrier_mp_mechanism} illustrates this effect for a representative single-reflection case, highlighting both the displacement of the DLL discriminator zero crossing and the rotation of the prompt-correlator phasor.
For a conventional noncoherent early--late power Delay-Locked Loop (DLL) with early--late spacing $d$, the early-minus-late power (EMLP) discriminator is
\begin{equation} 
D_{\mathrm{DLL}}(\epsilon) = \left| Z\!\left(\epsilon+\frac{d}{2}\right) \right|^2 - \left| Z\!\left(\epsilon-\frac{d}{2}\right) \right|^2 . 
\label{eq:dll_discriminator_intro} 
\end{equation} 
For an undistorted LOS signal, the discriminator zero crossing coincides with the direct-path delay, corresponding to $\epsilon=0$. In the presence of multipath, the delayed reflected components distort the composite correlation response and shift the stable DLL tracking point to $\widehat{\epsilon}_{\mathrm{MP}}$, satisfying $D_{\mathrm{DLL}} \!\left(\widehat{\epsilon}_{\mathrm{MP}}\right)=0$. 
The corresponding code pseudorange error is therefore 
\begin{equation} 
\Delta\rho_{\mathrm{MP}} = cT_c\,\widehat{\epsilon}_{\mathrm{MP}}, 
\label{eq:code_mp_bias_intro} 
\end{equation} 
where $c$ is the speed of light. Thus, under multipath reception, the pseudorange bias is fundamentally a \emph{tracking-point displacement}: depending on the relative delay, amplitude, and carrier phase of the reflected components, interference can shift the DLL equilibrium in either direction and produce rapidly varying errors as the receiver or surrounding reflectors move.
The carrier-phase observable is affected by the same direct--reflected signal superposition through the phase of the complex prompt correlator. Taking the direct-path carrier phase as the reference and evaluating the prompt correlator at the DLL tracking point, the multipath-induced carrier-phase perturbation is
\begin{equation} 
\Delta\phi_{\mathrm{MP}} = \arg \left\{ Z\!\left( \widehat{\epsilon}_{\mathrm{MP}} \right) \right\}. 
\label{eq:carrier_mp_phase_intro} 
\end{equation} 
Expressed as an equivalent range perturbation, this becomes 
\begin{equation} 
\Delta\Phi_{\mathrm{MP}} = \frac{\lambda}{2\pi} \Delta\phi_{\mathrm{MP}}, 
\label{eq:carrier_mp_range_intro} 
\end{equation}
where $\lambda$ is the carrier wavelength. Because both code and carrier errors depend on the evolving multipath geometry, their biases can vary rapidly with receiver or reflector motion, degrading carrier-phase continuity, ambiguity resolution, and high-precision positioning.

NLOS reception differs fundamentally from multipath contamination. When the direct path is blocked, the receiver may instead acquire and track a reflected or diffracted component whose propagation path exceeds the unobstructed satellite--receiver range. This introduces a positive excess propagation delay and hence a positively biased pseudorange component associated with the indirect path~\cite{hsu2018analysis}. Its magnitude depends strongly on the surrounding geometry, including the satellite direction and the location of the obstructing or reflecting surfaces. NLOS signals may also be attenuated or waveform-distorted, further increasing measurement and tracking uncertainty.

Unlike satellite-clock, orbital, and atmospheric errors, for which established models and external corrections are available, multipath and NLOS effects are strongly coupled to the receiver's immediate propagation environment and cannot generally be represented by a universal deterministic correction~\cite{misra2006global,hsu2018analysis}. Reliable urban GNSS therefore requires estimating the reception state or measurement reliability so that contaminated observations can be excluded, reweighted, or corrected before they adversely affect the navigation solution. This motivates the signal-domain classifiers, 3D-map-based visibility methods, and camera, light detection and ranging (LiDAR), and IMU-aided approaches reviewed in the following sections.

From this perspective, the survey considers the satellite-specific tracked reception state
\begin{equation}
S
\in
\left\{
\mathrm{LOS},
\mathrm{Multipath},
\mathrm{NLOS}
\right\},
\label{eq}
\end{equation}
as a latent state inferred from heterogeneous evidence: signal-domain methods use receiver-internal measurements, map-based methods predict geometric visibility, and auxiliary sensors provide complementary spatial or kinematic information. Here, \emph{LOS} denotes nominal tracked reception, whereas \emph{Multipath} and \emph{NLOS} denote anomalous tracked states; complete geometric blockage with no trackable component is treated separately as signal unavailability. This state-based view underpins the survey taxonomy.

\subsection{Related Works and Identified Gaps}
\label{subsec:related_works}

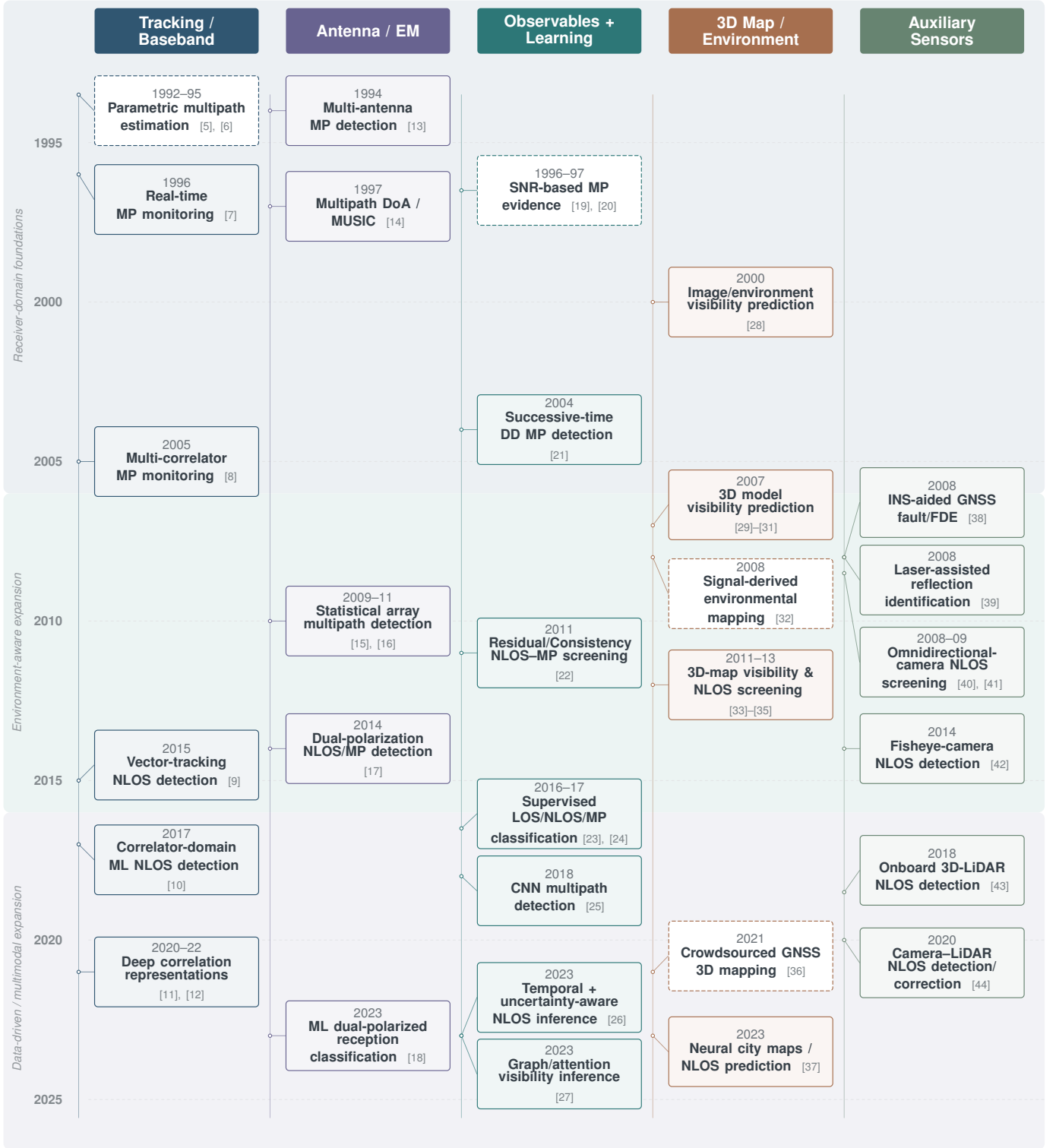
\begin{figure*}[!t]
\centering
\definecolor{timelineBlue}{RGB}{49,91,120}
\definecolor{timelineIndigo}{RGB}{113,106,156}
\definecolor{timelineTeal}{RGB}{46,128,127}
\definecolor{timelineAmber}{RGB}{183,119,82}
\definecolor{timelineSage}{RGB}{111,143,120}
\definecolor{timelineInk}{RGB}{41,51,59}

\resizebox{\textwidth}{!}{%
\begin{tikzpicture}[
    x=1.00cm,
    y=1.00cm,
    font=\sffamily,
    line cap=round,
    line join=round,
    fillA/.style={fill=timelineBlue!7},
    fillB/.style={fill=timelineIndigo!7},
    fillC/.style={fill=timelineTeal!7},
    fillD/.style={fill=timelineAmber!7},
    fillE/.style={fill=timelineSage!7},
    card/.style={
        rounded corners=2pt,
        line width=0.55pt,
        minimum width=3.60cm,
        minimum height=1.53cm,
        text width=3.20cm,
        align=center,
        inner xsep=3.2pt,
        inner ysep=1.7pt,
        font=\sffamily
    },
    direct/.style={card, solid},
    precursor/.style={
        card,
        dashed,
        dash pattern=on 2.2pt off 1.5pt,
        fill=white
    },
    colhdr/.style={
        rounded corners=2pt,
        minimum width=3.60cm,
        minimum height=1.0cm,
        text width=3.35cm,
        align=center,
        inner sep=3pt,
        font=\fontsize{10}{11.4}\selectfont\sffamily\bfseries,
        text=white
    },
    yearlbl/.style={
        font=\fontsize{8}{9.2}\selectfont\sffamily\bfseries,
        text=timelineInk!58,
        anchor=east
    },
    eralbl/.style={
        font=\fontsize{7.4}{8.5}\selectfont\sffamily\itshape,
        text=timelineInk!50,
        align=center,
        rotate=90
    },
    spine/.style={
        line width=0.45pt,
        opacity=0.42
    },
    tether/.style={
        line width=0.35pt,
        opacity=0.66
    },
    yearguide/.style={
        draw=timelineInk!13,
        line width=0.35pt,
        dash pattern=on 1.5pt off 2.0pt
    }
]

\def\xA{0}
\def\xB{4.2}
\def\xC{8.4}
\def\xD{12.6}
\def\xE{16.8}
\def\sA{-2.15}
\def\sB{2.05}
\def\sC{6.25}
\def\sD{10.45}
\def\sE{14.65}
\def\yearStart{1994}
\def\yearTop{-1.50}
\def\yearStep{0.70}
\def\histY#1{\yearTop-\yearStep*(#1-\yearStart)}

\def\histyear#1{{%
  \fontsize{8.0}{9.0}\selectfont
  \color{timelineInk!68}#1}}
\def\histmethod#1{{%
  \fontsize{8.6}{8.9}\selectfont
  \bfseries\color{timelineInk!94}#1}}
\def\histcite#1{{%
  \fontsize{6.8}{7.5}\selectfont
  \mdseries\color{timelineInk!62}%
  \NoHyper#1\endNoHyper}}

\newcommand{\histcardtext}[2]{%
  \begin{minipage}[c][1.40cm][c]{3.10cm}%
    \centering
    \setlength{\parskip}{0pt}%
    \histyear{#1}\par
    \histmethod{#2}\par
  \end{minipage}%
}

\def\histtether#1#2{%
  \draw[tether, draw=#1] (#2anchor) -- (#2.west);
  \filldraw[
    fill=white,
    draw=#1,
    line width=0.40pt
  ] (#2anchor) circle (1.05pt);
}

\begin{scope}[on background layer]
  \fill[
    timelineBlue!9,
    rounded corners=4pt
  ] (-3.80,0.92) rectangle (19.05,-9.90);

  \fill[
    timelineTeal!9,
    rounded corners=4pt
  ] (-3.80,-9.90) rectangle (19.05,-16.90);

  \fill[
    timelineIndigo!9,
    rounded corners=4pt
  ] (-3.80,-16.90) rectangle (19.05,-24.30);
\end{scope}

\node[eralbl] at (-3.48,-5.35)
  {Receiver-domain foundations};
\node[eralbl] at (-3.48,-13.40)
  {Environment-aware expansion};
\node[eralbl] at (-3.48,-20.55)
  {Data-driven / multimodal expansion};

\node[
  colhdr,
  fill=timelineBlue!88!timelineInk
] at (\xA,0.25) {Tracking /\\[-1pt]Baseband};

\node[
  colhdr,
  fill=timelineIndigo!88!timelineInk
] at (\xB,0.25) {Antenna / EM};

\node[
  colhdr,
  fill=timelineTeal!88!timelineInk
] at (\xC,0.25) {Observables +\\[-1pt]Learning};

\node[
  colhdr,
  fill=timelineAmber!88!timelineInk
] at (\xD,0.25) {3D Map /\\[-1pt]Environment};

\node[
  colhdr,
  fill=timelineSage!88!timelineInk
] at (\xE,0.25) {Auxiliary\\[-1pt]Sensors};

\foreach \yy/\lbl in {
    1995/1995,
    2000/2000,
    2005/2005,
    2010/2010,
    2015/2015,
    2020/2020,
    2025/2025%
}{
    \draw[
      yearguide
    ] (-1.93,{\histY{\yy}})
      -- (19.05,{\histY{\yy}});

    \node[
      yearlbl
    ] at (-2.38,{\histY{\yy}}) {\lbl};
}

\draw[
  spine,
  draw=timelineBlue!88!timelineInk
] (\sA,-1.15) -- (\sA,-23.60);

\draw[
  spine,
  draw=timelineIndigo!88!timelineInk
] (\sB,-1.15) -- (\sB,-23.60);

\draw[
  spine,
  draw=timelineTeal!88!timelineInk
] (\sC,-1.15) -- (\sC,-23.60);

\draw[
  spine,
  draw=timelineAmber!88!timelineInk
] (\sD,-1.15) -- (\sD,-23.60);

\draw[
  spine,
  draw=timelineSage!88!timelineInk
] (\sE,-1.15) -- (\sE,-23.60);

\coordinate (trk1anchor) at (\sA,{\histY{1993.5}});
\node[
  precursor,
  draw=timelineBlue!88!timelineInk
] (trk1) at (\xA,{\histY{1993.5}-0.35}) {%
  \histcardtext{1992--95}{%
    Parametric multipath\\[-1.0pt]%
    estimation~\histcite{%
      \cite{vannee1992medll,townsend1995performance}}}};

\coordinate (trk2anchor) at (\sA,{\histY{1996}});
\node[
  direct,
  fillA,
  draw=timelineBlue!88!timelineInk
] (trk2) at (\xA,{\histY{1996}-0.55}) {%
  \histcardtext{1996}{%
    Real-time\\[-1.0pt]%
    MP monitoring~\histcite{\cite{itani1996realtime}}}};

\coordinate (trk3anchor) at (\sA,{\histY{2005}});
\node[
  direct,
  fillA,
  draw=timelineBlue!88!timelineInk
] (trk3) at (\xA,{\histY{2005}}) {%
  \histcardtext{2005}{%
    Multi-correlator\\[-1.0pt]%
    MP monitoring~\histcite{\cite{irsigler2005realtime}}}};

\coordinate (trk4anchor) at (\sA,{\histY{2015}});
\node[
  direct,
  fillA,
  draw=timelineBlue!88!timelineInk
] (trk4) at (\xA,{\histY{2015}+0.34}) {%
  \histcardtext{2015}{%
    Vector-tracking\\[-1.0pt]%
    NLOS detection~\histcite{\cite{hsu2015multipath}}}};

\coordinate (trk5anchor) at (\sA,{\histY{2017}});
\node[
  direct,
  fillA,
  draw=timelineBlue!88!timelineInk
] (trk5) at (\xA,{\histY{2017}-0.34}) {%
  \histcardtext{2017}{%
    Correlator-domain\\[-1.0pt]%
    ML NLOS detection~\histcite{\cite{suzuki2017nlos}}}};

\coordinate (trk6anchor) at (\sA,{\histY{2021}});
\node[
  direct,
  fillA,
  draw=timelineBlue!88!timelineInk
] (trk6) at (\xA,{\histY{2021}}) {%
  \histcardtext{2020--22}{%
    Deep correlation\\[-1.0pt]%
    representations\\[-1.0pt]%
    ~\histcite{\cite{suzuki2020nlos,blais2022novel}}}};

\histtether{timelineBlue!88!timelineInk}{trk1}
\histtether{timelineBlue!88!timelineInk}{trk2}
\histtether{timelineBlue!88!timelineInk}{trk3}
\histtether{timelineBlue!88!timelineInk}{trk4}
\histtether{timelineBlue!88!timelineInk}{trk5}
\histtether{timelineBlue!88!timelineInk}{trk6}

\coordinate (ant1anchor) at (\sB,{\histY{1994}});
\node[
  direct,
  fillB,
  draw=timelineIndigo!88!timelineInk
] (ant1) at (\xB,{\histY{1994}}) {%
  \histcardtext{1994}{%
    Multi-antenna\\[-1.0pt]%
    MP detection~\histcite{\cite{becker1994special}}}};

\coordinate (ant2anchor) at (\sB,{\histY{1997}});
\node[
  direct,
  fillB,
  draw=timelineIndigo!88!timelineInk
] (ant2) at (\xB,{\histY{1997}}) {%
  \histcardtext{1997}{%
    Multipath DoA /\\[-1.0pt]%
    MUSIC~\histcite{\cite{moelker1997multiple}}}};

\coordinate (ant3anchor) at (\sB,{\histY{2010}});
\node[
  direct,
  fillB,
  draw=timelineIndigo!88!timelineInk
] (ant3) at (\xB,{\histY{2010}}) {%
  \histcardtext{2009--11}{%
    Statistical array\\[-1.0pt]%
    multipath detection\\[-1.0pt]%
    ~\histcite{%
      \cite{fernandezprades2009statistical,closas2011statistical}}}};

\coordinate (ant4anchor) at (\sB,{\histY{2014}});
\node[
  direct,
  fillB,
  draw=timelineIndigo!88!timelineInk
] (ant4) at (\xB,{\histY{2014}}) {%
  \histcardtext{2014}{%
    Dual-polarization\\[-1.0pt]%
    NLOS/MP detection\\[-1.0pt]%
    ~\histcite{\cite{jiang2014nlos}}}};

\coordinate (ant5anchor) at (\sB,{\histY{2023}});
\node[
  direct,
  fillB,
  draw=timelineIndigo!88!timelineInk
] (ant5) at (\xB,{\histY{2023}}) {%
  \histcardtext{2023}{%
    ML dual-polarized\\[-1.0pt]%
    reception\\[-1.0pt]%
    classification~\histcite{\cite{kim2023machine}}}};

\histtether{timelineIndigo!88!timelineInk}{ant1}
\histtether{timelineIndigo!88!timelineInk}{ant2}
\histtether{timelineIndigo!88!timelineInk}{ant3}
\histtether{timelineIndigo!88!timelineInk}{ant4}
\histtether{timelineIndigo!88!timelineInk}{ant5}

\coordinate (obs1anchor) at (\sC,{\histY{1996.5}});
\node[
  precursor,
  draw=timelineTeal!88!timelineInk
] (obs1) at (\xC,{\histY{1996.5}}) {%
  \histcardtext{1996--97}{%
    SNR-based MP\\[-1.0pt]%
    evidence~\histcite{\cite{comp1996adaptive,sleewaegen1997snr}}}};

\coordinate (obs2anchor) at (\sC,{\histY{2004}});
\node[
  direct,
  fillC,
  draw=timelineTeal!88!timelineInk
] (obs2) at (\xC,{\histY{2004}}) {%
  \histcardtext{2004}{%
    Successive-time\\[-1.0pt]%
    DD MP detection~\histcite{\cite{lee2004gpsmultipathdd}}}};

\coordinate (obs3anchor) at (\sC,{\histY{2011}});
\node[
  direct,
  fillC,
  draw=timelineTeal!88!timelineInk
] (obs3) at (\xC,{\histY{2011}}) {%
  \histcardtext{2011}{%
    Residual/Consistency\\[-1.0pt]%
    NLOS--MP screening\\[-1.0pt]%
    ~\histcite{\cite{jiang2011consistency}}}};

\coordinate (obs4anchor) at (\sC,{\histY{2016.5}});
\node[
  direct,
  fillC,
  draw=timelineTeal!88!timelineInk
] (obs4) at (\xC,{\histY{2016.5}+0.32}) {%
  \histcardtext{2016--17}{%
    Supervised\\[1.0pt]%
    LOS/NLOS/MP\\[1.0pt]%
    classification\histcite{\cite{yozevitch2016robust,hsu2017gnss}}}};

\coordinate (obs5anchor) at (\sC,{\histY{2018}});
\node[
  direct,
  fillC,
  draw=timelineTeal!88!timelineInk
] (obs5) at (\xC,{\histY{2018}-0.32}) {%
  \histcardtext{2018}{%
    CNN multipath\\[-1.0pt]%
    detection~\histcite{\cite{quan2018convolutional}}}};

\coordinate (obs6anchor) at (\sC,{\histY{2023}});
\node[
  direct,
  fillC,
  draw=timelineTeal!88!timelineInk
] (obs6) at (\xC,{\histY{2023}+0.84}) {%
  \histcardtext{2023}{%
    Temporal +\\[-1.0pt]%
    uncertainty-aware\\[-1.0pt]%
    NLOS inference~\histcite{\cite{zhang2023learning}}}};

\coordinate (obs7anchor) at (\sC,{\histY{2023}});
\node[
  direct,
  fillC,
  draw=timelineTeal!88!timelineInk
] (obs7) at (\xC,{\histY{2023}-0.84}) {%
  \histcardtext{2023}{%
    Graph/attention\\[-1.0pt]%
    visibility inference\\[-1.0pt]%
    ~\histcite{\cite{zheng2023graph}}}};

\histtether{timelineTeal!88!timelineInk}{obs1}
\histtether{timelineTeal!88!timelineInk}{obs2}
\histtether{timelineTeal!88!timelineInk}{obs3}
\histtether{timelineTeal!88!timelineInk}{obs4}
\histtether{timelineTeal!88!timelineInk}{obs5}
\histtether{timelineTeal!88!timelineInk}{obs6}
\histtether{timelineTeal!88!timelineInk}{obs7}

\coordinate (map1anchor) at (\sD,{\histY{2000}});
\node[
  direct,
  fillD,
  draw=timelineAmber!88!timelineInk
] (map1) at (\xD,{\histY{2000}}) {%
  \histcardtext{2000}{%
    Image/environment\\[-1.0pt]%
    visibility prediction\\[-1.0pt]%
    ~\histcite{\cite{marais2000evaluation}}}};

\coordinate (map2anchor) at (\sD,{\histY{2007}});
\node[
  direct,
  fillD,
  draw=timelineAmber!88!timelineInk
] (map2) at (\xD,{\histY{2007}+0.45}) {%
  \histcardtext{2007}{%
    3D model\\[-1.0pt]%
    visibility prediction\\[-1.0pt]%
    ~\histcite{\cite{bradbury2007code,suh2007evaluation,taylor2007modelling}}}};

\coordinate (map3anchor) at (\sD,{\histY{2008}});
\node[
  precursor,
  draw=timelineAmber!88!timelineInk
] (map3) at (\xD,{\histY{2008}-0.80}) {%
  \histcardtext{2008}{%
    Signal-derived\\[-1.0pt]%
    environmental mapping~\histcite{\cite{kim2008localization}}}};

\coordinate (map4anchor) at (\sD,{\histY{2012}});
\node[
  direct,
  fillD,
  draw=timelineAmber!88!timelineInk
] (map4) at (\xD,{\histY{2012}}) {%
  \histcardtext{2011--13}{%
    3D-map visibility \&\\[-1.0pt]%
    NLOS screening~\histcite{%
      \cite{groves2011shadow,obst2012urban,peyraud2013non}}}};

\coordinate (map5anchor) at (\sD,{\histY{2021}});
\node[
  precursor,
  draw=timelineAmber!88!timelineInk
] (map5) at (\xD,{\histY{2021}+0.35}) {%
  \histcardtext{2021}{%
    Crowdsourced GNSS\\[-1.0pt]%
    3D mapping~\histcite{\cite{lines20213d}}}};

\coordinate (map6anchor) at (\sD,{\histY{2023}});
\node[
  direct,
  fillD,
  draw=timelineAmber!88!timelineInk
] (map6) at (\xD,{\histY{2023}-0.35}) {%
  \histcardtext{2023}{%
    Neural city maps /\\[-1.0pt]%
    NLOS prediction~\histcite{\cite{neamati2023neural}}}};

\histtether{timelineAmber!88!timelineInk}{map1}
\histtether{timelineAmber!88!timelineInk}{map2}
\histtether{timelineAmber!88!timelineInk}{map3}
\histtether{timelineAmber!88!timelineInk}{map4}
\histtether{timelineAmber!88!timelineInk}{map5}
\histtether{timelineAmber!88!timelineInk}{map6}

\coordinate (aux1anchor) at (\sE,{\histY{2008}});
\node[
  direct,
  fillE,
  draw=timelineSage!88!timelineInk
] (aux1) at (\xE,{\histY{2008}+1.2}) {%
  \histcardtext{2008}{%
    INS-aided GNSS\\[-1.0pt]%
    fault/FDE~\histcite{\cite{clark2008gpsins}}}};

\coordinate (aux2anchor) at (\sE,{\histY{2008}});
\node[
  direct,
  fillE,
  draw=timelineSage!88!timelineInk
] (aux2) at (\xE,{\histY{2008}-0.50}) {%
  \histcardtext{2008}{%
    Laser-assisted reflection\\[-1.0pt]%
    identification~\histcite{\cite{soloviev2008multipath}}}};

\coordinate (aux3anchor) at (\sE,{\histY{2008.5}});
\node[
  direct,
  fillE,
  draw=timelineSage!88!timelineInk
] (aux3) at (\xE,{\histY{2008.5}-1.95}) {%
  \histcardtext{2008--09}{%
    Omnidirectional-\\[-1.0pt]%
    camera NLOS\\[-1.0pt]%
    screening~\histcite{\cite{meguro2008development,meguro2009gps}}}};

\coordinate (aux4anchor) at (\sE,{\histY{2014}});
\node[
  direct,
  fillE,
  draw=timelineSage!88!timelineInk
] (aux4) at (\xE,{\histY{2014}}) {%
  \histcardtext{2014}{%
    Fisheye-camera\\[-1.0pt]%
    NLOS detection~\histcite{\cite{suzuki2014n}}}};

\coordinate (aux5anchor) at (\sE,{\histY{2018.5}});
\node[
  direct,
  fillE,
  draw=timelineSage!88!timelineInk
] (aux5) at (\xE,{\histY{2018.5}+0.48}) {%
  \histcardtext{2018}{%
    Onboard 3D-LiDAR\\[-1.0pt]%
    NLOS detection~\histcite{\cite{wen2018correcting}}}};

\coordinate (aux6anchor) at (\sE,{\histY{2020}});
\node[
  direct,
  fillE,
  draw=timelineSage!88!timelineInk
] (aux6) at (\xE,{\histY{2020}-0.50}) {%
  \histcardtext{2020}{%
    Camera--LiDAR\\[-1.0pt]%
    NLOS detection/ \\[-1.0pt]%
    correction~\histcite{\cite{bai2020using}}}};

\histtether{timelineSage!88!timelineInk}{aux1}
\histtether{timelineSage!88!timelineInk}{aux2}
\histtether{timelineSage!88!timelineInk}{aux3}
\histtether{timelineSage!88!timelineInk}{aux4}
\histtether{timelineSage!88!timelineInk}{aux5}
\histtether{timelineSage!88!timelineInk}{aux6}

\end{tikzpicture}%
}

\caption{Historical evolution of GNSS multipath/NLOS reception-state detection and satellite-visibility inference. Solid boxes denote direct methods, while dashed boxes denote enabling developments.}

\label{fig:historical_timeline}
\end{figure*}

\begin{table*}[!t]
\centering
\caption{Comparison of representative reviews and overview works relevant to GNSS LOS/NLOS/multipath detection and satellite-visibility determination.}
\label{tab:survey_comparison}
\renewcommand{\arraystretch}{1.3}
\setlength{\tabcolsep}{4.5pt}
\begin{adjustbox}{max width=\textwidth}
\begin{tabular}{@{} l c c c c c c c @{}}
\toprule
\rowcolor{gray!30}
& \textbf{Signal} & \textbf{Classical/} & \textbf{Deep} & \textbf{3D Map /} & \textbf{Aux. Sensors} & \textbf{Datasets} & \textbf{Geometric} \\
\rowcolor{gray!30}
\multirow{-2}{*}{\textbf{Review / Overview}} & \textbf{Features} & \textbf{Statistical/ML} & \textbf{Learning} & \textbf{Env. Modeling} & \textbf{(Vision/LiDAR/IMU)} & \textbf{\& Metrics} & \textbf{Visibility} \\
\midrule

\textbf{Bre{\ss}ler} et al. (2016)~\cite{bressler2016gnss} & \checkmark & \checkmark & -- & Partial & Partial & Partial & Partial \\

\textbf{McGraw} et al. (2020)~\cite{mcgraw2020robust} & \checkmark & Partial & -- & Partial & Partial & -- & Partial \\

\textbf{Lau} et al. (2021)~\cite{lau2021gnss} & \checkmark & -- & -- & -- & -- & -- & -- \\

\textbf{Pereira} et al. (2025)~\cite{pereira2025advancements}
& \checkmark & Partial & -- & -- & -- & -- & -- \\

\addlinespace[4pt]

\textbf{Siemuri} et al. (2022)~\cite{siemuri2022systematic} & Partial & \checkmark & \checkmark & -- & -- & Partial & -- \\

\textbf{Xu} et al. (2024)~\cite{xu2024machine} & \checkmark & \checkmark & \checkmark & -- & -- & \checkmark & -- \\

\textbf{Mohanty and Gao} (2024)~\cite{mohanty2024survey}
& Partial & \checkmark & \checkmark & -- & Partial & Partial & -- \\

\textbf{Boumaraf} et al. (2026)~\cite{boumaraf2026advancing}
& Partial & \checkmark & \checkmark & -- & Partial & \checkmark & -- \\

\addlinespace[4pt]

\textbf{Groves} et al. (2019)~\cite{groves2019performance} & Partial & -- & -- & \checkmark & -- & -- & \checkmark \\

\textbf{Feriol} et al. (2020)~\cite{feriol2020review} & Partial & Partial & -- & Partial & \checkmark & -- & Partial \\

\textbf{Wei} et al. (2025)~\cite{wei2025high} & Partial & Partial & Partial & -- & \checkmark & -- & -- \\

\rowcolor{gray!15}
\textbf{This survey} &
\textcolor{blue}{\checkmark} &
\textcolor{blue}{\checkmark} &
\textcolor{blue}{\checkmark} &
\textcolor{blue}{\checkmark} &
\textcolor{blue}{\checkmark} &
\textcolor{blue}{\checkmark} &
\textcolor{blue}{\checkmark} \\
\midrule
\multicolumn{8}{l}{\footnotesize \textit{Legend:} \checkmark = Substantial coverage; \quad Partial = Limited or adjacent coverage; \quad -- = Not a main focus.} \\
\bottomrule
\end{tabular}
\end{adjustbox}
\end{table*}

A wide range of reviews and overview works have addressed GNSS positioning degradation in urban environments, with particular attention to multipath and NLOS reception. We consider this literature from two complementary perspectives: the methodological scope covered by existing reviews and the historical evolution of the underlying reception-state detection and satellite-visibility paradigms. Existing review coverage remains uneven across methodological domains: some works emphasize receiver- and signal-level processing, whereas others focus primarily on machine learning, environmental modeling, or multisensor integration. Even broader reviews that span several of these topics generally do not organize them under a common per-satellite reception-state and visibility framework, making cross-domain comparison difficult.
Representative receiver- and signal-oriented reviews discuss the physical behavior of multipath and NLOS together with signal-level indicators, tracking techniques, and mitigation strategies. Bre{\ss}ler et al.~\cite{bressler2016gnss} provide an early broad overview of GNSS degradation in challenging environments, covering observable-level indicators, environmental information, auxiliary sensing, and outlier-detection concepts. From an integrity perspective, Zhu et al.~\cite{zhu2018gnss} review the challenges of reliable GNSS positioning in urban environments, with particular attention to propagation-induced errors and the limitations of conventional integrity-monitoring assumptions. McGraw et al.~\cite{mcgraw2020robust} similarly discuss multipath and NLOS in the context of receiver processing, antenna techniques, robust positioning, sensor aiding, and 3D-mapping-aided GNSS. Lau et al.~\cite{lau2021gnss} focus more specifically on multipath in high-precision GNSS, while Pereira et al.~\cite{pereira2025advancements} review correlation- and channel-estimation-based multipath mitigation at the receiver level. These works provide important physical and systems-level insight, but they do not systematically connect receiver-domain evidence with modern learning-based reception-state inference and real-time environmental perception.

A second group focuses on machine learning and deep learning in GNSS. Siemuri et al.~\cite{siemuri2022systematic} provide a broad systematic review of machine learning (ML) applications across GNSS, whereas Xu et al.~\cite{xu2024machine} concentrate specifically on ML-based multipath/NLOS mitigation and provide an associated dataset and benchmark perspective. Mohanty and Gao~\cite{mohanty2024survey} further survey classical and deep learning techniques across GNSS signal analysis, positioning, anomaly detection, and multisensor integration, while Boumaraf et al.~\cite{boumaraf2026advancing} provide a recent systematic view of traditional, deep, and emerging ML paradigms together with commonly used GNSS datasets and deployment challenges. These studies provide extensive coverage of data-driven inference, but do not systematically integrate geometric visibility prediction, 3D environmental representations, and onboard camera/LiDAR screening within the same taxonomy.

A complementary body of work addresses environment-aware and 3D-mapping-aided GNSS. Groves et al.~\cite{groves2019performance} provide a foundational review and performance assessment of 3D-mapping-aided GNSS, demonstrating how building models support satellite-visibility prediction and urban positioning. Such work establishes the value of environmental geometry for predicting LOS blockage but generally places less emphasis on the evolution of the underlying mapping representations themselves. In contrast, the present survey also traces the progression of mapping modalities used for GNSS visibility determination, from conventional building models and LiDAR- or photogrammetry-derived maps to crowdsourced signal-based representations and more recent learned implicit environmental models. This broader view connects advances in environmental representation with the corresponding development of visibility-prediction methods and signal-domain classification.

Auxiliary-sensor reviews form an adjacent but distinct literature. Feriol et al.~\cite{feriol2020review} review environmental-context detection using GNSS and onboard vision, while broader multisensor reviews such as Wei et al.~\cite{wei2025high} examine integration with visual, LiDAR, and inertial sensing. McGraw et al.~\cite{mcgraw2020robust} also discuss sensor aiding within a broader framework for robust GNSS positioning. These sensors are typically treated as aids for improving navigation robustness. In contrast, the present survey emphasizes their role in reception-state verification: cameras provide angular sky/occlusion information, LiDAR supplies metric 3D geometry for blockage and reflected-path reasoning, and IMUs provide kinematic and attitude consistency information.

Taken together, the reviewed literature shows a progressive broadening of the information used to determine GNSS reception state. Initial approaches relied mainly on receiver-internal evidence, including tracking-loop behavior, antenna-domain spatial signatures, and observable-level quality or consistency measures. This receiver-centric view was later complemented by external knowledge of the surrounding environment through 3D models, cameras, LiDAR, and inertial sensing, providing geometric and kinematic cues that are largely independent of the GNSS measurements themselves. More recent methods increasingly exploit temporal and inter-satellite dependencies, learned environmental representations, and multimodal combinations of these complementary information sources. Figure~\ref{fig:historical_timeline} places representative developments from these methodological families in their historical context, while distinguishing methods that directly infer multipath, NLOS, or satellite visibility from foundational techniques that enabled subsequent reception-state detection.

Dataset-oriented discussion remains comparatively sparse. Xu et al.~\cite{xu2024machine} provide a public dataset and benchmark-oriented evaluation for ML-based multipath/NLOS mitigation, while Boumaraf et al.~\cite{boumaraf2026advancing} summarize datasets used across data-driven GNSS research. Nevertheless, there remains no unified benchmark that compares signal-level, map-based, camera-based, and LiDAR-based reception-state methods under common labels, receiver configurations, and urban scenarios. Accordingly, the present survey treats datasets, labeling strategies, and evaluation protocols as a distinct methodological component, without claiming to introduce a new quantitative cross-method benchmark.

Table~\ref{tab:survey_comparison} summarizes the coverage of representative reviews and overview works across seven methodological dimensions. 
Overall, the literature demonstrates substantial progress within and across individual methodological families, but no prior review considered here systematically unifies receiver-domain signal evidence, modern learning-based reception-state inference, 3D-map-based visibility determination, and onboard camera/LiDAR/IMU verification under a common per-satellite reception-state framework while also treating datasets and labeling as a dedicated methodological problem. This motivates the unified cross-domain synthesis developed in this survey.

\subsection{Contributions and Survey Structure}

\begin{figure*}[!t]
\centering
\begin{adjustbox}{max width=\textwidth}
\begin{tikzpicture}[
    font=\sffamily,
    rowtext/.style={
        align=left,
        anchor=west,
        text width=4.78cm,
        inner sep=0pt,
        font=\sffamily\fontsize{8}{9.5}\selectfont,
        text=surveyInk
    },
    headertext/.style={
        align=center,
        text width=4.88cm,
        inner sep=0pt,
        text=white,
        font=\sffamily\bfseries\footnotesize
    },
    calloutwire/.style={
        dash pattern=on 0.8pt off 1.7pt,
        line width=0.65pt
    },
    calloutarrow/.style={
        -{Latex[length=1.7mm,width=1.15mm]},
        dash pattern=on 0.8pt off 1.7pt,
        line width=0.65pt
    },
    flow/.style={
        -{Latex[length=1.9mm,width=1.3mm]},
        line width=0.75pt
    },
    calloutlabel/.style={
        fill=white,
        inner xsep=2pt,
        inner ysep=1pt,
        font=\sffamily\fontsize{7.5}{8.5}\selectfont,
        text=surveySlate
    },
    outcome/.style={
        rectangle,
        rounded corners=3pt,
        draw=surveyNavy,
        fill=surveyNavy,
        text=white,
        line width=0.75pt,
        minimum width=17.55cm,
        minimum height=1.48cm,
        text width=16.85cm,
        align=center,
        inner sep=5pt
    },
    synthesis/.style={
        rectangle,
        rounded corners=3pt,
        draw=surveySlate!70,
        fill=surveySlate!7,
        text=surveyInk,
        line width=0.65pt,
        minimum width=17.55cm,
        minimum height=1.42cm,
        text width=16.85cm,
        align=center,
        inner sep=5pt
    }
]

\definecolor{surveyNavy}{RGB}{35,55,74}
\definecolor{surveyBlue}{RGB}{31,90,122}
\definecolor{surveyTeal}{RGB}{22,115,106}
\definecolor{surveyOrange}{RGB}{165,75,22}
\definecolor{surveySlate}{RGB}{80,85,99}
\definecolor{surveyInk}{RGB}{40,46,52}

\begin{scope}

\path[
    rounded corners=4pt,
    fill=surveyBlue!4,
    draw=surveyBlue!75,
    line width=0.7pt
] (0,0) rectangle (5.42,-7.70);

\begin{scope}
\clip[rounded corners=4pt] (0,0) rectangle (5.42,-7.70);
\fill[surveyBlue] (0,0) rectangle (5.42,-1.10);
\end{scope}

\draw[surveyBlue!65, line width=0.5pt]
    (0,-1.10) -- (5.42,-1.10);
\draw[surveyBlue!38, line width=0.45pt]
    (0.18,-2.75) -- (5.24,-2.75);
\draw[surveyBlue!38, line width=0.45pt]
    (0.18,-4.40) -- (5.24,-4.40);
\draw[surveyBlue!38, line width=0.45pt]
    (0.18,-6.05) -- (5.24,-6.05);

\node[headertext] (receiverHead) at (2.71,-0.55) {
    RECEIVER-SIDE GNSS EVIDENCE\\[-0.12ex]
    {\mdseries Sections~\ref{sec:features}--\ref{sec:algorithms}}
};

\node[rowtext] at (0.33,-1.93) {
    {\bfseries Antenna/EM}\\[-0.12ex]
    DoA, polarization, spatial covariance;\\[-0.12ex]
    multi-antenna phase/geometry
};

\node[rowtext] at (0.33,-3.58) {
    {\bfseries Tracking/baseband}\\[-0.12ex]
    correlation distortion and SQM;\\[-0.12ex]
    multi-correlator/channel-response cues
};

\node[rowtext] at (0.33,-5.23) {
    {\bfseries Observables}\\[-0.12ex]
    C/N$_0$, residuals/innovations, CMC;\\[-0.12ex]
    Doppler and temporal behavior
};

\node[rowtext] (receiverInference) at (0.33,-6.88) {
    {\bfseries Inference + data}\\[-0.12ex]
    statistical tests; ML models;\\[-0.12ex]
    datasets and visibility labeling
};

\end{scope}

\begin{scope}

\path[
    rounded corners=4pt,
    fill=surveyTeal!4,
    draw=surveyTeal!75,
    line width=0.7pt
] (6.065,0) rectangle (11.485,-7.70);

\begin{scope}
\clip[rounded corners=4pt] (6.065,0) rectangle (11.485,-7.70);
\fill[surveyTeal] (6.065,0) rectangle (11.485,-1.10);
\end{scope}

\draw[surveyTeal!65, line width=0.5pt]
    (6.065,-1.10) -- (11.485,-1.10);
\draw[surveyTeal!38, line width=0.45pt]
    (6.245,-3.30) -- (11.305,-3.30);
\draw[surveyTeal!38, line width=0.45pt]
    (6.245,-5.50) -- (11.305,-5.50);

\node[headertext] (mapHead) at (8.775,-0.55) {
    3D MAP / ENVIRONMENT EVIDENCE\\[-0.12ex]
    {\mdseries Section~\ref{sec:3d_mapping}}
};

\node[rowtext] at (6.395,-2.20) {
    {\bfseries Environmental representations}\\[-0.12ex]
    3D city and reconstructed maps;\\[-0.12ex]
    neural implicit and crowdsourced maps
};

\node[rowtext] at (6.395,-4.40) {
    {\bfseries A priori spatial predictors}\\[-0.12ex]
    geometric screening/sky masks;\\[-0.12ex]
    neural visibility fields; 3D ray tracing
};

\node[rowtext] at (6.395,-6.60) {
    {\bfseries Reception-state verifiers}\\[-0.12ex]
    shadow matching and 3DMA ranging;\\[-0.12ex]
    live GNSS--map consistency
};

\end{scope}

\begin{scope}

\path[
    rounded corners=4pt,
    fill=surveyOrange!4,
    draw=surveyOrange!80,
    line width=0.7pt
] (12.13,0) rectangle (17.55,-7.70);

\begin{scope}
\clip[rounded corners=4pt] (12.13,0) rectangle (17.55,-7.70);
\fill[surveyOrange] (12.13,0) rectangle (17.55,-1.10);
\end{scope}

\draw[surveyOrange!65, line width=0.5pt]
    (12.13,-1.10) -- (17.55,-1.10);
\draw[surveyOrange!38, line width=0.45pt]
    (12.31,-2.73) -- (17.37,-2.73);
\draw[surveyOrange!38, line width=0.45pt]
    (12.31,-4.36) -- (17.37,-4.36);
\draw[surveyOrange!38, line width=0.45pt]
    (12.31,-5.99) -- (17.37,-5.99);

\node[headertext] (sensorHead) at (14.84,-0.55) {
    AUXILIARY SENSORS\\[-0.12ex]
    {\mdseries Section~\ref{sec:auxiliary_sensors}}
};

\node[rowtext] at (12.46,-1.92) {
    {\bfseries IMU}\\[-0.12ex]
    motion propagation and\\[-0.12ex]
    kinematic consistency
};

\node[rowtext] at (12.46,-3.55) {
    {\bfseries Camera}\\[-0.12ex]
    sky segmentation and\\[-0.12ex]
    satellite/occluder visibility
};

\node[rowtext] at (12.46,-5.18) {
    {\bfseries LiDAR}\\[-0.12ex]
    local 3D geometry and\\[-0.12ex]
    ray-based occlusion testing
};

\node[
    rowtext,
    anchor=center,
    align=center,
    text width=4.92cm,
    font=\sffamily\itshape\fontsize{8}{9.5}\selectfont
] at (14.84,-6.85) {
    real-time environmental and\\[-0.12ex]
    kinematic validation
};

\end{scope}


\draw[calloutwire, draw=black!80]
    (8.775,0) -- (8.775,0.42)
    -- node[calloutlabel, above] {offline labels / supervision}
    (5.73,0.42);

\draw[calloutwire, draw=black!80]
    (14.84,0) -- (14.84,0.42) -- (8.775,0.42);

\draw[calloutarrow, draw=black!80]
    (5.73,0.42) -- (5.73,-6.88) -- (receiverInference.east);


\node[outcome] (outcome) at (8.775,-9.12) {
    {\footnotesize\bfseries
    Per-satellite reception state + confidence}\\[-0.02ex]
    {\footnotesize
    LOS \enspace\textbar\enspace
    Multipath-contaminated \enspace\textbar\enspace
    NLOS}\\[-0.02ex]
    {\footnotesize
    retain \enspace\textbar\enspace
    down-weight \enspace\textbar\enspace
    exclude \enspace\textbar\enspace
    correct}
};

\draw[flow, draw=surveyBlue]
    (2.71,-7.70) -- (2.71,-8.38);

\draw[calloutarrow, draw=surveyTeal!85]
    (8.775,-7.70) -- (8.775,-8.38);

\draw[calloutarrow, draw=surveyOrange!85]
    (14.84,-7.70) -- (14.84,-8.38);


\node[synthesis] (synthesis) at (8.775,-11.18) {
    {\bfseries Comparative Insights, Design Trade-offs, and Open Directions}\\[-0.02ex]
    {\fontsize{8}{9.5}\selectfont
    Sections~\ref{sec:tradeoffs} and~\ref{sec:future}}\\[0.10ex]
    {\fontsize{7.8}{9.2}\selectfont
evidence quality \enspace\textbullet\enspace
system complexity \enspace\textbullet\enspace
generalization \enspace\textbullet\enspace
uncertainty \enspace\textbullet\enspace
map/data reliability}
};

\draw[flow, draw=surveySlate!85]
    (outcome.south) -- (synthesis.north);

\end{tikzpicture}
\end{adjustbox}
\caption{Organization and information flow of the survey. Receiver-side GNSS evidence, map-derived environmental information, and auxiliary sensing provide complementary inputs for per-satellite reception-state inference and measurement handling. The subsequent comparative synthesis evaluates their practical trade-offs and motivates the open research directions.}
\label{fig:sections}
\end{figure*}
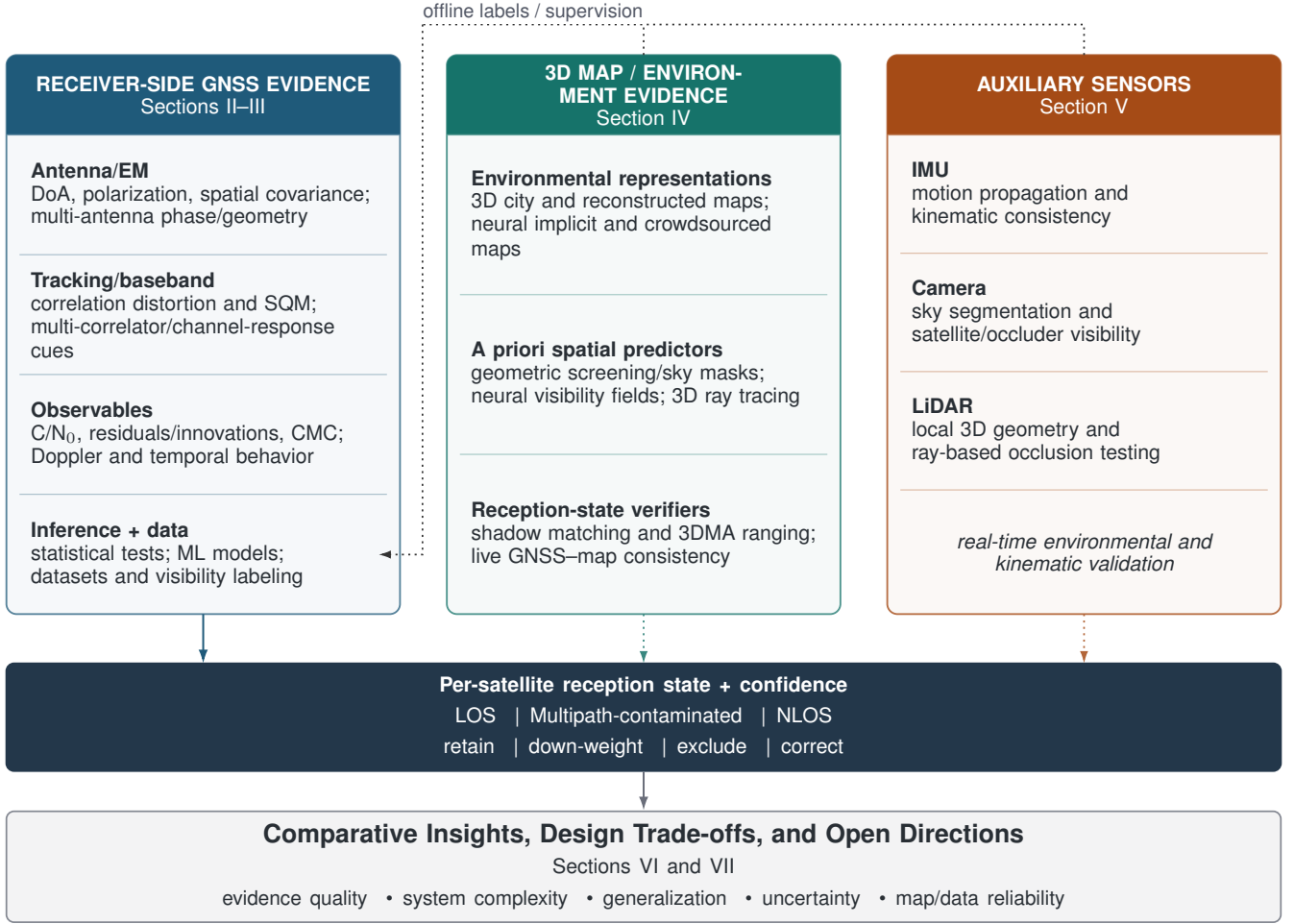

Building on the gap identified above, this survey makes the following contributions:
\begin{itemize}
    \item It organizes GNSS reception-state classification around three complementary information sources---receiver-side signal evidence, 3D map-based environmental information, and real-time auxiliary sensing---and situates the principal methodological developments within their historical evolution.

    \item It separately reviews the signal features used for reception-state characterization and the algorithmic paradigms used to classify them, including threshold-based rules, classical machine learning, deep learning, sequence models, and multimodal strategies.

    \item It reviews 3D mapping modalities for GNSS visibility analysis, including static city models, LiDAR-derived point clouds, vision-based reconstruction, neural implicit maps, and crowdsourced signal-based mapping, with attention to level of detail, map freshness, memory footprint, updateability, and real-time query cost.

    \item It synthesizes map-based visibility determination algorithms, including geometric LOS screening, explicit ray tracing, shadow matching, three-dimensional mapping-aided (3DMA) ranging, and neural implicit visibility inference.
    
    \item It examines real-time auxiliary-sensor anomaly detection using IMU, camera, and LiDAR, emphasizing their role in signal-reception state determination before measurement integration.
    
    \item It compares practical design trade-offs across method families and identifies open challenges in datasets, domain generalization, edge deployment, calibration, and trustworthy learning-based detection.
\end{itemize}

The remainder of this article follows this cross-domain organization. 
Section~\ref{sec:features} and~\ref{sec:algorithms} review receiver-side evidence for reception-state characterization, the algorithms that transform this evidence into decisions, and the datasets and labeling procedures used to develop and evaluate them. 
Section~\ref{sec:3d_mapping} examines environmental representations and map-based satellite-visibility inference, while Section~\ref{sec:auxiliary_sensors} reviews real-time verification using auxiliary sensing.
Section~\ref{sec:tradeoffs} then synthesizes the practical trade-offs across these methodological families, and Section~\ref{sec:future} outlines open research directions and concludes the survey. 
Fig.~\ref{fig:sections} summarizes this organization and the information flow among the reviewed evidence sources.

\section{GNSS Signal Features and Observables}
\label{sec:features}

While systematic errors such as atmospheric delays and receiver clock drift can be effectively mitigated using established analytical models or differential corrections, NLOS and multipath interference are fundamentally governed by the localized physical environment and the instantaneous geometric relationship between the satellites, the receiver, and surrounding urban structures. Because these effects are highly localized and spatially decorrelated, purely analytical models that ignore the immediate physical environment are largely ineffective in deep urban canyons. Consequently, modern GNSS navigation engines increasingly rely on direct signal-based detection. Rather than treating signal degradation as generic white noise, these systems exploit the distinctive physical and statistical signatures that structural reflections imprint upon the incoming waveform. 

To systematically categorize these distortion signatures, contemporary literature structures signal-based feature extraction across three distinct stages of the receiver architecture: (i) the antenna and electromagnetic (EM) domain, (ii) the tracking and baseband domain, and (iii) the observables or data domain~\cite{xu2024machine,yin2023features}. Each of these processing depths offers unique feature modalities, distinct hardware accessibility constraints, and specific sensitivities to signal deformation.
By examining the GNSS signal across these disparate representation levels, researchers have identified a vast taxonomy of degradation indicators. These range from physical polarization reversals and spatial arrival angles to complex distortions in the baseband correlation peaks and statistical anomalies in the final pseudorange measurements. The remainder of this section sequentially examines how structural reflections encode themselves within these three specific pipeline levels. By outlining the characteristic features and data-access requirements of the EM, baseband, and observable domains, this section establishes the physical and phenomenological foundation necessary for the algorithmic classification paradigms and machine learning methodologies subsequently detailed in Section~\ref{sec:algorithms}.

\begin{figure*}[t]
\centering
\begin{adjustbox}{max width=\textwidth}
\begin{tikzpicture}[
    x=1cm,
    y=1cm,
    font=\sffamily
]
\definecolor{schemCharcoal}{RGB}{42,46,53}
\definecolor{schemInk}{RGB}{57,65,73}
\definecolor{schemSlate}{RGB}{97,106,116}
\definecolor{schemIndigo}{RGB}{70,67,132}
\definecolor{schemAmber}{RGB}{177,118,39}
\definecolor{schemThreshold}{RGB}{154,79,77}
\definecolor{schemRule}{RGB}{207,213,218}
\definecolor{schemPaper}{RGB}{255,255,255}

\tikzset{
    schemstation/.style={
        circle,
        draw=schemIndigo,
        text=schemIndigo,
        line width=0.70pt,
        minimum size=0.43cm,
        inner sep=0pt
    },
    schemtitle/.style={
        anchor=center,
        inner sep=0pt,
        text=schemCharcoal,
        font=\sffamily\fontsize{9.2}{10.7}\selectfont\bfseries
    },
    schemsubtitle/.style={
        anchor=center,
        inner sep=0pt,
        text=schemSlate,
        font=\sffamily\fontsize{7.7}{9.1}\selectfont\itshape
    },
    schemgroup/.style={
        anchor=north west,
        inner sep=0pt,
        text=schemIndigo,
        font=\sffamily\fontsize{7.35}{8.5}\selectfont\bfseries
    },
    schemtext/.style={
        anchor=north west,
        align=left,
        text width=4.95cm,
        inner sep=0pt,
        text=schemCharcoal,
        font=\sffamily\fontsize{7.6}{9.0}\selectfont
    },
    schemprofile/.style={
        anchor=north west,
        align=left,
        text width=5.05cm,
        inner sep=0pt,
        text=schemSlate,
        font=\sffamily\fontsize{7.0}{8.1}\selectfont\itshape
    }
}

\draw[draw=schemSlate, line width=0.65pt] (0.00,8.38) -- (17.55,8.38);
\foreach \x in {5.75,11.80} {
    \draw[draw=schemSlate, line width=0.65pt] (\x-0.14,8.23) -- (\x+0.14,8.38) -- (\x-0.14,8.53);
}
\node[anchor=center, inner sep=0pt, text=schemSlate,
      font=\sffamily\fontsize{7.25}{8.4}\selectfont] at (8.775,8.63) {Receiver representation depth};
\draw[draw=schemRule, dash pattern=on 1.0pt off 2.0pt, line width=0.55pt] (5.75,0.68) -- (5.75,7.95);
\draw[draw=schemRule, dash pattern=on 1.0pt off 2.0pt, line width=0.55pt] (11.80,0.68) -- (11.80,7.95);

\node[schemstation, font=\sffamily\fontsize{8.0}{8.0}\selectfont\bfseries] at (2.675,7.64) {1};
\node[schemtitle] at (2.675,7.14) {Antenna \& EM Domain};

\draw[draw=schemInk, line width=0.70pt] (0.80,5.18) -- (4.55,5.18);
\draw[draw=schemInk, line width=0.65pt] (1.45,5.18) -- (1.45,5.56);
\draw[draw=schemInk, line width=0.65pt] (2.675,5.18) -- (2.675,5.56);
\draw[draw=schemInk, line width=0.65pt] (3.90,5.18) -- (3.90,5.56);
\filldraw[fill=schemPaper, draw=schemIndigo, line width=0.65pt] (1.45,5.66) circle (0.105);
\filldraw[fill=schemPaper, draw=schemIndigo, line width=0.65pt] (2.675,5.66) circle (0.105);
\filldraw[fill=schemPaper, draw=schemIndigo, line width=0.65pt] (3.90,5.66) circle (0.105);
\draw[draw=orange, line width=0.75pt, decorate,
      decoration={snake, amplitude=0.60mm, segment length=2.9mm}] (0.50,6.45) -- (1.34,5.74);
\draw[draw=orange, line width=0.75pt, decorate,
      decoration={snake, amplitude=0.60mm, segment length=2.9mm}] (1.72,6.45) -- (2.56,5.74);
\draw[draw=orange, line width=0.75pt, decorate,
      decoration={snake, amplitude=0.60mm, segment length=2.9mm}] (2.94,6.45) -- (3.78,5.74);

\node[schemgroup] at (0.45,4.53) {Features};
\node[schemtext] at (0.45,4.27) {Direction of arrival / polarization\\spatial covariance};
\draw[draw=schemRule, line width=0.45pt] (0.45,3.49) -- (1.17,3.49);
\node[schemgroup] at (0.45,3.30) {Receiver access};
\node[schemtext] at (0.45,3.04) {Calibrated, synchronized array RF samples\\geometry};
\draw[draw=schemRule, line width=0.45pt] (0.45,2.26) -- (1.17,2.26);
\node[schemgroup] at (0.45,2.07) {Typical analysis};
\node[schemtext] at (0.45,1.81) {Subspace estimation / beamforming\\geometric checks};
\draw[draw=schemRule, line width=0.45pt] (0.45,1.03) -- (1.17,1.03);
\node[schemprofile] at (0.45,0.86) {Specialized hardware};

\node[schemstation, font=\sffamily\fontsize{8.0}{8.0}\selectfont\bfseries] at (8.775,7.64) {2};
\node[schemtitle] at (8.775,7.14) {Tracking \& Baseband};
\draw[draw=schemSlate,
      dash pattern=on 2.0pt off 1.6pt,
      line width=0.60pt]
      (6.15,6.52) -- (6.47,6.52);
\node[anchor=west, inner sep=0pt, text=schemSlate,
      font=\sffamily\fontsize{6.8}{7.6}\selectfont]
      at (6.55,6.52) {direct};

\draw[draw=schemAmber,
      dash pattern=on 3.0pt off 1.2pt on 0.8pt off 1.2pt,
      line width=0.72pt]
      (7.35,6.52) -- (7.67,6.52);
\node[anchor=west, inner sep=0pt, text=schemAmber,
      font=\sffamily\fontsize{6.8}{7.6}\selectfont]
      at (7.75,6.52) {reflected};

\draw[draw=schemIndigo,
      line width=0.80pt]
      (9.05,6.52) -- (9.37,6.52);
\node[anchor=west, inner sep=0pt, text=schemIndigo,
      font=\sffamily\fontsize{6.8}{7.6}\selectfont]
      at (9.45,6.52) {composite};

\draw[draw=schemInk, line width=0.55pt] (6.47,5.16) -- (11.06,5.16);
\draw[draw=schemInk, line width=0.55pt] (6.47,5.16) -- (6.47,6.23);
\node[anchor=west, inner sep=0pt, text=schemInk,
      font=\sffamily\fontsize{7.0}{8.0}\selectfont] at (11.10,5.16) {$\tau$};
\draw[draw=schemSlate, dash pattern=on 2.0pt off 1.6pt, line width=0.65pt]
    (6.65,5.16) -- (8.75,5.88) -- (10.85,5.16);
\draw[draw=schemIndigo, line width=0.85pt]
    (6.65,5.16) -- (7.50,5.4514286) -- (8.25,5.8664662) -- (8.75,6.1431579) -- (9.25,6.0769925) -- (9.40,6.0571429) -- (10.85,5.1733333) -- (10.90,5.16);
\draw[draw=schemAmber, dash pattern=on 3.0pt off 1.2pt on 0.8pt off 1.2pt, line width=0.72pt]
    (7.50,5.16) -- (9.40,5.56) -- (10.90,5.16);
\draw[draw=schemRule, dash pattern=on 1.0pt off 1.3pt, line width=0.50pt] (8.25,5.16) -- (8.25,5.8664662);
\draw[draw=schemRule, dash pattern=on 1.0pt off 1.3pt, line width=0.50pt] (8.75,5.16) -- (8.75,6.1431579);
\draw[draw=schemRule, dash pattern=on 1.0pt off 1.3pt, line width=0.50pt] (9.25,5.16) -- (9.25,6.0769925);
\fill[schemIndigo] (8.25,5.8664662) circle (0.047);
\fill[schemIndigo] (8.75,6.1431579) circle (0.047);
\fill[schemIndigo] (9.25,6.0769925) circle (0.047);
\node[anchor=north, inner sep=0pt, text=schemIndigo,
      font=\sffamily\fontsize{6.8}{7.6}\selectfont\bfseries] at (8.25,5.08) {E};
\node[anchor=north, inner sep=0pt, text=schemIndigo,
      font=\sffamily\fontsize{6.8}{7.6}\selectfont\bfseries] at (8.75,5.08) {P};
\node[anchor=north, inner sep=0pt, text=schemIndigo,
      font=\sffamily\fontsize{6.8}{7.6}\selectfont\bfseries] at (9.25,5.08) {L};

\node[schemgroup] at (6.18,4.53) {Features};
\node[schemtext] at (6.18,4.27) {Correlation asymmetry / peak shape\\early--prompt--late / fitted delay};
\draw[draw=schemRule, line width=0.45pt] (6.18,3.49) -- (6.90,3.49);
\node[schemgroup] at (6.18,3.30) {Receiver access};
\node[schemtext] at (6.18,3.04) {Complex multi-correlator taps\\loop / front-end settings};
\draw[draw=schemRule, line width=0.45pt] (6.18,2.26) -- (6.90,2.26);
\node[schemgroup] at (6.18,2.07) {Typical analysis};
\node[schemtext] at (6.18,1.81) {Signal-quality / distortion metrics\\parametric fitting / anomaly tests};
\draw[draw=schemRule, line width=0.45pt] (6.18,1.03) -- (6.90,1.03);
\node[schemprofile, text width=5.35cm] at (6.18,0.86) {Receiver-internal access};

\node[schemstation, font=\sffamily\fontsize{8.0}{8.0}\selectfont\bfseries] at (14.875,7.64) {3};
\node[schemtitle] at (14.875,7.14) {Observables};

\draw[draw=schemInk, line width=0.55pt] (13.13,5.12) -- (16.85,5.12);
\draw[draw=schemInk, line width=0.55pt] (13.13,5.12) -- (13.13,6.30);
\node[anchor=south, inner sep=0pt, text=schemInk,
      font=\sffamily\fontsize{7.0}{8.0}\selectfont] at (13.13,6.32) {$C/N_0$};
\node[anchor=west, inner sep=0pt, text=schemInk,
      font=\sffamily\fontsize{7.0}{8.0}\selectfont] at (16.90,5.12) {$t$};
\draw[draw=schemThreshold, dash pattern=on 2.0pt off 1.6pt, line width=0.70pt] (13.15,5.55) -- (16.83,5.55);
\node[anchor=west, inner xsep=1.8pt, inner ysep=0.8pt, fill=schemPaper, text=schemThreshold,
      font=\sffamily\fontsize{6.8}{7.6}\selectfont] at (13.30,5.61) {threshold};
\draw[draw=schemIndigo, line width=0.85pt]
    plot coordinates {(13.18,6.08) (13.65,6.02) (14.10,6.11) (14.55,5.96)
                      (15.00,5.77) (15.45,5.39) (15.88,5.31) (16.28,5.41) (16.75,5.52)};
\fill[schemThreshold] (15.45,5.39) circle (0.045);

\node[schemgroup] at (12.30,4.53) {Features};
\node[schemtext] at (12.30,4.27) {$C/N_0$ / residuals / code-minus-carrier\\Doppler / geometry / temporal checks};
\draw[draw=schemRule, line width=0.45pt] (12.30,3.49) -- (13.02,3.49);
\node[schemgroup] at (12.30,3.30) {Receiver access};
\node[schemtext] at (12.30,3.04) {Standard observables + ephemerides\\time history / navigation state};
\draw[draw=schemRule, line width=0.45pt] (12.30,2.26) -- (13.02,2.26);
\node[schemgroup] at (12.30,2.07) {Typical analysis};
\node[schemtext] at (12.30,1.81) {Robust statistics / consistency checks\\temporal / machine-learning inference};
\draw[draw=schemRule, line width=0.45pt] (12.30,1.03) -- (13.02,1.03);
\node[schemprofile] at (12.30,0.86) {Broad availability};

\draw[draw=schemSlate, line width=0.70pt, -{Latex[length=1.8mm,width=1.2mm]}] (0.30,0.16) -- (17.25,0.16);
\node[anchor=east, inner sep=0pt, text=schemSlate,
      font=\sffamily\fontsize{7.15}{8.3}\selectfont] at (17.25,-0.05) {Increasing abstraction and typical data accessibility};
\end{tikzpicture}
\end{adjustbox}
\caption{Complementary GNSS receiver-processing evidence across representation depth. Antenna/EM features characterize spatial wavefields, tracking/baseband features characterize correlation responses, and observable-domain features characterize measurement, geometric, and temporal consistency. The domains can be fused; their left-to-right order indicates increasing abstraction and typical data accessibility rather than a strictly sequential detector.}
\label{fig:feature_hierarchy}
\end{figure*}
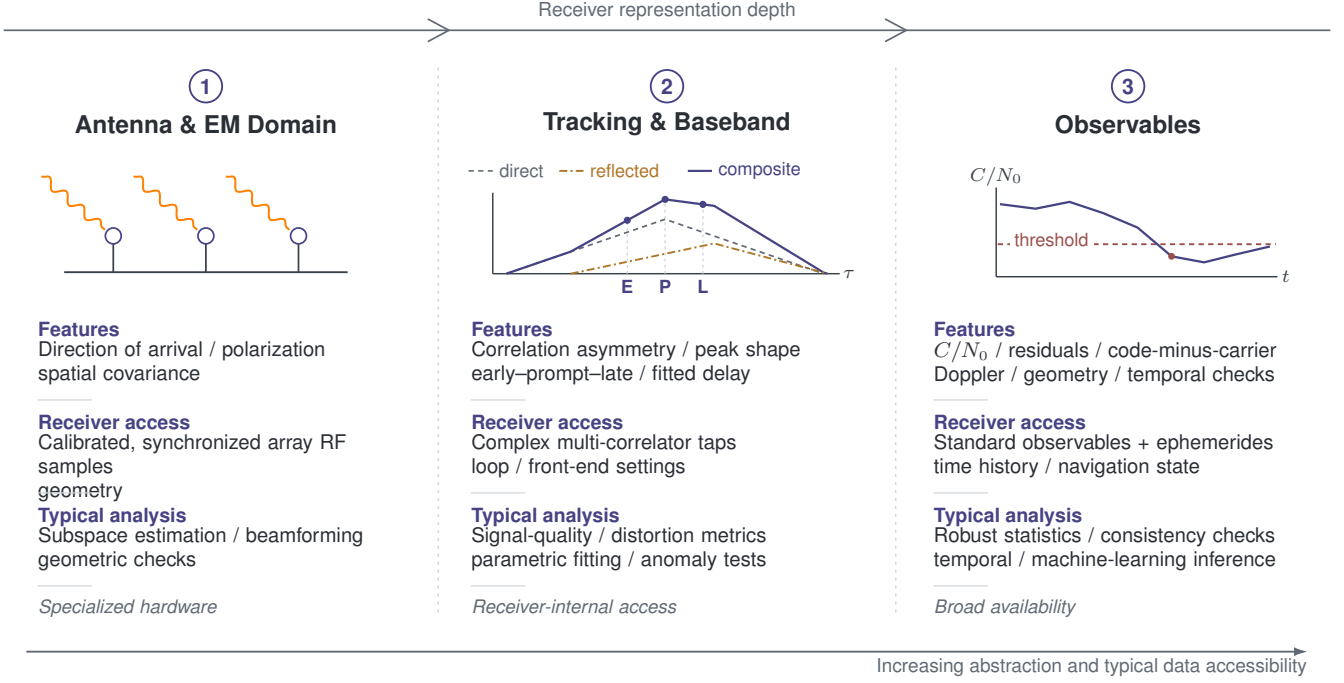

To contextualize these degradation indicators, Fig.~\ref{fig:feature_hierarchy} illustrates the hierarchical distribution of GNSS classification features across the sequential stages of the receiver pipeline. The subsequent subsections systematically dissect the primary feature families isolated within the antenna, baseband, and observable domains. For each operational stage, this review details the underlying physical interpretations, core mathematical formulations, specific hardware accessibility constraints, and the inherent trade-offs these features present for robust satellite visibility classification.

\subsection{Antenna / EM Level}
At the antenna and EM level, array-based processing exploits the spatial and polarization structure of the incident radio-frequency wavefield across the antenna aperture to support discrimination and separation of LOS and reflected signal components. By sampling the same GNSS signal at multiple spatially separated antenna elements, the receiver gains access to inter-element phase relationships, direction-of-arrival (DoA) information, spatial correlation properties and, for polarization-sensitive architectures, polarization diversity. 
It is important to note within this survey's taxonomy that while these spatial signatures provide the physical basis for identifying and suppressing multipath and NLOS components, the associated array processing may be implemented either at the intermediate-frequency stage or after signal de-spreading and correlation, depending on the receiver architecture~\cite{fernandezprades2009statistical,vagle2016performance,fohlmeister2017dual}. Consequently, these methods are classified here based on their fundamental reliance on the physical aperture, distinguishing them from the purely temporal, single-antenna tracking loop architectures discussed in subsequent sections.

\subsubsection{Direction of Arrival (DoA)}
Among array-based techniques, DoA estimation exploits the spatial response of a multi-element antenna aperture to incident GNSS wavefronts. For a far-field component arriving from direction $\mathbf{u}$, the relative phase progression across the array is determined by the element locations and the propagation direction. A corresponding steering vector can be expressed as
\begin{equation}
    \mathbf{a}(\mathbf{u})
    =
    \left[
    e^{-j\frac{2\pi}{\lambda}\mathbf{r}_1^{T}\mathbf{u}},
    \ldots,
    e^{-j\frac{2\pi}{\lambda}\mathbf{r}_M^{T}\mathbf{u}}
    \right]^{T},
\end{equation}
where $\mathbf{r}_m$ denotes the position of the $m$th antenna element relative to a common array reference, $\lambda$ is the carrier wavelength, and $\mathbf{u}$ is the incident-wave direction~\cite{vagle2016performance}. 
When the expected LOS steering vector is derived from satellite geometry, the satellite direction must be expressed in the array frame, requiring accurate array calibration and attitude knowledge. The attitude may be provided by an inertial attitude solution or a GNSS/IMU fusion estimator~\cite{vagle2016performance,ahmed2025gnss}.

The discriminative value of the azimuth and elevation components of DoA depends on both the array geometry and the surrounding propagation environment. Rectangular GNSS arrays characterize the spatial response jointly in elevation and azimuth, and practical two-dimensional array configurations have demonstrated multipath detection and filtering in both angular planes~\cite{vagle2016performance,razgunas2023gnss}. In street canyons dominated by approximately vertical building facades, azimuthal deviations can be particularly informative because specular reflections may substantially alter the horizontal propagation direction. Elevation remains complementary for ground, rooftop, sloped-surface, and other reflection geometries. 

LOS and reflected components generally arrive from different directions and therefore induce distinct spatial signatures across the antenna aperture. Early multi-antenna methods exploited known array geometry for multipath detection~\cite{becker1994special}, while subsequent approaches explicitly estimated multipath arrival directions using high-resolution techniques such as multiple signal classification (MUSIC)~\cite{moelker1997multiple}. These spatial differences can also be exploited through beamforming or spatial filtering to enhance the expected LOS component and suppress reflected energy. However, strong coherence between LOS and reflected signals can hinder spatial separation, motivating techniques such as spatial smoothing and polarization diversity~\cite{vagle2016performance,fohlmeister2017dual}.

\subsubsection{Polarization Diversity}
Polarization offers an additional physical dimension for characterizing GNSS signal propagation conditions. GNSS signals are natively transmitted with Right-Hand Circular Polarization (RHCP), whereas reflection and diffraction can alter the polarization state, often introducing stronger Left-Hand Circular Polarization (LHCP) or elliptical components depending on the propagation geometry and reflecting surface~\cite{braasch2017multipath}. This contrast between direct and reflected propagation motivates polarization-sensitive reception as an indicator of multipath and NLOS conditions. 
Dual-polarized antenna systems enable simultaneous observation of RHCP and LHCP signal components, allowing polarization-dependent metrics---such as RHCP--LHCP power imbalance or carrier-to-noise density ratio ($C/N_0$) differences---to be exploited for multipath and NLOS identification. Dual-polarization measurements have been used directly to detect NLOS reception and severe multipath~\cite{jiang2014nlos}, as well as within polarization-aware multipath mitigation schemes~\cite{fohlmeister2017dual}. More recent studies leverage features derived from dual-polarized measurements within machine-learning frameworks for reception-condition classification in urban environments~\cite{kim2023machine}.

\subsubsection{Spatial Correlation}
Spatial correlation, as captured by the array covariance matrix, provides a powerful mechanism for detecting and characterizing multipath and NLOS propagation at the antenna front end. Under nominal conditions, the spatial data follow the covariance structure expected from the dominant signal and receiver noise, whereas multipath introduces spatially correlated contributions that cause departures from this hypothesis. Statistical antenna-array detectors have therefore exploited covariance-matrix structure and the dispersion of its eigenvalues to detect multipath without explicitly estimating its direction of arrival~\cite{fernandezprades2009statistical,closas2011statistical}. 
Eigenstructure, subspace-based, and sparse Bayesian analyses of the spatial (and dual-polarized) data further exploit these deviations to resolve indirect signal components, estimate their directions and polarization characteristics, and enable spatial filtering or beamforming strategies that suppress multipath while preserving the LOS contribution~\cite{fohlmeister2017dual,chang2024joint}.

\vspace{1em} 

The aforementioned features---DoA, polarization, and spatial correlation---form the technical foundation for active mitigation strategies within the spatial domain. 
More sophisticated array architectures leverage high-resolution DoA estimation and beamforming---such as Delay-and-Sum or Minimum-Power Distortionless Response (MPDR) with spatial smoothing---to attenuate energy arriving from directions inconsistent with the nominal LOS spatial signature. By incorporating polarization-diverse filtering and joint spatio-temporal subspace or sparse processing, these systems can resolve and suppress multipath components that are otherwise difficult to separate in the time domain, thereby reducing code- and carrier-tracking errors through processing integrated within or after the initial tracking and correlation stages~\cite{vagle2016performance,fohlmeister2017dual,hong2020subspace}.

Despite their high physical discrimination performance, antenna-level processing introduces significant implementation overhead. The efficacy of spatial filtering depends heavily on accurate array calibration, inter-element synchronization, and stable geometry. Furthermore, the increased power consumption, computational cost, and hardware form-factor associated with multi-element apertures often conflict with the constraints of mass-market mobile devices. 
Nevertheless, demonstrations using relatively compact or low-cost multi-antenna and dual-polarized configurations indicate increasing practical potential for spatial and polarization discrimination in robust urban navigation~\cite{chen2021performance,kim2023machine}.

\subsection{Tracking Level}
At the tracking (baseband) level, signal classification exploits distortions in the code correlation function observed by the DLL. Standard GNSS receivers estimate the signal time-of-arrival by aligning a locally generated pseudorandom noise (PRN) replica with the incoming signal through code tracking, typically using prompt, early, and late correlator outputs within conventional tracking-loop architectures~\cite{gong2025review}. For binary phase-shift keying (BPSK)-like spreading signals, the corresponding code autocorrelation function (ACF) is ideally symmetric about the correct code delay. The prompt correlator is centered near the estimated peak, while the early and late correlators are offset by a prescribed chip spacing.
In nominal LOS conditions, the ACF remains symmetric around the prompt peak, resulting in approximately equal energy in the early and late correlators. In contrast, multipath propagation introduces delayed replicas of the transmitted signal that superimpose with the direct component. Depending on the relative delay, phase, and amplitude of the reflected components, the resulting composite ACF may exhibit peak broadening, shifts in the peak position, and skewness~\cite{xu2019intelligent}. While a theoretically pure, single-ray NLOS signal merely shifts and attenuates a symmetric ACF~\cite{xu2019intelligent}, practical urban NLOS conditions typically involve complex combinations of multiple diffracted and reflected rays. This breaks the inherent symmetry of the ideal ACF, often leading to the appearance of multiple local maxima and severe imbalances between the early and late correlator outputs~\cite{suzuki2021nlos}.
A canonical metric capturing this distortion is the normalized early--late slope (often referred to as the symmetric ratio test), defined as
\begin{equation}
    \Delta = \frac{I_{\mathrm{early}} - I_{\mathrm{late}}}{I_{\mathrm{prompt}}},
\end{equation}
where $I_{\mathrm{early}}$, $I_{\mathrm{late}}$, and $I_{\mathrm{prompt}}$ denote the in-phase correlator outputs at the early, late, and prompt code offsets, respectively~\cite{pirsiavash2017characterization}. 
Under ideal, multipath-free LOS conditions, the BPSK ACF is symmetric around the prompt peak, yielding approximately equal early and late correlator outputs and a theoretical expected value of zero for $\Delta$~\cite{xu2019intelligent,pirsiavash2017characterization}.
Conversely, multipath conditions induce asymmetric correlation peaks, producing significant non-zero deviations in $\Delta$. To prevent noise-induced false alarms, robust frameworks employ dynamic statistical thresholds rather than static limits. 
Because the theoretical variance of $\Delta$ fluctuates with instantaneous parameters like $C/N_0$, signal degradation can be flagged using a calibrated confidence envelope---typically $\pm3$ standard deviations---corresponding to a nominal per-trial false-alarm probability of 0.0027 under a Gaussian assumption~\cite{pirsiavash2017characterization}.

Beyond this basic formulation, a broader class of multi‑correlator and signal‑quality‑monitoring (SQM) metrics has been developed to capture ACF distortions more comprehensively, progressing from early real-time multipath monitoring to multi-correlator implementations~\cite{itani1996realtime,irsigler2005realtime}. 
These include Ratio, Delta, and Double-Delta metrics, often evaluated with narrow- or high-resolution correlator configurations, as well as area‑based indicators such as the Normalized Early–Prompt–Late Area (NEPLA), which exploit multiple correlator samples or integrated correlation regions to increase sensitivity to multipath~\cite{pirsiavash2017characterization,xu2025nepla}. More recently, multi-frequency correlation-domain detectors have exploited the complementary response of different GNSS bands to the same propagation conditions. For example, the multi-frequency cross-product area (MF-CPA) detector combines standard early--prompt--late in-phase and quadrature correlator outputs across frequencies to reduce detection blind zones that can occur in single-frequency correlator metrics~\cite{yang2026multifrequency}.
While highly informative, these metrics require access to correlator outputs at the tracking level and are influenced by thermal noise, front‑end bandwidth, correlator spacing, and receiver‑specific tracking architectures~\cite{pirsiavash2017characterization,xu2025nepla}. In particular, short-delay multipath may remain difficult to detect when correlator spacings are symmetrically positioned around a distorted peak---causing the distortion to cancel out in the discriminator---or when bandwidth limitations inherently smooth the ACF~\cite{pirsiavash2017characterization}.  Multi-frequency diversity provides one recent means of reducing such single-frequency detection ambiguities~\cite{yang2026multifrequency}.

Tracking-level detection can also exploit information shared across receiver channels rather than treating each DLL independently. In particular, vector-tracking-loop formulations have been used to perform NLOS detection directly at the baseband processing level by exploiting tracking information shared across multiple satellite channels~\cite{hsu2015multipath}.
More sophisticated tracking-level approaches explicitly model the received correlation response as a superposition of direct and reflected components.
Consistent with the correlator-domain model in \eqref{eq:mp_correlator_model}, estimating the path amplitudes $g_i$, relative delays $\Delta_i$, and phases $\theta_i$ from multiple correlator samples~\cite{vannee1992medll,townsend1995performance}. 
By fitting these parameters, Multipath Estimating Delay Lock Loop (MEDLL) yields physically interpretable estimates of the underlying multipath components, including their relative delays and powers. More recently, data-driven estimators have inferred such propagation parameters directly from sampled correlation functions; for example, random-forest regression has been used to estimate multipath amplitudes and relative code delays and benchmarked against MEDLL~\cite{qi2024random}.

Despite their strong interpretability and sensitivity to multipath, parametric correlation-domain estimators require access to raw or multi-correlator outputs and can incur a substantial computational burden as the number of estimated propagation paths increases~\cite{qi2024random}. 
Detailed tracking outputs are also often unavailable through mass-market receiver interfaces, restricting many correlation-domain approaches to software-defined receivers (SDRs), research platforms, and configurable receiver hardware.
Nevertheless, recent work demonstrates that useful signal-degradation detection can be constructed from standard early--prompt--late correlator outputs when these are accessible~\cite{yang2026multifrequency}. 
In parallel, data-driven approaches have progressed from machine-learning classification using multi-correlator information~\cite{suzuki2017nlos} toward richer learned correlation-domain representations. Correlation images or delay--time representations formed from multiple correlator outputs have been used with convolutional neural networks for direct multipath detection~\cite{suzuki2020nlos,blais2022novel,guillard2023cnn}.
Recent multimodal classification further illustrates this complementarity: correlation-domain representations are used to identify multipath distortion, whereas observable-level information is subsequently used to distinguish LOS from NLOS signals~\cite{wei2025multimodal}. Thus, whether utilizing simple early--late asymmetries, parametric path estimation, or learned correlation representations, tracking-level features provide direct evidence of non-ideal signal propagation while remaining complementary to the higher-level observables considered next.

\subsection{Observables Level}
At the observables level, GNSS receivers provide standard measurement products, such as $C/N_0$, pseudorange, carrier phase, Doppler frequency, and derived geometrical quantities that can be exploited for visibility assessment without requiring access to raw baseband data. As these measurements are widely accessible across receiver classes, observable-domain features constitute one of the most widely used foundations for LOS, NLOS, and multipath detection.

\subsubsection{Carrier-to-Noise Ratio}
Among observable-level features, $C/N_0$ and its temporal variability are the most widely used indicators of GNSS signal quality. The instantaneous $C/N_0$ is defined as the ratio between the received carrier power $C$ and the one-sided noise power spectral density $N_0$, and is commonly expressed in dB-Hz as
\begin{equation}
    C/N_0 = 10 \log_{10}\!\left(\frac{C}{N_0}\right).
\end{equation}
To characterize signal fading and attenuation, the variability of $C/N_0$ over a sliding window of $N$ epochs is commonly summarized by the windowed standard deviation
\begin{equation}
    \sigma_{C/N_0} = \sqrt{\frac{1}{N}\sum_{i=1}^N \bigl(C/N_0(i) - \overline{C/N_0}\bigr)^2},
\end{equation}
where $C/N_0(i)$ denotes the instantaneous value at the $i$-th epoch and $\overline{C/N_0}$ is the mean over the same window.

Under open-sky LOS conditions, GNSS signals generally exhibit high and stable $C/N_0$ levels. Conversely, multipath and NLOS reception can produce attenuation and pronounced temporal fluctuations, often accompanied by increased ranging errors~\cite{kubo2020gnss,hussain2021complexity,li2024characterization}.
The sensitivity of signal-strength observables to propagation effects was recognized early: SNR-based studies related multipath-induced power variations to code and carrier errors and exploited them for mitigation~\cite{comp1996adaptive,sleewaegen1997snr}, while Brunner et al.~\cite{brunner1999gps} incorporated measured $C/N_0$ into the stochastic SIGMA-$\Delta$ model to adapt the uncertainty assigned to diffraction-affected GPS observations. For multipath reception, the observed fluctuations arise from the time-varying superposition of direct and reflected components. Relative motion changes the path-dependent Doppler shifts and hence their relative carrier phases, producing alternating constructive and destructive interference. Consequently, the fluctuation rate increases as the receiver--satellite--reflector geometry evolves more rapidly~\cite{braasch2017multipath}.
However, a clear separation is not always guaranteed; LOS and NLOS $C/N_0$ distributions frequently overlap, particularly in dense urban canyons where NLOS signals may transiently exhibit LOS-like power levels or where constructive multipath interference artificially boosts the observed signal strength~\cite{kubo2020gnss,li2023machine,wang2015smartphone,hsu2018analysis}. Moreover, absolute $C/N_0$ statistics are heavily influenced by hardware-specific factors, including receiver design, antenna gain patterns, and local environmental conditions~\cite{groves2013height,paziewski2019signal,wanninger2020gnss}. Therefore, $C/N_0$ statistics are typically integrated with other observables within threshold-based, machine-learning, or time-series classification frameworks to ensure robust discrimination between LOS, NLOS, and multipath conditions~\cite{kubo2020gnss,li2023machine,wang2022multipath,ozeki2022gnss}.

\subsubsection{Pseudorange Residuals and Double-Differencing}
While $C/N_0$ characterizes the power and fading of the received signal, pseudorange residuals and double-differenced (DD) observables provide direct measures of geometric range inconsistency and serve as critical complementary indicators of signal quality in GNSS processing. In over-determined positioning solutions, residuals quantify the mismatch between individual measurements and the estimated satellite--receiver geometry. For a pseudorange observation $\rho_i$ and its predicted value $\hat{\rho}_i$ derived from a position solution, the residual is defined as
\begin{equation}
    r_i = \rho_i - \hat{\rho}_i ,
\end{equation}
and reflects unmodeled effects such as multipath and NLOS biases~\cite{wang2022multipath,suzuki2019pseudorange,ozeki2022gnss}. 
This residual-consistency principle has also been used directly for satellite screening: single-epoch least-squares residuals can be compared across measurements to identify NLOS and strongly multipath-contaminated signals~\cite{jiang2011consistency}.

In relative or multi-receiver configurations, DD combinations are commonly employed to suppress common-mode errors. The double-differenced pseudorange between satellites $i$ and $j$ observed at receivers $1$ and $2$ is given by
\begin{equation}
    \Delta\Delta\rho_{ij}^{1,2} = (\rho_{i,1} - \rho_{j,1}) - (\rho_{i,2} - \rho_{j,2}) .
\end{equation}

For sufficiently short baselines, this differencing strategy largely cancels receiver and satellite clock biases and suppresses atmospheric errors common to the two receivers, thereby isolating local, receiver-specific effects. As a result, DD residuals are particularly sensitive to multipath and NLOS contamination at the rover antenna~\cite{wang2022multipath,xu2019intelligent}. A related receiver-only approach forms successive-time double differences of pseudorange and carrier-phase observations to construct a channel-wise multipath test statistic without requiring a precomputed receiver position~\cite{lee2004gpsmultipathdd}.

However, the effectiveness of these residual-based features depends critically on measurement redundancy and favorable geometry. When a large fraction of satellites is affected by NLOS---such as in deep urban canyons---standard consistency checks and fault detection and exclusion (FDE) algorithms become significantly less discriminative. 
When several NLOS measurements are simultaneously present, their biases can be partially absorbed into the estimated position and clock states, reducing residual separability and weakening conventional consistency checks~\cite{groves2013height,ozeki2022gnss}.
Furthermore, conventional between-satellite DD formulations require the selection of a suitable reference or master satellite to anchor the differences~\cite{xu2019intelligent}. Strategies that rely on simple $C/N_0$ thresholds or highest-elevation rules to select this reference degrade rapidly in urban environments, because, as established previously, signal strength does not reliably correspond to signal type when strong specular reflections are present. Consequently, pseudorange and DD residuals are most effective when sufficient LOS observations remain to securely anchor both the positioning solution and the reference satellite selection.

\subsubsection{Code-Minus-Carrier Divergence}
Another widely used metric is the Code-Minus-Carrier (CMC) divergence, which serves as a standard observable for isolating code multipath, particularly at reference stations and in multi-frequency setups~\cite{wanninger2020gnss}. The CMC observable compares the code pseudorange $\rho$ with the carrier-phase-based range $\lambda\phi$ and is defined as:
\begin{equation}
    \text{CMC} = \rho - \lambda\phi
\end{equation}
where $\phi$ is the carrier phase in cycles and $\lambda$ is the wavelength. Because code measurements are generally more susceptible to multipath and noise than carrier-phase measurements, variations in the CMC observable provide a useful indicator of code distortion. However, absolute CMC also contains the carrier-phase ambiguity and ionospheric terms, which must be removed or accounted for before the observable can be interpreted in terms of code multipath. In urban and dynamic scenarios, time-differenced CMC ($\Delta$CMC) provides a more practical alternative by cancelling the constant carrier-phase ambiguity and emphasizing short-term variations associated with multipath, thereby avoiding the long averaging intervals otherwise required to estimate the ambiguity-related offset~\cite{caamano2020detection}.

However, the practical utility of absolute CMC in urban environments is severely constrained by its dependence on continuous phase tracking. The metric is highly sensitive to cycle slips, and in urban environments, frequent signal blockages and loss-of-lock events render continuous CMC time series fragmented and difficult to use reliably~\cite{caamano2020detection,shinghal2021conditioning}. In deep urban canyons, these frequent cycle slips induce repeated resets and spurious ambiguity jumps, which mask actual signal features and prevent the long-term convergence required for effective mean-removal and multipath characterization~\cite{shinghal2021conditioning}. Consequently, phase tracking discontinuities break the metric's consistency, necessitating robust slip detection, re-initialization algorithms, or time-differencing strategies to distinguish true code multipath distortions from ambiguity shifts~\cite{caamano2020detection,shinghal2021conditioning}.

\subsubsection{Pseudorange--Doppler Consistency}
Doppler-based consistency checks rely on the fundamental kinematic link between the rate of change of the code pseudorange and the observed Doppler frequency shift. Theoretically, under unobstructed LOS conditions, the time derivative of the pseudorange $\Delta \rho / \Delta t$ should tightly align with the range rate deduced from the Doppler measurement $-\lambda f_D$. This relationship can be expressed as:
\begin{equation}
    \frac{\Delta \rho}{\Delta t} + \lambda f_D \approx 0
\end{equation}
where $f_D$ represents the measured Doppler frequency. Under nominal LOS reception, the pseudorange change and Doppler-derived range rate describe the same underlying range evolution and should therefore remain mutually consistent. Because pseudorange and Doppler are estimated through distinct code- and frequency/carrier-tracking processes, multipath and NLOS propagation can perturb them differently, producing a non-zero consistency error~\cite{hsu2017gnss,ozeki2022gnss}. Hsu~\cite{hsu2017gnss} explicitly introduced this pseudorange--Doppler consistency feature and combined it with received signal strength within a support vector machine (SVM) classifier to distinguish clean, multipath, and NLOS measurements.

Despite its utility, the reliability of this metric depends on accurate time alignment, sufficiently short differencing intervals, and stable code and Doppler tracking. Rapid receiver dynamics, tracking noise, or loss-of-lock events can increase the apparent inconsistency independently of multipath. This limitation is particularly relevant for consumer-grade receivers and dense urban environments, where noisy Doppler observations, tracking instability, and frequent loss-of-lock events can obscure the true Doppler-rate inconsistency and reduce the reliability of the metric~\cite{shinghal2021conditioning,wang2022multipath}.

\subsubsection{Geometric and Satellite Ephemeris Indicators}
Satellite geometry parameters computed from ephemeris data, particularly elevation and azimuth angles, supply essential spatial context for evaluating signal reception quality. Although applying an elevation mask and weighting measurements by satellite altitude are conventional receiver strategies, extensive research illustrates their limitations in heavily built-up areas. In deep urban canyons, signals from high-elevation satellites can still be blocked and received only as NLOS reflections off adjacent skyscrapers, while some low-elevation signals may successfully propagate through street corridors to remain LOS~\cite{wang2022multipath,hsu2018analysis}.

As a result, relying solely on elevation-driven rules proves inadequate. Empirical evaluations utilizing modern multi-GNSS smartphone receivers have documented that satellite elevation alone does not uniquely determine either the reception state or the resulting pseudorange error in urban settings~\cite{guo2020characteristics,paziewski2019signal,wanninger2020gnss}. In such challenging conditions, observation weighting models that rely on measured signal strength ($C/N_0$) have been shown to outperform purely elevation-based weighting in evaluated urban scenarios~\cite{groves2013height,guo2020characteristics}. Consequently, geometric parameters are seldom deployed in isolation for modern signal classification. Instead, contemporary data-driven and machine-learning frameworks structurally fuse satellite azimuth and elevation with a suite of signal-quality and other observable-domain metrics to comprehensively identify multipath and NLOS contamination~\cite{wang2022multipath,ozeki2022gnss,xu2024machine}.

\subsubsection{Temporal and Multivariate Sequence Features}
Transitioning from epoch-specific evaluations, recent studies increasingly emphasize the \emph{temporal evolution} of GNSS observables. By framing signal metrics as continuous time series, it becomes possible to detect short-lived obstructions, fading behaviors, and tracking stability trends that single-epoch snapshots inherently obscure.
Fundamental temporal features map scalar variables---including $C/N_0$, Doppler shifts, and pseudorange residuals---over a sliding observation window comprising $N$ epochs. The observation timescale is method dependent, but representative implementations typically span tens of seconds. Quan et al.~\cite{quan2018convolutional} used a 16-s window for CNN-based multipath detection, while Kubo et al.~\cite{kubo2020gnss} considered a one-minute guard interval for continuous $C/N_0$ screening. These durations reflect a trade-off between temporal context and detection latency. Within this timeframe, signal degradation is measured via statistical metrics such as variance, amplitude range, and fluctuation intensity. These parameters can capture sustained power loss and short-term fluctuations associated with multipath and NLOS conditions. For example, continuous monitoring shows that NLOS signals regularly exhibit prolonged attenuation combined with severe, high-frequency variations, even when absolute $C/N_0$ values briefly peak at LOS levels. Detection algorithms built on this premise use temporal consistency constraints, routinely discarding satellites that fail to hold elevation-appropriate $C/N_0$ baselines over a set duration~\cite{kubo2020gnss}. This progression also motivated learned temporal representations: Quan et al.~\cite{quan2018convolutional} applied a convolutional neural network (CNN)-based framework to detect carrier-phase multipath using $C/N_0$ and pseudorange-multipath combinations over successive observations.

Beyond purely temporal statistics, observable-domain classifiers increasingly exploit multi-frequency feature sets. Branched models can accommodate intermittently available frequency channels, while factor-analysis-based schemes compress correlated multi-frequency signal-quality indicators before classification~\cite{garcia2023robust,li2024efficient}.
More sophisticated frameworks treat these indicators as \emph{multivariate sequences}, aggregating measurements such as geometric angles, $C/N_0$, CMC, and residuals into joint epoch-wise feature vectors. Such models assume that the synchronized behavior of multiple features provides a more robust indicator of the propagation state than independent metrics. 
Recent architectures are explicitly designed to capture these complex cross-variable and temporal dynamics. Transformer-enhanced long short-term memory (LSTM) models have combined geometric and observable features for joint NLOS detection and observation-uncertainty prediction~\cite{zhang2023learning}. At the cross-satellite level, graph-transformer approaches construct a sky-satellite graph from the satellites observed at each epoch and use multihead attention to learn environment-aware relationships for visibility prediction~\cite{zheng2023graph}. More recent multivariate architectures further combine temporal feature dynamics with spatial satellite relationships~\cite{li2025multivariate}.

The formulation of these temporal features necessitates careful design compromises. Widening the observation window grants the classifier deeper historical context---which mitigates transient noise---but simultaneously increases detection latency and computational demands, potentially hindering real-time receiver operations~\cite{li2024efficient,li2025multivariate}. Conversely, smaller windows offer rapid responsiveness but leave statistical summaries more vulnerable to receiver tracking jitter. Another core trade-off exists between model transparency and predictive capacity. While simple moving-average checks are highly interpretable and straightforward to configure~\cite{kubo2020gnss}, deep multivariate sequence models map complex, non-linear relationships that typically function as opaque black boxes. To improve interpretability, Layer-wise Relevance Propagation (LRP) has been applied to LSTM-based multipath models, revealing increased feature relevance during introduced anomalous signal intervals, while attention-based architectures provide insight into learned cross-variable and spatial dependencies~\cite{wang2025towards,li2025multivariate}.

\subsubsection{Multi-Antenna and Spatial Observables}
Beyond classical single-antenna observables, recent studies demonstrate that multi-antenna GNSS arrays can be exploited at the observable level to enhance multipath and NLOS detection. 
Unlike the antenna/EM-level methods discussed previously, the emphasis here is on consistency among tracked carrier-phase observables across antennas rather than direct processing of the incident wavefield or array manifold.
Both~\cite{zorn2020accurate} and~\cite{ahmed2023multipath} exploit antenna-array spatial diversity and cross-antenna carrier-phase relationships to identify multipath-contaminated observations, relying on the principle that true LOS signals must satisfy rigid baseline geometry constraints, whereas NLOS and multipath reflections typically violate these spatial relationships.
The work in~\cite{zorn2020accurate} approaches this problem through a joint estimation framework designed specifically for uncalibrated systems. This method treats attitude determination, array calibration, and multipath mitigation as a coupled problem, simultaneously estimating multipath states and hardware biases within the filtering loop to maintain robustness without prior calibration.
Conversely, a \emph{detection-and-exclusion} strategy is adopted in~\cite{ahmed2023multipath}. Building on the instantaneous attitude determination framework of~\cite{ballal2014gnss}, this scheme utilizes externally supplied attitude estimates to strictly enforce baseline--geometry relationships. Instead of absorbing multipath into a joint state estimator, satellite observations that deviate from the expected carrier-phase differences are identified as geometric violations and excluded from the navigation solution.
A more recent extension integrates GNSS array measurements with inertial sensing through an Unscented Kalman Filter (UKF) on manifolds, enabling continuous attitude tracking and dynamic multipath detection via satellite-subset-specific attitude inconsistency with the filter-derived attitude estimate~\cite{ahmed2025gnss}. To preserve computational efficiency in scenarios with a large number of visible satellites, a random sample consensus (RANSAC)-based satellite-selection strategy is employed to reject outliers before the update step.

Together, these approaches illustrate a shift toward exploiting spatial diversity not only at the EM level but also through observable-domain consistency checks. While such methods require multi-antenna hardware and increased system complexity, they provide enhanced robustness for multipath and NLOS detection in dense urban environments where a substantial fraction of observations may be corrupted.

\section{Algorithmic Detection and Multimodal Datasets}
\label{sec:algorithms}

Having established the physical features and distortion signatures extracted across the GNSS receiver pipeline in Section~\ref{sec:features}, this section turns to the computational challenge: mapping these multidimensional feature spaces into reliable LOS, NLOS, and multipath classifications. Translating raw signal indicators into actionable visibility states relies on a broad spectrum of computational paradigms. This section first synthesizes these frameworks in \S~\ref{subsec:algorithmic_approaches}, detailing the field's evolution from static statistical boundaries to highly complex, data-driven deep neural networks. Subsequently, \S~\ref{subsec:datasets_labeling} addresses the fundamental challenge for learning-based models: the acquisition of high-fidelity ground truth. We review the prominent multimodal datasets and the retrospective offline labeling pipelines required to train these signal-level classifiers.

\subsection{Algorithmic Approaches}
\label{subsec:algorithmic_approaches}

\begin{table*}[htbp]
\centering
\caption{Algorithmic taxonomy for GNSS LOS/NLOS and multipath discrimination across signal-processing domains. The table summarizes the dominant computational paradigm, operating signal domain, representative evidence, and practical constraints. Detailed numerical performance should be interpreted in the context of each study's receiver hardware and dataset.}
\label{tab:algorithm_taxonomy_compact}

\small
\renewcommand{\arraystretch}{1.22}
\setlength{\tabcolsep}{3.5pt}

\definecolor{taxHeader}{RGB}{40,62,82}
\definecolor{taxRule}{RGB}{190,199,207}
\definecolor{taxAntenna}{RGB}{247,250,252}
\definecolor{taxTrack}{RGB}{244,249,248}
\definecolor{taxObs}{RGB}{250,247,243}

\arrayrulecolor{taxRule}

\renewcommand{\tabularxcolumn}[1]{>{\raggedright\arraybackslash}m{#1}}

\begin{tabularx}{\textwidth}{@{}
>{\raggedright\arraybackslash}m{2.5cm}
>{\raggedright\arraybackslash}m{2.85cm}
>{\raggedright\arraybackslash}X
>{\raggedright\arraybackslash}X
>{\raggedright\arraybackslash}X
@{}}

\toprule

\rowcolor{taxHeader}
\textcolor{white}{\textbf{Signal Domain}} &
\textcolor{white}{\textbf{Paradigm}} &
\textcolor{white}{\textbf{Core Principle}} &
\textcolor{white}{\textbf{Reported Finding}} &
\textcolor{white}{\textbf{Main Constraint}} \\

\midrule

\rowcolor{taxAntenna}
Antenna / Electromagnetic
& Spatial and Polarization Methods
& DoA, polarization, and spatial signal characteristics are exploited to distinguish direct and reflected propagation.
& Multipath detection and reception-state classification have been demonstrated using antenna-array spatial processing and dual-polarized measurements~\cite{razgunas2023gnss,kim2023machine}.
& Requires specialized RF hardware, antenna calibration, and additional spatial processing. \\

\cmidrule(lr){1-5}

\rowcolor{taxTrack}
Tracking / Baseband
& Correlation-Domain ML / Deep Learning
& Supervised ML and deep models learn distortions encoded in multi-correlator or delay--Doppler representations.
& Correlation-domain methods demonstrate strong multipath and NLOS discrimination from richer tracking-level signal representations~\cite{blais2022novel,suzuki2021nlos,guillard2023cnn}.
& Requires access to tracking-level correlator outputs and representative labeled or simulated training data. \\

\cmidrule(lr){1-5}

\rowcolor{taxObs}
&
Threshold / Statistical
& Hand-crafted rules on $C/N_0$, elevation angle, pseudorange/code--carrier consistency, and temporal stability metrics.
& Lightweight signal screening without offline training~\cite{kubo2020gnss,lee2004gpsmultipathdd}.
& Fixed thresholds or statistical assumptions can lose discrimination across changing receivers and propagation environments. \\

\hhline{~----}

\rowcolor{taxObs}
&
Classical Machine Learning
& Supervised classifiers, such as SVM, decision trees, and random forests, trained on engineered signal-quality and geometry features.
& Supervised classifiers can learn nonlinear LOS/NLOS/MP decision boundaries from engineered observables, but performance remains sensitive to the training environment~\cite{ozeki2022gnss,li2024efficient,yozevitch2016robust}.
& Requires reliable labels; limited robustness to cross-city or cross-scenario domain shift. \\

\hhline{~----}

\rowcolor{taxObs}
&
Sequence Models
& Sequence models exploit temporal evolution in $C/N_0$, Doppler, residuals, geometry, or related observables.
& Temporal models capture transient NLOS and multipath behavior that single-epoch classifiers may miss~\cite{zhang2023learning,geragersian2022multipath}.
& Requires labeled sequences; observation windows introduce latency and additional computation. \\

\hhline{~----}

\rowcolor{taxObs}
\raisebox{2.8em}[0pt][0pt]{Observables}
&
Graph Models
& Satellite measurements are represented as graph nodes, with spatial relationships encoded through graph connectivity; graph attention or transformer mechanisms aggregate information across simultaneously observed satellites.
& Graph-based models improve satellite-visibility classification and generalization by exploiting cross-satellite spatial relationships~\cite{zheng2023graph,li2025multivariate}.
& Requires graph construction and increased model complexity; performance remains dataset- and representation-dependent. \\

\hhline{~----}

\rowcolor{taxObs}
&
Unsupervised / Domain-Adaptive Learning
& Latent-space clustering and domain adaptation reduce dependence on manually labeled target-domain measurements.
& Synthetic-to-real adaptation has substantially improved experimental multipath detection when labeled real data are scarce~\cite{zawislak2022gnss}.
& Performance depends on learned representation quality and the remaining mismatch between source and target domains. \\

\hhline{~----}

\rowcolor{taxObs}
&
Hybrid / Branched / Ensemble Models
& Branched feature models, bagging, boosting, or voting combine complementary feature groups or classifiers.
& Such architectures improve robustness against single-feature failure and intermittent measurement loss~\cite{koiloth2024ml,garcia2023robust}.
& More complex training, hyperparameter tuning, and deployment engineering. \\

\arrayrulecolor{taxRule}
\bottomrule

\end{tabularx}

\arrayrulecolor{black}
\end{table*}

The algorithmic literature has evolved from fixed statistical decision rules toward increasingly flexible data-driven classifiers. Threshold-based methods remain attractive because of their transparency and negligible training requirements, but their performance depends strongly on whether the assumed signal statistics remain valid in the deployment environment~\cite{kubo2020gnss,lee2004gpsmultipathdd}. Classical supervised learning relaxes these fixed decision boundaries by learning nonlinear relationships among multiple features; nevertheless, several studies demonstrate that strong in-domain performance can deteriorate substantially when the receiver, trajectory, or urban environment differs from the training data~\cite{yozevitch2016robust,ozeki2022gnss,li2024efficient}. Recent comparative work has also evaluated portfolios of supervised learning models specifically for urban GNSS LOS/NLOS classification~\cite{zhu2023portfolio}. Consequently, cross-environment generalization has become a central concern in GNSS reception-state classification.

More recent architectures exploit dependencies that conventional single-epoch classifiers neglect. Sequence models incorporate temporal context to distinguish persistent propagation effects from short-term measurement fluctuations~\cite{geragersian2022multipath,zhang2023learning}, whereas graph-based methods explicitly model relationships among simultaneously observed satellites and can incorporate their evolving spatial configuration~\cite{zheng2023graph,li2025multivariate}. In parallel, unsupervised domain adaptation reduces reliance on labeled measurements from every new environment by transferring learned representations between source and target domains~\cite{zawislak2022gnss}. Branched and ensemble architectures provide another form of robustness by combining complementary feature subsets or maintaining operation when some measurements are intermittently unavailable~\cite{garcia2023robust,koiloth2024ml}.

Overall, increasing model flexibility generally improves the ability to represent complex propagation-dependent decision boundaries, but introduces greater dependence on representative training data, model complexity, and validation across unseen conditions. No single algorithmic family is therefore universally preferable. The principal remaining challenges are cross-environment generalization, label efficiency, interpretability, and real-time deployment under the computational and data-access constraints of practical GNSS receivers.

\subsection{Labeling Strategies and Multimodal Datasets}
\label{subsec:datasets_labeling}

Datasets and labeling strategies are central to the development and benchmarking of GNSS reception-state classifiers. Unlike many standard machine-learning problems, relatively few GNSS datasets provide native per-satellite LOS/NLOS or multipath labels. Most widely used navigation datasets were collected primarily for localization and sensor-fusion research; reception-state labels must therefore be reconstructed by aligning raw GNSS measurements with accurate reference trajectories and independent environmental information, such as 3D city models, LiDAR geometry, or sky-view imagery~\cite{beaucamp2025comparison}. Table~\ref{tab:gnss_datasets_taxonomy} summarizes representative resources according to the information available for reception-state classification and the way in which reference labels are obtained. Broadly, these resources span four complementary regimes: multimodal datasets that support geometry- or perception-based annotation, GNSS-centric datasets and prepared classification benchmarks, raw-IF datasets that expose tracking-level signal features, and controlled synthetic datasets in which propagation states are known by construction. These regimes trade off label fidelity, receiver and scenario diversity, signal-domain richness, and access to explicit propagation-state ground truth.

\textbf{Navigation-Centric Multimodal Corpora:}
Datasets such as \emph{UrbanLoco}~\cite{wen2020urbanloco}, \emph{UrbanNav}~\cite{hsu2023hong}, \emph{SensNav}~\cite{gao2025sensnav}, and \emph{M2DGR}~\cite{yin2021m2dgr} are particularly valuable because their perception sensors and accurate reference poses permit satellite visibility to be inferred independently of the GNSS observables subsequently used for classification. The KLT dataset further reduces the annotation burden by providing per-satellite LOS/NLOS labels derived from fisheye imagery together with the corresponding GNSS measurements~\cite{hu2023fisheye}. Such geometry- and perception-based supervision is attractive because it avoids defining reception state from the same signal features used by the classifier. Its accuracy nevertheless depends on pose synchronization, sensor calibration, environmental-model completeness, and the treatment of dynamic or boundary occlusions~\cite{beaucamp2025comparison}.

\textbf{GNSS-Centric Datasets and Classification Benchmarks:}
GNSS-centric resources emphasize measurement volume, receiver diversity, or directly prepared learning targets rather than dense environmental sensing. Large collections such as \emph{GSDC} and \emph{WHU-Smartphone} are therefore useful for studying hardware variability and cross-device generalization~\cite{fu2022summary,wang2024whu}, but an accurate reference trajectory alone does not determine whether an individual satellite was received through LOS, LOS with multipath, or NLOS propagation. In the absence of independent environmental information, reception-state supervision may consequently rely on residual- or consistency-based proxies and should be regarded as weaker ground truth. Datasets such as smartLoc~\cite{reisdorf2016comparable} reduce this burden by supplying reception-state indicators alongside the raw GNSS measurements, whereas dedicated benchmarks such as Xu et al.~\cite{xu2024machine} provide prepared classification targets and standardized evaluation tasks. These resources improve experimental reproducibility, although the fidelity and independence of the supplied labels remain important when interpreting reported classifier performance.

\textbf{Raw-IF and Tracking-Level Resources:} A complementary class of datasets exposes the received signal below the conventional measurement level. TEX-CUP~\cite{narula2020texcup} and the SJTU GNSS test dataset~\cite{sjtu2018gnssdataset} provide raw or digital intermediate-frequency (IF) data together with navigation reference information, enabling analysis of correlator distortion, tracking-loop behavior, carrier/code dynamics, frequency diversity, and antenna-dependent signatures that are unavailable in standard Receiver Independent Exchange Format (RINEX)-level datasets. More recently, CardiffNav~\cite{cardiffnav2026} extends this capability by combining multi-channel raw GNSS RF/in-phase and quadrature (I/Q) recordings with synchronized LiDAR, camera, inertial, and reference-navigation data across deliberately challenging propagation environments. This richer sensing context is particularly valuable for relating tracking-level signal distortions to independently observed environmental geometry. The principal limitation of these resources remains labeling: native per-satellite LOS/NLOS/multipath truth is generally unavailable, so geometry-, imagery-, or path-based annotation is still required before the recordings can serve as supervised reception-state benchmarks.

\textbf{Controlled Propagation Datasets:}
Controlled synthetic resources address the opposite problem. In datasets such as MVRGD-MUE~\cite{silva2026multipath}, direct and reflected propagation paths are generated explicitly, allowing reception states and multipath quantities to be assigned without the ambiguity inherent in field annotation. Such data are especially useful for controlled ablation studies and for distinguishing LOS, multipath, and NLOS conditions that are difficult to label independently in real measurements. Their limitation is the corresponding simulation-to-reality gap: propagation geometry, material properties, reflection models, antenna response, and receiver tracking behavior cannot reproduce the full diversity of real urban channels.

\begin{table*}[h!]
\centering
\caption{Representative datasets supporting GNSS LOS/NLOS and multipath reception-state classification in urban environments.}
\label{tab:gnss_datasets_taxonomy}

\footnotesize
\renewcommand{\arraystretch}{1.02}
\setlength{\tabcolsep}{3.2pt}

\definecolor{dataHeader}{RGB}{40,62,82}
\definecolor{dataRule}{RGB}{190,199,207}

\definecolor{dataNav}{RGB}{247,250,252}
\definecolor{dataNavBand}{RGB}{231,238,244}
\definecolor{dataNavText}{RGB}{40,62,82}

\definecolor{dataGNSS}{RGB}{244,249,248}
\definecolor{dataGNSSBand}{RGB}{227,240,237}
\definecolor{dataGNSSText}{RGB}{22,105,96}

\definecolor{dataTrack}{RGB}{249,247,252}
\definecolor{dataTrackBand}{RGB}{236,232,244}
\definecolor{dataTrackText}{RGB}{86,67,116}

\definecolor{dataSim}{RGB}{252,249,243}
\definecolor{dataSimBand}{RGB}{244,236,219}
\definecolor{dataSimText}{RGB}{124,84,34}

\arrayrulecolor{dataRule}

\renewcommand{\tabularxcolumn}[1]{>{\raggedright\arraybackslash}m{#1}}

\begin{tabularx}{\textwidth}{@{}
>{\raggedright\arraybackslash}m{2.25cm}
>{\raggedright\arraybackslash}m{2.75cm}
>{\raggedright\arraybackslash}m{3.00cm}
>{\raggedright\arraybackslash}X
>{\raggedright\arraybackslash}X
@{}}

\toprule

\rowcolor{dataHeader}
\textcolor{white}{\textbf{Dataset}} &
\textcolor{white}{\textbf{Data Modalities}} &
\textcolor{white}{\textbf{Labeling Route}} &
\textcolor{white}{\textbf{Best Use}} &
\textcolor{white}{\textbf{Main Caveat}} \\

\midrule

\rowcolor{dataNavBand}
\multicolumn{5}{c}{
\textcolor{dataNavText}{\textbf{Navigation-centric multimodal datasets}}
} \\

\arrayrulecolor{dataNavText!30}
\specialrule{0.45pt}{0pt}{0pt}
\arrayrulecolor{dataRule}

\rowcolor{dataNav}
UrbanLoco~\cite{wen2020urbanloco}
& Raw GNSS, LiDAR, cameras, IMU, reference GNSS/INS
& Derivable from synchronized LiDAR/camera geometry and reference pose
& Geometry-assisted LOS/NLOS classification in dense urban canyons
& Reception-state labels must be reconstructed from synchronized environmental sensing and reference pose. \\

\cmidrule(lr){1-5}

\rowcolor{dataNav}
UrbanNav~\cite{hsu2023hong}
& Multi-grade raw GNSS, LiDAR, cameras, IMU, reference GNSS/INS
& Derivable from reference pose and synchronized LiDAR/camera context
& LOS/NLOS classification in medium-to-harsh urban canyons and tunnel transitions
& No native reception-state labels; accuracy depends on environmental-model quality and pose/time alignment. \\

\cmidrule(lr){1-5}

\rowcolor{dataNav}
KLT~\cite{hu2023fisheye}
& Raw multi-GNSS, reference GNSS/INS, fisheye camera, LiDAR, IMU
& Provided per-satellite LOS/NLOS labels from fisheye-based environmental labeling
& Direct training and benchmarking of satellite-level LOS/NLOS classifiers
& Only GPS/BeiDou are labeled; very-low-light imagery can reduce labeling reliability. \\

\cmidrule(lr){1-5}

\rowcolor{dataNav}
SensNav~\cite{gao2025sensnav}
& High-rate raw GNSS, multiple IMUs, RGB and sky-fisheye cameras (campus subset), reference GNSS/INS
& Derivable via sky-view projection in camera-equipped campus sequences
& Temporal LOS/NLOS classification with high-rate GNSS and sky-view context
& Sky-view imagery is not available for all sequences; labels require synchronized projection. \\

\cmidrule(lr){1-5}

\rowcolor{dataNav}
M2DGR~\cite{yin2021m2dgr}
& GNSS, multi-camera, sky camera, LiDAR, IMU, event/IR sensors
& No native labels; derivable from sky-camera imagery and synchronized reference pose in outdoor sequences
& Vision-assisted LOS/NLOS labeling and multimodal reception-state classification
& GNSS is low-rate and outdoor-limited; ground-robot geometry differs from road vehicles. \\

\arrayrulecolor{dataRule}
\specialrule{0.7pt}{0pt}{0pt}

\rowcolor{dataGNSSBand}
\multicolumn{5}{c}{
\textcolor{dataGNSSText}{\textbf{GNSS-centric observable datasets and classification benchmarks}}
} \\

\arrayrulecolor{dataGNSSText!30}
\specialrule{0.45pt}{0pt}{0pt}
\arrayrulecolor{dataRule}

\rowcolor{dataGNSS}
smartLoc~\cite{reisdorf2016comparable}
& Raw u-blox GNSS, NovAtel reference GNSS/INS, vehicle motion
& Provided LOS/NLOS indicator sourced from the NovAtel reference receiver
& Observable-level LOS/NLOS classification from code, carrier, Doppler, and $C/N_0$
& Receiver-derived labels; LOS does not distinguish clean LOS from LOS+MP. \\

\cmidrule(lr){1-5}

\rowcolor{dataGNSS}
GSDC~\cite{fu2022summary}
& Android raw GNSS and smartphone IMU
& Reference trajectories provided; no native per-satellite reception-state labels
& Weakly supervised reception-state modeling across diverse smartphone receivers
& No environmental sensing or native per-satellite reception-state labels. \\

\cmidrule(lr){1-5}

\rowcolor{dataGNSS}
WHU-Smartphone~\cite{wang2024whu}
& Multiple Android smartphones, low-cost GNSS module, reference GNSS/INS
& High-precision reference trajectory provided; no native per-satellite reception-state labels
& Receiver-diversity and cross-device reception-state studies
& High-precision reference data do not provide direct propagation-state ground truth. \\

\cmidrule(lr){1-5}

\rowcolor{dataGNSS}
Xu et al. Benchmark~\cite{xu2024machine}
& Measurement-level GNSS features across four scenarios
& Provided LOS/multipath/NLOS benchmark targets across four scenarios
& Standardized evaluation of learning-based multipath/NLOS classification and mitigation
& Evaluation is confined to four benchmark scenarios. \\

\arrayrulecolor{dataRule}
\specialrule{0.7pt}{0pt}{0pt}

\rowcolor{dataTrackBand}
\multicolumn{5}{c}{
\textcolor{dataTrackText}{\textbf{Raw-IF datasets for tracking-level reception-state classification}}
} \\

\arrayrulecolor{dataTrackText!30}
\specialrule{0.45pt}{0pt}{0pt}
\arrayrulecolor{dataRule}

\rowcolor{dataTrack}
TEX-CUP~\cite{narula2020texcup}
& Wideband multi-frequency GNSS IF, dual antennas, SDR observables, stereo cameras, IMUs, reference GNSS/INS
& Reference trajectory provided; no native satellite-state labels
& Correlator-, loop-, frequency-, and antenna-level reception-state analysis
& LOS/NLOS/LOS+MP truth requires independent scene/path annotation. \\

\cmidrule(lr){1-5}

\rowcolor{dataTrack}
SJTU GNSS Test Dataset~\cite{sjtu2018gnssdataset}
& Digital GNSS IF, raw observables, dual antennas, IMUs, calibrated trajectory, ephemerides
& Reference trajectory provided; no native satellite-state labels
& Tracking- and observable-level classification across blockage/MP scenarios
& Scenario classes are not per-satellite LOS/NLOS ground truth. \\

\cmidrule(lr){1-5}

\rowcolor{dataTrack}
CardiffNav~\cite{cardiffnav2026}
& Multi-channel raw GNSS RF/IQ, receiver GNSS, 3D LiDAR, RGB/stereo cameras, IMU, reference navigation
& No native satellite-state labels; derivable from environmental sensing and reference pose
& Raw-signal and multimodal LOS/NLOS/MP studies in canyon, obstruction, and GNSS-denied transitions
& RF--scene alignment and independent geometry-based labeling are required. \\

\arrayrulecolor{dataRule}
\specialrule{0.7pt}{0pt}{0pt}

\rowcolor{dataSimBand}
\multicolumn{5}{c}{
\textcolor{dataSimText}{\textbf{Controlled synthetic propagation datasets}}
} \\

\arrayrulecolor{dataSimText!30}
\specialrule{0.45pt}{0pt}{0pt}
\arrayrulecolor{dataRule}

\rowcolor{dataSim}
MVRGD-MUE~\cite{silva2026multipath}
& Simulated raw GNSS, vehicle truth, IMU/odometry, 3D ray-traced propagation
& LOS/LOS+MP/NLOS states known from simulated propagation paths
& Controlled classification with explicit propagation/path truth
& Fidelity depends on geometry, propagation, and receiver models. \\
\bottomrule

\end{tabularx}

\arrayrulecolor{black}
\end{table*}

The literature strongly discourages relying on rudimentary signal thresholds (e.g., static 35 dB-Hz limits),    as these hardware-dependent heuristics routinely misclassify strong reflections or mildly attenuated LOS signals~\cite{kubo2020gnss}. 
To ensure algorithms learn propagation-related signal characteristics rather than artifacts of the labeling procedure, researchers advocate for a reproducible labeling pipeline:
\begin{enumerate}
    \item Derive deterministic, geometry-based LOS/NLOS labels using 3D ray-tracing or sky-masking when accurate poses and perception data are available~\cite{beaucamp2025comparison}.
    \item Cross-check geometric labels against independent semantic imagery and, where appropriate, strictly corrected signal residuals as quality-control indicators to identify map gaps, dynamic occlusions, or vegetation~\cite{zhang2023learning}.
    \item Exclude ambiguous epochs (e.g., satellites exactly on building mask boundaries) to prevent injecting noisy labels into the training corpus~\cite{beaucamp2025comparison,zhang2023learning}.
\end{enumerate}

Within the dataset-construction pipeline, 3D maps and auxiliary sensors often function as offline annotation mechanisms that generate the reference labels required to train signal-based classifiers. In this role, they provide geometry- or perception-derived supervision rather than serving directly as online navigation sensors. However, the boundary between offline annotation and online operation is increasingly narrowing: the same 3D map, camera, and LiDAR mechanisms used to label datasets can also be integrated into live navigation systems for real-time satellite visibility prediction, measurement screening, and multipath mitigation. The following sections therefore examine how 3D map-based frameworks (Section~\ref{sec:3d_mapping}) and auxiliary-sensor strategies (Section~\ref{sec:auxiliary_sensors}) are deployed beyond dataset labeling as online GNSS anomaly-detection tools.

\section{3D Map-Based Visibility Determination}
\label{sec:3d_mapping}

This section investigates the integration of 3D environmental models to deterministically predict GNSS satellite visibility and identify NLOS and multipath conditions in urban canyons. Accurate spatial representations of buildings, vegetation, and infrastructure allow positioning engines to explicitly 
determine whether the direct satellite path is blocked or geometrically available and, with sufficiently detailed propagation models, to evaluate candidate reflected paths. This paradigm, widely formalized as \emph{3D-mapping-aided GNSS (3DMA GNSS)}, demonstrably enhances localization accuracy. By isolating LOS signals from degraded receptions, 3DMA GNSS enables the targeted exclusion, dynamic down-weighting, or explicit geometric correction of erroneous measurements~\cite{groves2019performance,zhong2023optimizing,wang2015smartphone,hsu2015nlos}.

The algorithmic approaches discussed in Section~\ref{sec:algorithms} primarily infer signal propagation states directly from the received waveform or measurement anomalies. In stark contrast, 3DMA frameworks predict visibility states deterministically based on the surrounding spatial geometry. 
Receivers subsequently fuse these geometry-driven visibility priors with GNSS measurements or signal-derived observables to mitigate multipath-induced biases and fortify the overall positioning geometry.

\begin{figure*}[t!] 
\centering 
\includegraphics[width=0.84\textwidth] {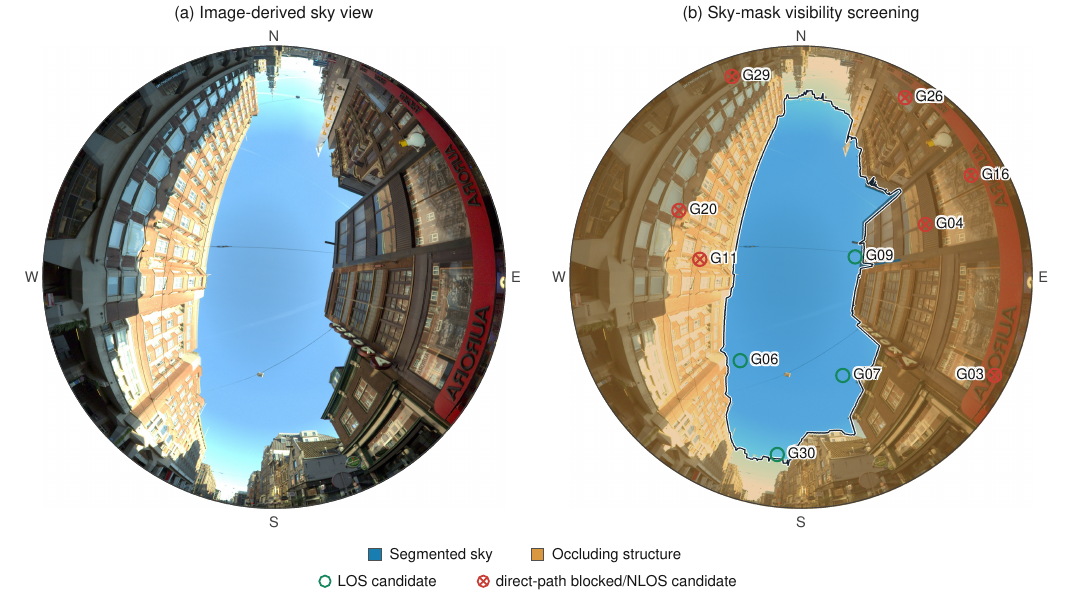}
\caption{Illustration of image-derived GNSS visibility screening from georeferenced urban imagery. (\textbf{a}) A City of Amsterdam Open Panorama image at Muntplein, Amsterdam, reprojected into a north-up hemispherical sky view. (\textbf{b}) The resulting sky/occluder mask is queried with representative satellite directions to classify candidate direct-path visibility and blockage. Imagery: Gemeente Amsterdam, Kernregistratie Panoramabeelden, CC BY 4.0; reprojected and processed by the authors.}
\label{fig:image_derived_sky_visibility} 
\vspace{-0.3cm} 
\end{figure*} 

To systematically navigate this domain, the remainder of this section is partitioned into two distinct focal areas:
\begin{itemize}
    \item \textbf{3D Mapping Modalities:} We examine the acquisition, representation, and continuous maintenance of the spatial databases required for urban environmental awareness.
    \item \textbf{Algorithmic Utilizations:} We categorize the computational deployment of these maps, analyzing geometric screening, shadow matching, ray-tracing, and neural occlusion models. We evaluate their respective computational trade-offs and strategies for fusion with the signal-based detection frameworks established in Section~\ref{sec:algorithms}.
\end{itemize}

\subsection{3D Mapping Modalities for GNSS Visibility Analysis}

Map-based GNSS visibility determination requires an environmental representation from which satellite obstruction or reflection geometry can be queried. The required representation depends primarily on the intended propagation model. Extruded building models are generally sufficient for building-boundary prediction and shadow matching~\cite{wang2013gnss,hsu2015nlos}, whereas explicit multipath ray tracing benefits from more detailed roof and façade geometry and, for physically based reflection modeling, additional surface properties~\cite{li2024characterization}. Map completeness and temporal validity are equally important: structures or occluders missing from the stored representation can produce discrepancies between predicted and actual satellite visibility~\cite{beaucamp2025comparison,xu2024urban}. The principal environmental representations used in this context are reviewed below.

\subsubsection{Explicit Geometric Maps}

The conventional foundation of 3DMA GNSS is an explicit geometric representation of the surrounding environment, ranging from extruded building footprints and City Geography Markup Language (CityGML) models to polygonal surface representations~\cite{kbayer2018performances,groves2019performance,schon2022integrity}. Their principal advantage is direct geometric interpretability: given a receiver candidate position and satellite direction, building boundaries or surface intersections can be queried explicitly to predict obstruction and, with sufficiently detailed geometry, potential reflection paths. Early explicit-map studies used virtual city and 3D GIS models to characterize urban LOS and multipath propagation, while photogrammetric and LiDAR-derived surface models were also used for satellite-visibility prediction~\cite{bradbury2007code,suh2007evaluation,taylor2007modelling}.
The appropriate geometric fidelity depends on the visibility model; Level of Detail 1 (LoD1) building models provide a scalable representation for building-boundary and LOS/NLOS screening, whereas explicit multipath simulation generally requires richer façade and roof geometry~\cite{li2024characterization}. Large-scale products are increasingly reducing the coverage limitation of such models; for example, the GlobalBuildingAtlas provides billions of height-augmented LoD1 building instances at global scale~\cite{zhu2025globalbuildingatlas}. Nevertheless, map completeness, height accuracy, and temporal freshness directly affect visibility predictions.

Where greater local geometric fidelity is required, pre-surveyed LiDAR point clouds or photogrammetric meshes can represent façades, vegetation, and other structures omitted by coarse building models~\cite{ma2024solid,lv2025accelerated}. These representations provide a richer basis for ray-intersection and reflection modeling, but increase acquisition, storage, preprocessing, and map-maintenance requirements. Accordingly, explicit maps exhibit a fundamental trade-off between geographic scalability and the geometric detail available to the subsequent GNSS visibility algorithm.

\subsubsection{Image-Derived Sky Representations} 
For satellite-visibility determination, a complete volumetric reconstruction is not always required. 
Early transport-oriented work combined environmental imagery with predicted satellite trajectories to estimate LOS/blocked satellite availability, with subsequent extensions incorporating multipath effects in land-mobile environments~\cite{marais2000evaluation,marais2005land}. Georeferenced panoramic imagery or virtual views rendered from a 3D city model can similarly be transformed into hemispherical sky representations, expressing the surrounding scene directly in the azimuth--elevation domain. Suzuki et al.~\cite{suzuki2015simulation} later demonstrated virtual fisheye generation from photo-realistic city models, while subsequent work exploited panoramic and street-view imagery for local and large-scale GNSS visibility mapping~\cite{tongleamnak2017simulation,kou2021mapping,gu2025sva}. 
After sky/occluder segmentation, the resulting angular mask can be queried directly using satellite azimuth and elevation. 
Fig.~\ref{fig:image_derived_sky_visibility} illustrates this process using a real urban panorama and Global Positioning System (GPS) satellite geometry. Satellite directions are projected onto the segmented hemispherical mask, with sky projections treated as candidate LOS and projections onto occluding structures as direct-path blocked or candidate NLOS. These remain geometry-derived visibility labels rather than radio frequency (RF) ground truth, since sky-visible signals may still contain multipath and blocked directions do not necessarily imply that a reflected signal is being tracked. Such angular representations provide a compact and computationally efficient alternative to full 3D models for direct-path visibility prediction and labeling, but discard scene depth and surface orientation and therefore cannot reconstruct explicit reflection paths or detailed multipath delays.

\subsubsection{Neural Scene Representations}

Neural scene representations provide an alternative to explicitly storing building surfaces or dense point clouds. Radiance-field methods such as Neural Radiance Fields (NeRF) encode scene structure as a continuous learned function~\cite{mildenhall2021nerf}, enabling visibility-related quantities to be queried along arbitrary spatial rays.
This representation has recently been adapted directly to GNSS through \emph{Neural City Maps}, which learn urban structure from imagery and use the resulting continuous field for satellite NLOS prediction and shadow matching~\cite{neamati2023neural,neamati2024neural}. Compared with coarse building models, the learned field can represent scene structures present in the training imagery without requiring an explicit LoD1/LoD2 mesh. However, its reliability depends on spatial coverage and quality of the training observations, and poorly observed regions can produce uncertain or erroneous visibility predictions. Neural representations therefore offer a flexible continuous alternative to conventional maps, while introducing additional training cost and reduced geometric transparency.

\subsubsection{Crowdsourced Signal-Based Mapping}
\label{subsec:crowdsourced_mapping}

\begin{figure*}[t!]
\centering
\includegraphics[width=0.85\linewidth]{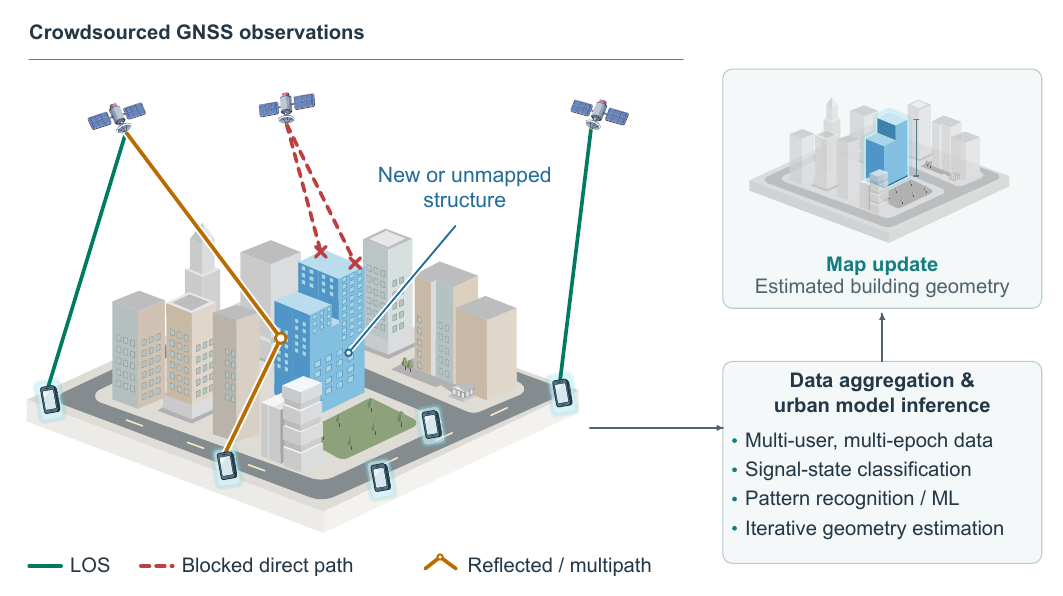}
\caption{Conceptual pipeline of crowdsourced signal-based mapping. In contrast to conventional 3DMA GNSS, which uses an existing environmental model to predict satellite visibility, the inverted paradigm aggregates observed GNSS reception states (e.g., LOS, NLOS, or blocked) from multiple users to infer or update the surrounding urban geometry.}
\label{fig:crowdsourced_mapping}
\vspace{-0.3cm}
\end{figure*}

A complementary paradigm inverts the conventional 3DMA relationship by using GNSS reception itself to infer environmental structure, as conceptually illustrated in Fig.~\ref{fig:crowdsourced_mapping}. Because transitions between LOS, NLOS, and blocked reception depend on satellite direction and surrounding obstacles, repeated observations from different locations and satellite geometries can reveal building boundaries or sky-obstruction profiles. An early demonstration of this inverse relationship used reductions in GPS signal-to-noise ratio caused by building obstruction to reconstruct urban building geometry~\cite{kim2008localization}.
Lines and Basiri~\cite{lines20213d}, for example, use crowdsourced GNSS signal patterns together with known building footprints to estimate LoD1 building heights without dedicated LiDAR or visual mapping sensors. 
Subsequent work extends this concept from map construction toward environmental monitoring. GNSS satellite visibility has been used to estimate local sky boundaries~\cite{xu2020sky}, while discrepancies between observed and map-predicted visibility can reveal changes to existing building models~\cite{xu2024urban}. More recently, Jia et al.~\cite{jia2025buildingheight} estimate building height from the transition of learned smartphone-GNSS LOS/NLOS probabilities, further illustrating how signal-domain classification can be converted into geometric map information.

Learning-based representations extend this inversion beyond explicit building-height estimation. Zhao et al.~\cite{zhao2024crowdsourced} introduced Satellitic Radiance Fields (SaRF), in which large-scale crowdsourced GPS measurements train a differentiable volumetric representation of urban structure and radio propagation. Such approaches provide a potential mechanism for inexpensive and frequently refreshed environmental mapping, but their reliability depends on the spatial diversity and density of contributed measurements, the accuracy of inferred reception states, and sufficient satellite-view geometry. Crowdsourced signal-derived maps are therefore best viewed as a complementary mechanism for constructing or updating conventional environmental models rather than as guaranteed replacements for surveyed geometry.

\subsection{Map-Based Visibility Prediction and Detection}
\label{subsec:3d_map_detection_algs}

\begin{figure*}[t!]
    \centering
    \resizebox{\textwidth}{!}{%
    \begin{tikzpicture}[
        >=Latex,
        font=\sffamily\footnotesize,
        logicbox/.style={draw=gray!55, fill=gray!3, rounded corners=2pt,
            line width=0.7pt, align=center, minimum height=0.82cm,
            inner xsep=3mm, inner ysep=1.5mm},
        panel/.style={draw=#1!72!black, fill=#1!4, rounded corners=4pt,
            line width=0.8pt, minimum width=7.35cm, minimum height=4.65cm},
        paneltitle/.style={align=center, text width=6.55cm,
            font=\sffamily\bfseries\normalsize},
        predictor/.style={draw=blue!62!black, fill=blue!7, rounded corners=2.5pt,
            line width=0.7pt, align=center, text width=5.55cm,
            minimum height=0.90cm, inner sep=2.5pt},
        verifier/.style={draw=orange!78!black, fill=orange!8,
            rounded corners=2.5pt, line width=0.7pt, align=center,
            text width=5.55cm, minimum height=0.96cm, inner sep=2.5pt},
        flowarrow/.style={->, line width=0.8pt, draw=gray!72!black},
        linkarrow/.style={->, line width=0.8pt, draw=gray!75!black},
        linklabel/.style={fill=white, inner xsep=2pt, inner ysep=1pt,
            font=\sffamily\scriptsize, text=black!82}
    ]

    \node[panel=blue] (predictorpanel) at (-4.90,1.05) {};
    \node[panel=orange] (verifierpanel) at (4.90,1.05) {};

    \node[paneltitle, text=blue!58!black] at (-4.90,2.85)
        {Map-Derived Spatial Predictors\\[-1pt]
        \scriptsize\mdseries\itshape
        3D map $\rightarrow$ expected visibility / propagation prior};

    \node[paneltitle, text=orange!76!black] at (4.90,2.85)
        {Reception-State Verifiers\\[-1pt]
        \scriptsize\mdseries\itshape
        map prior + GNSS observation $\rightarrow$ consistency evidence};

    \node[predictor] (screen) at (-4.90,1.90)
        {\textbf{Geometric LOS Screening}\\[-1pt]
        \scriptsize Binary skymask / visibility boundary};

    \node[predictor] (neural) at (-4.90,0.73)
        {\textbf{Neural-Field Visibility Prediction}\\[-1pt]
        \scriptsize Soft LOS probability / transmittance field};

    \node[predictor] (raytrace) at (-4.90,-0.44)
        {\textbf{Explicit 3D Ray-Tracing}\\[-1pt]
        \scriptsize Excess delay, Doppler, attenuation priors};

    \node[verifier] (shadow) at (4.90,1.52)
        {\textbf{Shadow Matching}\\[-1pt]
        \scriptsize Visibility / signal consistency};

    \node[verifier] (ranging) at (4.90,-0.17)
        {\textbf{3DMA Ranging}\\[-1pt]
        \scriptsize Pseudorange-consistency evidence};

    \coordinate (shadowupper) at ($(shadow.west)+(0,0.20)$);
    \coordinate (shadowlower) at ($(shadow.west)+(0,-0.20)$);
    \coordinate (rangingin) at (ranging.west);

    \draw[linkarrow] (screen.east) --
        node[linklabel, above, sloped]
        {binary visibility prior} (shadowupper);

    \draw[linkarrow] (neural.east) --
        node[linklabel, below, sloped]
        {soft $p_{\mathrm{LOS}}$} (shadowlower);

    \draw[linkarrow] (raytrace.east) --
        node[linklabel, above, sloped]
        {$\Delta\tau$, $\Delta f_d$, $\alpha$} (rangingin);

    \end{tikzpicture}%
    }
    \caption{Taxonomy of map-based algorithms for GNSS visibility analysis.
    \emph{Map-derived spatial predictors} query an environmental representation
    to generate expected visibility or propagation priors, whereas
    \emph{reception-state verifiers} compare these priors with GNSS observations
    to evaluate map--measurement consistency and provide evidence about anomalous
    reception states.}
    \label{fig:algorithm_taxonomy_tree}
    \vspace{-0.3cm}
\end{figure*}
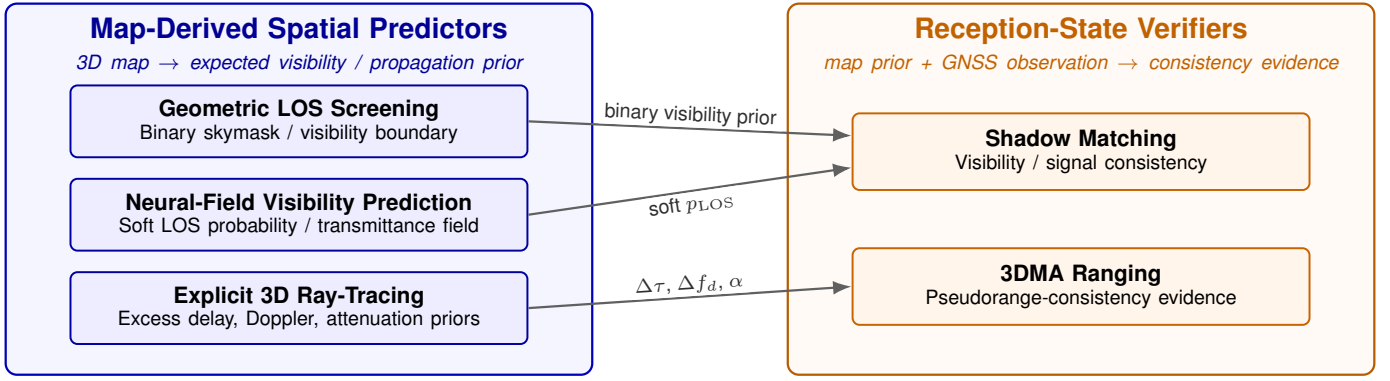

Building upon the environmental representations introduced in the preceding review of 3D mapping modalities, this subsection examines how 3D maps are computationally operationalized for GNSS reception-state inference. A useful distinction is between map-based \emph{prediction} and consistency-based \emph{detection}. Signal- and observable-domain methods infer reception states from effects already manifested in the received signal or derived measurements. In contrast, map-based predictors infer the expected visibility or propagation state from the receiver position, satellite geometry, and environmental model, without requiring the current GNSS observation being classified. Such predictions may be precomputed offline or evaluated online at each navigation epoch~\cite{groves2019performance,karimidoona2026advanced}. When the map-derived expectation is subsequently compared or fused with the actual GNSS observation, the spatial prior becomes a source of reception-state consistency evidence. Agreement supports the predicted state, whereas disagreement may indicate anomalous reception, map incompleteness, dynamic occlusion, or receiver-pose uncertainty. Static maps are particularly susceptible to unmodeled temporary obstacles and environmental changes~\cite{beaucamp2025comparison,xu2024urban}, which motivates fusion with the signal- and observable-domain detectors discussed in \S~\ref{sec:algorithms}.

These methods produce different forms of spatial prior and map--measurement consistency evidence. Because this survey focuses on reception-state inference rather than downstream coordinate estimation, we organize them into the functional taxonomy shown in Fig.~\ref{fig:algorithm_taxonomy_tree}: \emph{map-derived spatial predictors} generate binary visibility, propagation-path, or probabilistic visibility information, whereas \emph{reception-state verifiers} combine such predictions with GNSS measurements to assess consistency and provide additional evidence about the reception state.

\subsubsection{Geometric LOS Screening}

\begin{figure}[t]
\centering

\includegraphics[width=0.96\linewidth]{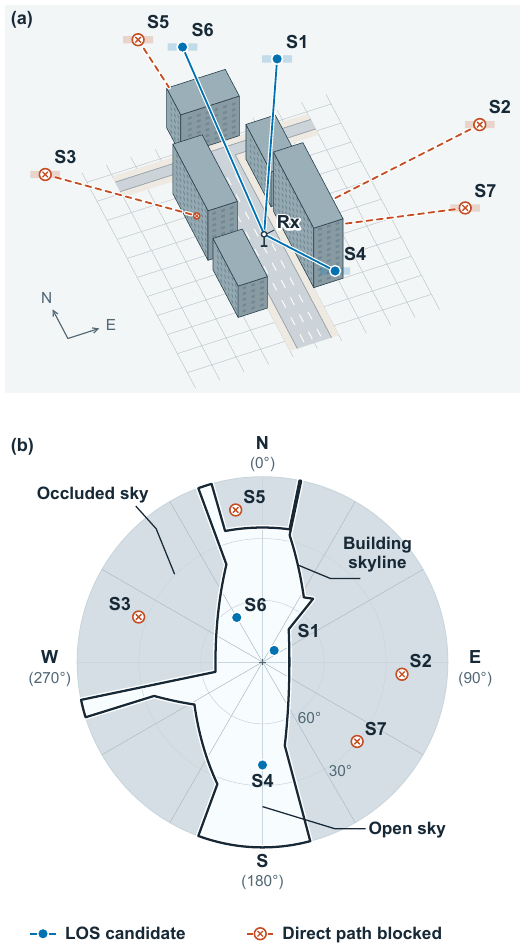}
\caption{Conceptual illustration of geometric GNSS visibility screening. (\textbf{a}) Satellite--receiver rays are tested for intersections with a 3D urban model. (\textbf{b}) The same geometry defines a sky mask, where projected satellite directions are classified relative to the building skyline.}
\label{fig:geom_screening_skymask}
\end{figure}

Geometric LOS screening is the most direct form of map-based visibility prediction. Given an assumed receiver position and satellite direction, a ray is tested against the surrounding building geometry; an unobstructed ray predicts direct-path visibility, whereas an intersection
predicts direct-path blockage~\cite{peyraud2013non,hsu2015nlos}. As illustrated in Fig.~\ref{fig:geom_screening_skymask}, the same geometric relationship can be represented either through explicit 3D intersection tests or through a building-boundary (skymask) representation in the azimuth--elevation domain. For repeated real-time queries, building boundaries may be precomputed over a spatial grid so that online visibility prediction reduces to comparing the satellite elevation with the local skyline at the corresponding azimuth. Alternative implementations further accelerate this operation; for example, texture-based satellite visibility detection performs ray marching on a 2D height-map texture and was developed specifically as an efficient alternative to conventional 3D ray tracing for mobile and automotive platforms ~\cite{acarman2017texture}. The resulting prediction is primarily binary and therefore highly interpretable, but its reliability remains coupled to the assumed receiver position and the completeness of the environmental model. Missing vegetation, temporary structures, dynamic obstacles, or geometric errors near building boundaries can cause disagreement between predicted and actual reception states~\cite{beaucamp2025comparison,zhong2023optimizing}. Geometric screening therefore provides an efficient spatial prior for identifying likely obstruction, while more detailed propagation or measurement-consistency models are required when the indirect signal path itself must be characterized.

\subsubsection{Explicit 3D Ray-Tracing}

Explicit 3D ray-tracing mathematically simulates GNSS signal propagation by modeling geometric interactions between satellite transmissions and the 3D urban environment. Given a sufficiently detailed city model---incorporating wall geometries, roof topologies, and, where required, surface material properties---the algorithm evaluates both direct and indirect propagation routes. By mapping these propagation paths to the receiver's coordinates, the algorithm can characterize the expected signal state: an unobstructed direct path without significant indirect components corresponds to clean LOS reception; when the direct path is blocked but an indirect reflected or diffracted component reaches the receiver, NLOS reception is predicted; and the coexistence of multiple significant propagation components is characterized as multipath~\cite{li2024characterization,pisova2015detection}.

Unlike basic geometric screening, which outputs only binary visibility masks, ray-tracing functions as a higher-fidelity physical propagation simulator. It can predict critical propagation metrics, including excess path delay ($\Delta\tau$), relative Doppler frequency shifts ($\Delta f_d$), phase rotations ($\Delta\phi$), and signal attenuation ($\alpha$)~\cite{li2024characterization}. By extracting these map-derived propagation features, the ray-tracer supplies physically informed spatial priors for downstream delay-based verification (3DMA ranging) or for constructing detailed spatial fingerprints.

Historically, the computational burden of evaluating multiple reflections has limited the use of detailed ray-tracing in real-time applications. However, recent acceleration techniques---most notably pixel shader-based ray-tracing (PSBRT)~\cite{lyu2023efficient}---substantially reduce this burden. By projecting testing rays through an omnidirectional camera model and exploiting highly parallel Graphics Processing Unit (GPU) processing, PSBRT reduces the computational growth from exponential in conventional mirror-based implementations to approximately linear with the maximum considered number of reflections. This enables real-time-capable NLOS correction and environmental rendering on parallel GPU hardware, including cellphone-embedded GPUs~\cite{lyu2023efficient}.

Nevertheless, the fidelity of deterministic ray-tracing depends strongly on the geometric accuracy and completeness of the 3D map, the assumed electromagnetic properties of reflecting surfaces, and the propagation mechanisms represented by the model. Detailed roof and facade geometries also increase mapping and computational requirements, while dynamic objects and unmodeled higher-order interactions can introduce discrepancies between simulated and actual reception conditions~\cite{lyu2023efficient,hsu2015nlos}. Errors or omissions in the building model can lead to incorrect visibility and propagation predictions and may consequently degrade 3DMA GNSS positioning performance~\cite{zhong2022outlier}. Uncertainty in surface material properties can likewise affect the predicted attenuation and phase of reflected components, limiting the accuracy of ray-tracing-based multipath modeling and correction~\cite{bijjahalli2019gnss}. Thus, when erroneous map-derived predictions are used for measurement exclusion, weighting, or correction, the resulting position solution may be degraded rather than improved. Hybrid approaches can partially mitigate these limitations by combining ray-traced environmental information with independent consistency checks; for example, Zidan et al.~\cite{zidan2020mapping} integrate ray-tracing-derived information with consistency checking and a decision-tree stage to identify and exclude multipath/NLOS-affected GNSS measurements.

\subsubsection{Shadow Matching}
\label{sec:shadow_likelihood}

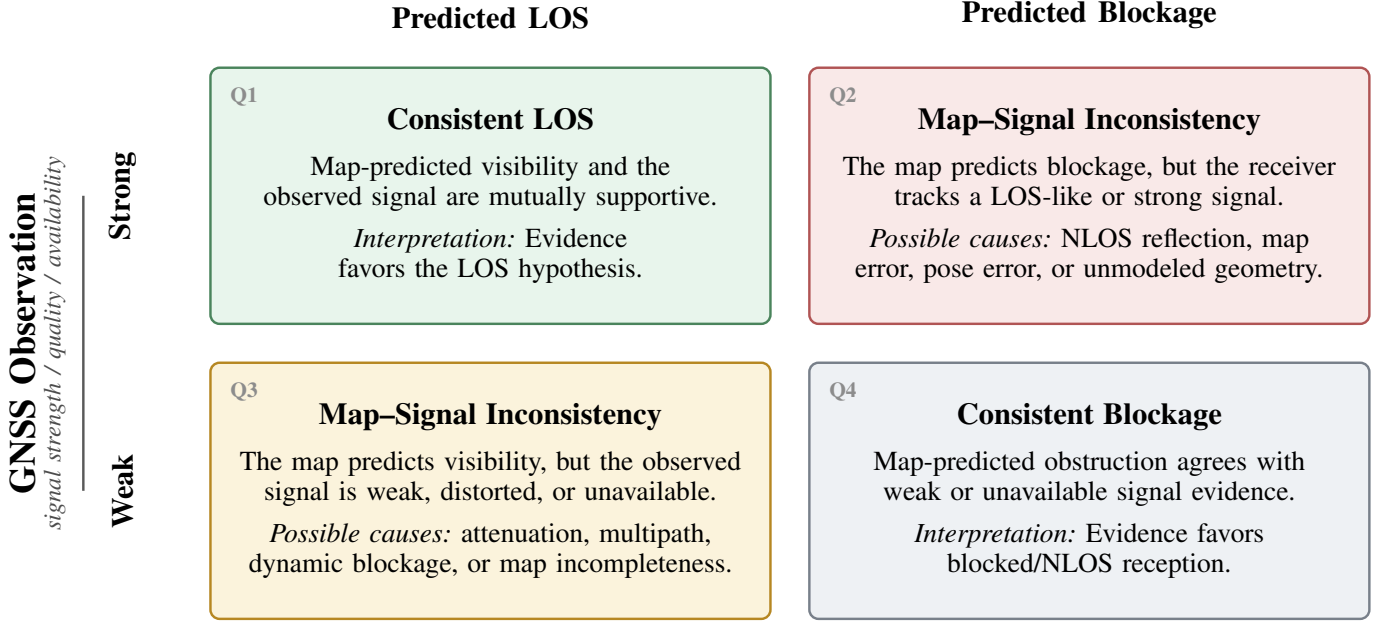
\begin{figure*}[htbp]
\centering
\resizebox{\textwidth}{!}{%
\begin{tikzpicture}[
    >=latex,
    node distance=0.42cm and 0.42cm,
    cell/.style={
        draw,
        thick,
        rounded corners=3pt,
        minimum width=5.0cm,
        minimum height=2.95cm,
        text width=5.95cm,
        align=center,
        inner sep=2.5mm,
        font=\small
    },
    header/.style={
        font=\normalsize\bfseries,
        align=center,
        text=black
    },
    axislabel/.style={
        font=\large\bfseries,
        align=center,
        text=black
    },
    subnote/.style={
        font=\footnotesize\itshape,
        text=black!70,
        align=center
    },
    qtag/.style={
        anchor=north west,
        font=\scriptsize\bfseries,
        text=black!45
    }
]

\definecolor{AgreeGreen}{RGB}{232,245,236}
\definecolor{AgreeBorder}{RGB}{74,132,92}

\definecolor{NeutralGray}{RGB}{238,241,244}
\definecolor{NeutralBorder}{RGB}{120,128,138}

\definecolor{WarnAmber}{RGB}{251,243,220}
\definecolor{WarnBorder}{RGB}{182,135,34}

\definecolor{AlertRose}{RGB}{249,233,232}
\definecolor{AlertBorder}{RGB}{176,84,84}


\node[cell, fill=AgreeGreen, draw=AgreeBorder] (Q1) {%
{\normalsize\bfseries Consistent LOS}\\[1.2ex]
Map-predicted visibility and the observed signal are mutually supportive.\\[0.9ex]
\textit{Interpretation:} Evidence favors the LOS hypothesis.
};

\node[cell, fill=AlertRose, draw=AlertBorder, right=of Q1] (Q2) {%
{\normalsize\bfseries Map--Signal Inconsistency}\\[1.2ex]
The map predicts blockage, but the receiver tracks a LOS-like or strong signal.\\[0.9ex]
\textit{Possible causes:} NLOS reflection, map error, pose error, or unmodeled geometry.
};

\node[cell, fill=WarnAmber, draw=WarnBorder, below=of Q1] (Q3) {%
{\normalsize\bfseries Map--Signal Inconsistency}\\[1.2ex]
The map predicts visibility, but the observed signal is weak, distorted, or unavailable.\\[0.9ex]
\textit{Possible causes:} attenuation, multipath, dynamic blockage, or map incompleteness.
};

\node[cell, fill=NeutralGray, draw=NeutralBorder, right=of Q3] (Q4) {%
{\normalsize\bfseries Consistent Blockage}\\[1.2ex]
Map-predicted obstruction agrees with weak or unavailable signal evidence.\\[0.9ex]
\textit{Interpretation:} Evidence favors blocked/NLOS reception.
};

\node[qtag] at ([xshift=1.2mm,yshift=-1.2mm]Q1.north west) {Q1};
\node[qtag] at ([xshift=1.2mm,yshift=-1.2mm]Q2.north west) {Q2};
\node[qtag] at ([xshift=1.2mm,yshift=-1.2mm]Q3.north west) {Q3};
\node[qtag] at ([xshift=1.2mm,yshift=-1.2mm]Q4.north west) {Q4};

\node[header, above=0.32cm of Q1] (H1) {Predicted LOS};
\node[header, above=0.32cm of Q2] (H2) {Predicted Blockage};

\coordinate (TopRowCenter) at ($(Q1.center)!0.5!(Q2.center)$);
\coordinate (BottomRowCenter) at ($(Q3.center)!0.5!(Q4.center)$);

\node[header, rotate=90] (V1)
at ([xshift=-1.0cm]Q1.west |- TopRowCenter)
{Strong};

\node[header, rotate=90] (V2)
at ([xshift=-1.0cm]Q3.west |- BottomRowCenter)
{Weak};

\path (H1) -- coordinate[midway] (Xmid) (H2);

\node[axislabel, above=0.78cm of Xmid] (Xtitle)
{Map-Derived Visibility Prediction};

\node[subnote, below= -0.2 cm of Xtitle] at (Xtitle.south)
{3D map, receiver pose, and satellite geometry};

\draw[thick, draw=black!65]
([xshift=-0.2cm,yshift=0.28cm]H1.north west) --
([xshift=0.2cm,yshift=0.28cm]H2.north east);

\coordinate (MatrixCenterY)
at ($(TopRowCenter)!0.5!(BottomRowCenter)$);

\node[axislabel, rotate=90] (Ytitle)
at ([xshift=-2.15cm]Q1.west |- MatrixCenterY)
{GNSS Observation};

\node[subnote, rotate=90]
at ([xshift=0.1cm]Ytitle.south)
{signal strength / quality / availability};

\draw[thick, draw=black!65]
([xshift=-1.45cm]Q1.west |- TopRowCenter) --
([xshift=-1.45cm]Q3.west |- BottomRowCenter);

\end{tikzpicture}%
}
\caption{Map--measurement consistency interpretation of shadow matching. Agreement between predicted visibility and GNSS observations supports the LOS or blockage hypothesis, while disagreement indicates anomalous reception or uncertainty due to reflections, multipath, dynamic occlusion, map errors, or receiver-pose errors.}
\label{fig:shadow_matching_matrix}
\vspace{-0.3cm}
\end{figure*}

Introduced by Groves~\cite{groves2011shadow}, shadow matching is fundamentally a 3DMA positioning technique: satellite visibility predicted from a 3D city model is compared with visibility evidence obtained from the receiver, and candidate positions are scored according to their consistency. Subsequent implementations refine this scoring using improved visibility models and signal-derived LOS/NLOS information~\cite{wang2013gnss,wang2015smartphone, xu2020machine,zhong2023optimizing}. From the reception-state perspective adopted in this survey, the same comparison can be interpreted as a geometry-aware consistency test when the receiver pose is sufficiently constrained. Agreement between map-predicted and measurement-derived visibility supports the corresponding reception hypothesis, whereas disagreement flags an inconsistency requiring further interpretation. Importantly, such disagreement does not uniquely imply NLOS reception: map incompleteness, pose uncertainty, dynamic occluders, or measurement errors can produce the same outcome. This distinction motivates probabilistic rather than rigid binary scoring. Recent 3DMA frameworks use likelihood or Bayesian formulations to accommodate uncertainty in visibility prediction and observed signal state ~\cite{zhong2023optimizing}. Set-valued shadow matching further represents building and shadow regions using constrained zonotopes, reducing sensitivity to spatial discretization and boundary uncertainty~\cite{bhamidipati2022set}. Neural City Maps provide another extension by replacing discrete building-only visibility predictions with learned continuous scene representations for shadow prediction~\cite{neamati2023neural,neamati2024neural}. These developments extend shadow matching toward increasingly uncertainty-aware map--measurement consistency inference.

\subsubsection{3DMA Ranging}
\label{sec:delay_verification}
Range-based 3DMA GNSS uses the environmental model together with pseudorange measurements to evaluate the compatibility between map-predicted reception conditions and the observed ranges~\cite{hsu2015nlos,ng2020computation, ng2021urban}. Although these methods were developed primarily for map-aided positioning and NLOS correction, their ranging likelihoods can also be interpreted from a reception-state perspective as \emph{pseudorange-consistency evidence}. Geometrical ranging methods explicitly estimate the additional propagation distance associated with an NLOS path. Ray-tracing approaches reconstruct candidate reflected paths and their excess delays~\cite{hsu2015nlos}, whereas skymask-based methods approximate the reflection geometry using building boundaries to reduce the computational burden~\cite{ng2020computation}. The resulting map-predicted pseudorange can then be compared with the receiver measurement: agreement supports the candidate propagation hypothesis, whereas a large inconsistency may indicate an incorrect reflection model, additional propagation effects, map error, or measurement contamination. A complementary family, commonly termed \emph{likelihood-based ranging} (LBR), avoids explicitly reconstructing the reflection path. Instead, the pseudorange innovation is evaluated statistically according to the map-predicted LOS/NLOS condition, providing a map-conditioned range-consistency likelihood~\cite{ng2021urban,zhong2023optimizing}. Thus, ranging-based 3DMA spans a continuum from explicit geometric delay prediction to statistical, map-conditioned pseudorange modeling. In the context of this survey, its key contribution is the use of map-conditioned pseudorange information as complementary evidence for reception-state inference, while subsequent measurement correction, weighting, or exclusion is regarded as downstream mitigation.

\subsubsection{Neural-Field Visibility Prediction} 
\label{sec:neural_prediction} 
Building on the neural scene representations introduced in the preceding mapping-modality review, satellite visibility is inferred by querying the learned field along the satellite-directed ray rather than through the explicit ray--polygon intersections used by conventional 3D city models. Let $\mathbf{x}_r$ denote the receiver position and $\mathbf{u}_s$ the unit vector toward satellite $s$. The candidate propagation path is parameterized as 
\begin{equation} 
\mathbf{x}(t)=\mathbf{x}_r+t\mathbf{u}_s,\qquad t\geq0 , 
\end{equation} 
where $t$ denotes distance along the ray. Samples evaluated along this path provide learned density- or occupancy-related quantities that can be aggregated into a visibility feature or soft LOS/NLOS score. For density-based neural fields, the corresponding volumetric transmittance can be expressed generically as~\cite{mildenhall2021nerf} 
\begin{equation} 
T(t)=\exp\!\left( -\int_0^t \sigma(\mathbf{x}(s))\,ds \right), 
\end{equation} 
where $\sigma(\mathbf{x})$ denotes the learned field density. This ray-integrated quantity represents the transmittance implied by the accumulated learned density along the satellite direction and can be mapped to a soft visibility prediction. This contrasts with the hard LOS/NLOS decision produced by conventional geometric ray casting. This formulation is the algorithmic counterpart of the neural mapping representations introduced earlier. \emph{Neural City Maps} demonstrate the principle directly for GNSS NLOS prediction and subsequently for shadow matching~\cite{neamati2023neural,neamati2024neural}. 

By replacing discrete building boundaries with a continuous learned scene field, these methods can represent environmental structures present in the training imagery that may be absent from simplified building-only models and provide spatially smooth visibility evidence near complex scene boundaries. The resulting flexibility, however, introduces a different set of dependencies. Prediction quality is governed by the spatial coverage and fidelity of the training observations, and poorly observed regions or learned scene uncertainty can propagate directly into the satellite-visibility estimate~\cite{neamati2023neural,neamati2024neural}. Neural-field visibility prediction should therefore be regarded as a soft, learned spatial prior. Compared with explicit 3D maps, it offers continuous querying and potentially richer scene representation at the cost of offline scene learning, dependence on training coverage, and reduced geometric transparency. 

Overall, 3DMA visibility methods span complementary levels of environmental reasoning. Geometric screening provides efficient binary visibility priors, whereas explicit ray tracing adds richer propagation-path information; shadow matching and 3DMA ranging subsequently exploit map--measurement consistency as reception-state evidence. Neural-field methods extend this spectrum toward continuous, learned visibility priors. Nevertheless, all map-based approaches remain dependent on the fidelity and temporal validity of their environmental representation, motivating the use of auxiliary sensors that can provide complementary information about the current receiver dynamics and surrounding scene.

\section{Auxiliary Sensors for Urban NLOS and Multipath Detection}
\label{sec:auxiliary_sensors}

While 3D city models provide map-derived spatial priors, their static nature limits their ability to represent transient occluders and short-term environmental changes. Auxiliary sensors provide complementary, time-varying information that can be used to test the consistency of GNSS measurements against receiver motion or the locally observed environment. 
Crucially, while these sensors are standard components for downstream navigation and state estimation, this review frames them through a specific functional lens: as auxiliary sources of evidence for GNSS reception-state verification. By providing kinematic or geometric information that is structurally decoupled from the GNSS RF chain, these sensors enable the navigation engine to assess whether GNSS measurements are consistent with the expected receiver motion or surrounding environment before they are fully incorporated into the navigation solution. The following subsections detail how inertial, visual, and laser-based data are utilized to refine signal-state classification, shifting the focus from their traditional role in coordinate estimation toward their complementary role in GNSS anomaly detection and verification.

\subsection{IMU-Aided Kinematic Consistency Checking}
\label{subsec:imu_consistency}
Inertial Measurement Units (IMUs) operate at high sampling rates, typically tens to hundreds of hertz, and provide continuous measurements of specific force and angular rate and, over short intervals, support prediction of receiver motion and attitude. Within the signal-classification taxonomy of this survey, the IMU therefore acts primarily as a short-term kinematic consistency checker. Unlike cameras or LiDAR, it does not observe satellite blockage or reflecting surfaces; instead, it provides an independent motion reference against which GNSS measurements can be tested.

Early closely coupled GPS/INS work combined raw-GNSS fault detection and exclusion with inertial integration to maintain navigation integrity in shadowed environments~\cite{clark2008gpsins}.
Sun et al.~\cite{sun2021imu} exploit this principle in an IMU-aided multiple-fault detection and exclusion scheme for urban GNSS/IMU integration. Their pseudorange-comparison method combines IMU data with GNSS pseudoranges in a sliding-window framework to detect and exclude multiple simultaneous faulty measurements. 
At the tracking level, Yan et al.~\cite{yan2023insmultipath} integrate inertial navigation system (INS) aiding with vector tracking and a multipath-error estimator to estimate and mitigate dynamic short-delay multipath in dense urban environments. These approaches illustrate how inertial prediction can provide additional temporal or kinematic constraints when GNSS measurements become inconsistent with expected receiver motion.

In multi-antenna systems, IMU-based attitude tracking can further assess whether the spatial phase geometry of received signals is consistent with rigid-body antenna-array motion. Ahmed et al. combine GNSS array measurements with IMU data using a UKF, detecting inconsistent satellite combinations affected by multipath~\cite{ahmed2025gnss}. The resulting method provides an IMU-aided geometric consistency check for multipath and NLOS detection.

\subsection{Camera-Aided NLOS Detection}
\label{sec:vision_based_approaches}
Upward-looking imaging sensors have a long history in GNSS visibility analysis and NLOS mitigation. Early work combined an omnidirectional infrared (IR) camera with a digital surface model to extract urban sky/building boundaries under satellite-blocked conditions~\cite{meguro2008development}. This concept was subsequently used explicitly to identify satellites without direct visibility and exclude their measurements from urban GPS positioning~\cite{meguro2009gps}.
In that framework, infrared contrast between sky and surrounding buildings was exploited to identify satellites without direct visibility and prevent their measurements from contaminating the positioning solution.
Building on this principle, upward-pointing fisheye and omnidirectional cameras were subsequently used to estimate sky and obstacle regions and project satellite directions into the image for real-time visibility screening~\cite{suzuki2014n,sanchez2016use,kato2016nlos}.
Within the taxonomy of this survey, the camera is not used for visual odometry, visual simultaneous localization and mapping (SLAM), or final pose estimation. 
Instead, it provides real-time environment-derived evidence about whether the direct satellite--receiver path is visually unobstructed. This information is largely independent of GNSS signal-strength or residual statistics, although reliable screening still requires camera calibration, pose or orientation information, time synchronization, and robust sky segmentation. Importantly, visual direct-path availability does not imply multipath-free reception: a satellite projected onto the visible sky may still be affected by reflected components. Camera screening should therefore be interpreted primarily as direct-path visibility evidence rather than as a complete multipath classifier.

It is important to distinguish these real-time onboard methods from the panoramic reprojections, street-view visibility grids, and image-derived sky representations reviewed in Section~\ref{sec:3d_mapping}. Although these representations provide useful \emph{a priori} visibility information, they are generally constructed from previously acquired environmental data and may therefore miss transient or recently changed occluders such as buses, trucks, temporary scaffolding, pedestrians, or vegetation. Onboard upward-looking cameras instead observe the local sky and surrounding occlusion geometry at the current measurement epoch, providing time-varying environmental evidence for satellite-visibility screening.
The underlying image-domain screening principle is nevertheless the same as that illustrated in Fig.~\ref{fig:image_derived_sky_visibility}. Following intrinsic camera calibration and extrinsic alignment between the GNSS antenna, camera, and local navigation frames, the azimuth and elevation of each tracked satellite are transformed into the camera frame and projected onto the segmented sky-view image. Satellites projected onto visible sky regions are treated as candidate LOS measurements, whereas those projected onto buildings, tree canopies, bridges, vehicles, or other occluding structures are identified as direct-path blocked and therefore candidate NLOS measurements. In the onboard case, however, the sky mask is generated from imagery acquired at the current epoch, allowing the same geometric visibility test to respond to short-term changes in the local environment.

To systematically review this visual verification pipeline, the remainder of this subsection is organized into three functional components. First, we examine classical \emph{sky-mask construction} techniques based on deterministic image-processing. Second, we explore the transition toward \emph{learning-based semantic sky segmentation}, including CNN-, Fully Convolutional Network (FCN)-, and transformer-based approaches. Finally, we summarize the geometric \emph{satellite projection} step that converts the resulting sky mask into per-satellite visibility evidence for online screening or offline supervision.

\subsubsection{Sky-Mask Construction}
The foundational stage of camera-based GNSS visibility analysis is the construction of a binary mask separating open sky from surrounding occluding structures. Early approaches relied on hand-crafted image-processing rules, using visual appearance, local regions, or skyline boundaries to distinguish sky from non-sky regions. Representative implementations include superpixel-based segmentation of fisheye imagery~\cite{suzuki2014n}, appearance- and clustering-based sky-region extraction~\cite{kato2016nlos}, and image-boundary extraction within fisheye-based GNSS/INS screening frameworks~\cite{sanchez2016use}. More refined deterministic pipelines incorporate gradient information and energy-based optimization to recover complex skyline contours from omnidirectional imagery ~\cite{yao2022vision}. These methods are generally interpretable and computationally efficient, but their performance can be sensitive to illumination, cloud and sky appearance, reflective surfaces, vegetation, and fine occlusion boundaries. Such limitations motivated the increasing use of learning-based segmentation.

\subsubsection{Learning-Based Semantic Sky Segmentation}
Learning-based methods replace hand-crafted thresholds and boundary rules with pixel-wise models trained to distinguish sky from urban occluders under more diverse visual conditions. General sky-segmentation studies highlight the sensitivity of this task to illumination and weather variation ~\cite{mihail2016sky}, while recent fisheye-oriented work further addresses wide-angle distortion and hemispherical imagery~\cite{bouillon2025pixel}. Within GNSS-specific visual screening, Lee et al.~\cite{lee2020skymask} introduced CNN-based sky/building segmentation for sky-pointing fisheye imagery. Subsequent systems adopted dense fully convolutional segmentation; for example, the vision front end of Sky-GVIO employs an FCN to generate sky/non-sky masks for GNSS NLOS screening~\cite{wang2024sky}. More recently, Swin Transformer models have been applied to fisheye sky segmentation before satellite visibility determination~\cite{hu2023fisheye}, reflecting the broader transition toward attention-based visual representations. Deep segmentation is not required in every implementation. Some recent sensor-fusion systems retain lightweight classical processing when computational efficiency is prioritized; Sky-GVINS, for example, evaluates several segmentation strategies and adopts Otsu-based global thresholding ~\cite{yin2024sky}. From the reception-state perspective, the specific segmentation architecture is therefore secondary to the resulting sky/occluder mask, which provides the geometric support for subsequent satellite visibility screening.

\subsubsection{Satellite Projection and NLOS Detection}
Once the sky mask is available, each tracked satellite must be geometrically associated with a location in the image. As illustrated previously in Fig.~\ref{fig:image_derived_sky_visibility}, the underlying operation is common to both archived image-derived visibility maps and real-time onboard cameras. GNSS ephemerides and an approximate receiver state provide satellite azimuth and elevation, while calibrated camera extrinsics and receiver attitude align these directions with the camera frame~\cite{sanchez2016use}. A calibrated fisheye projection model then maps each satellite direction onto the segmented image~\cite{hu2023fisheye}. A projection within the sky region supports candidate LOS reception, whereas a projection onto a building, vegetation, vehicle, bridge, or other occluding region indicates direct-path blockage and therefore candidate NLOS reception. For onboard cameras, this test is performed using imagery acquired at the current epoch and can therefore capture transient occluders that are absent from stored environmental representations. The resulting classification remains geometric visibility evidence: a visually unobstructed direction may still contain multipath. The same projection process is also useful for offline supervision. Camera- and INS-derived visibility labels have been used to train receiver-side classification models from GNSS observables, allowing the resulting detector to operate without a live camera during deployment ~\cite{li2024efficient,zhang2024lightgbm}. Such labels should be treated as geometry-derived supervision, since segmentation, calibration, synchronization, attitude, and near-boundary errors can propagate into the assigned reception state. Recent work has further extended camera-aided GNSS toward direct multimodal learning. Hu et al.~\cite{hu2026fisheye} jointly exploit fisheye-image features, GNSS measurements, and satellite-position information within a deep multimodal architecture for NLOS/multipath-aware pseudorange correction in urban canyons.

Overall, camera-aided methods provide real-time angular evidence of direct-path satellite visibility from the locally observed scene. Their effectiveness depends primarily on reliable sky/occluder segmentation, camera calibration, receiver attitude, and synchronization, while the absence of direct depth information limits explicit reconstruction of blocking and reflecting geometry.

\subsection{LiDAR-Based GNSS Anomaly Detection}
\label{subsec:lidar_based_approaches}
As discussed in Section~\ref{sec:3d_mapping}, LiDAR is widely used in offline 3D mapping, where dense point clouds are converted into geometric models for prior satellite-visibility analysis. In the auxiliary-sensor taxonomy of this survey, however, the emphasis is different: onboard LiDAR is treated as a real-time exteroceptive sensor for GNSS reception-state verification. 
An early precursor combined a laser scanner, GPS receiver, and INS to identify multipath reflections and infer the line of sight from the receiver toward the reflecting object~\cite{soloviev2008multipath}.
By measuring the three-dimensional structure surrounding the receiver, LiDAR provides viewpoint-specific information about buildings, facades, vegetation, vehicles, and other occluding or reflecting objects that may not be represented in static maps~\cite{wen2021gnss,wen20223d}.
Within the signal-classification pipeline, LiDAR-based methods operate as geometry-driven consistency checks. Local point-cloud geometry can be used to test whether a satellite direction is compatible with direct LOS propagation or is obstructed by the surrounding scene. With additional surface reconstruction or signal-side information, the same geometry can also support hypotheses about reflected NLOS or multipath propagation ~\cite{pugliese2023nlos,liu20253d}. This is fundamentally different from the traditional use of LiDAR in localization or SLAM: here, LiDAR is not reviewed primarily as a pose-estimation sensor, but as a source of geometric evidence for screening GNSS measurements before they are trusted by the navigation backend.

A key distinction from the sky-mask camera methods reviewed above is the availability of metric three-dimensional scene geometry. Camera sky masks primarily provide angular direct-path visibility, whereas LiDAR additionally provides depth and local surface geometry that can support the localization of candidate blockers or reflecting surfaces~\cite{bai2020using}. The reflected propagation path itself nevertheless remains inferred from the environmental geometry and, where available, GNSS measurements.
The following discussion focuses on two core components of this process: \emph{point-cloud scene construction} and \emph{visibility and reflected-path inference}.

\subsubsection{Point-Cloud Scene Construction}
The first stage in LiDAR-aided GNSS anomaly detection is the construction of a local geometric representation around the receiver. A 3D LiDAR produces a point cloud describing nearby facades, road boundaries, poles, trees, vehicles, and other structures. These raw points are commonly filtered, segmented, and organized into structural representations such as building surfaces and top edges, reconstructed planes or facades, and local or accumulated point-cloud maps~\cite{wen2018correcting,wen20223d,pugliese2023nlos}. This representation forms the LiDAR analogue of the camera sky mask: it is the environmental surrogate against which satellite visibility is evaluated, while additionally retaining metric depth and three-dimensional geometry.
Because LiDAR sensing is viewpoint-limited, point-cloud completeness is a central practical issue. Urban facades may be partially observed, rooftop edges may fall outside the vertical field of view, and highly reflective or absorptive materials may produce sparse or noisy returns. Early work addressed this limitation by combining onboard 3D LiDAR measurements with coarse building-height information, allowing the detected building top edges to be extended in the skyplot for NLOS identification and pseudorange correction~\cite{wen2018correcting}. 
Later methods improved this process by maintaining local sliding-window point-cloud maps that update as the receiver moves~\cite{wen20223d}, reconstructing facade surfaces from online LiDAR observations~\cite{pugliese2023nlos}, or accumulating richer point-cloud representations for integrated LiDAR/IMU/GNSS processing~\cite{zhang2025lidarnlos}.

Compared with static 3D city models, onboard LiDAR provides two important advantages for anomaly detection. First, it observes the current environment rather than relying solely on potentially outdated map geometry. Second, it can capture transient objects such as buses, trucks, parked vehicles, and temporary roadside structures. Such objects may block LOS signals or contribute to reflected paths even when the pre-built map predicts open sky~\cite{wen2021gnss}. Thus, LiDAR fills a critical gap between purely static 3D-map visibility prediction and purely signal-domain GNSS classification.

\begin{figure*}[t]
\centering
\begin{tikzpicture}[
    x=0.68cm,
    y=0.72cm,
    >=Latex,
    font=\sffamily,
    line cap=round,
    line join=round,
    box/.style={
        draw=black!40,
        rounded corners=1.5pt,
        line width=0.58pt,
        minimum width=3.1cm,
        minimum height=0.94cm,
        text width=2.78cm,
        align=center,
        inner sep=3.2pt,
        font=\sffamily\scriptsize
    },
    signal/.style={box, fill=blue!7, draw=blue!50!black},
    learn/.style={box, fill=violet!8, draw=violet!50!black},
    mapm/.style={box, fill=teal!10, draw=teal!55!black},
    sensor/.style={box, fill=orange!9, draw=orange!60!black},
    hybrid/.style={box, fill=gray!11, draw=gray!58!black}
]

\fill[gray!1.5, rounded corners=3pt]
    (1.25,0.95) rectangle (19.8,11.45);
\draw[dashed, gray!19, line width=0.4pt] (6.0,0.95) -- (6.0,11.45);
\draw[dashed, gray!19, line width=0.4pt] (10.75,0.95) -- (10.75,11.45);
\draw[dashed, gray!19, line width=0.4pt] (15.5,0.95) -- (15.5,11.45);
\draw[dashed, gray!19, line width=0.4pt] (1.25,3.55) -- (19.8,3.55);
\draw[dashed, gray!19, line width=0.4pt] (1.25,6.20) -- (19.8,6.20);
\draw[dashed, gray!19, line width=0.4pt] (1.25,8.85) -- (19.8,8.85);

\draw[-{Latex[length=3.4mm,width=2.4mm]}, line width=0.82pt, draw=black!70]
    (1.25,0.95) -- (20.0,0.95);
\draw[-{Latex[length=3.4mm,width=2.4mm]}, line width=0.82pt, draw=black!70]
    (1.25,0.95) -- (1.25,11.55);

\node[below=5pt, align=center, text=black!68] at (2.35,0.95)
    {\footnotesize\textbf{Low}\\[-2pt]\scriptsize standard receiver outputs};
\node[below=5pt, align=center, text=black!68] at (18.85,0.95)
    {\footnotesize\textbf{High}\\[-2pt]\scriptsize specialized RF / auxiliary sensing};
\node[below=27pt, font=\scriptsize\bfseries, text=black!76] at (10.65,0.95)
    {Receiver / Sensing / Integration Burden};

\node[left=18pt, align=right, text=black!68] at (1.25,1.55)
    {\footnotesize\textbf{Low}\\[-2pt]\scriptsize self-contained};
\node[left=18pt, align=right, text=black!68] at (1.25,11.05)
    {\footnotesize\textbf{High}\\[-2pt]\scriptsize data- / prior-dependent};
\node[rotate=90, font=\scriptsize\bfseries, text=black!76] at (-0.80,6.25)
    {External Data / Prior Dependence};

\draw[fill=blue!7, draw=blue!50!black, rounded corners=1pt, line width=0.4pt]
    (1.55,12.08) rectangle (1.90,12.30);
\node[anchor=west, font=\sffamily\fontsize{6.2}{7.0}\selectfont, text=black!62] at (2.08,12.19)
    {Signal / model-based};
\draw[fill=violet!8, draw=violet!50!black, rounded corners=1pt, line width=0.4pt]
    (5.50,12.08) rectangle (5.85,12.30);
\node[anchor=west, font=\sffamily\fontsize{6.2}{7.0}\selectfont, text=black!62] at (6.03,12.19)
    {Learning-based};
\draw[fill=teal!10, draw=teal!55!black, rounded corners=1pt, line width=0.4pt]
    (8.70,12.08) rectangle (9.05,12.30);
\node[anchor=west, font=\sffamily\fontsize{6.2}{7.0}\selectfont, text=black!62] at (9.23,12.19)
    {3D map};
\draw[fill=orange!9, draw=orange!60!black, rounded corners=1pt, line width=0.4pt]
    (11.30,12.08) rectangle (11.65,12.30);
\node[anchor=west, font=\sffamily\fontsize{6.2}{7.0}\selectfont, text=black!62] at (11.83,12.19)
    {Auxiliary sensor};
\draw[fill=gray!11, draw=gray!58!black, rounded corners=1pt, line width=0.4pt]
    (15.15,12.08) rectangle (15.50,12.30);
\node[anchor=west, font=\sffamily\fontsize{6.2}{7.0}\selectfont, text=black!62] at (15.68,12.19)
    {Fusion};

\node[signal] at (3.65,1.90)
{
    {\sffamily\fontsize{7.1}{7.8}\selectfont\bfseries Observable-Level\\Rules}\\[-1pt]
    {\sffamily\fontsize{6.2}{7.0}\selectfont\color{black!58} C/N0, residual, CMC,\\Doppler tests}
};
\node[signal] at (6.45,3.80)
{
    {\sffamily\fontsize{7.1}{7.8}\selectfont\bfseries Tracking / Baseband\\Screening}\\[-1pt]
    {\sffamily\fontsize{6.2}{7.0}\selectfont\color{black!58} ACF, SQM, and /\\multi-correlator features}
};
\node[signal] at (17.00,1.85)
{
    {\sffamily\fontsize{7.1}{7.8}\selectfont\bfseries Antenna Array\\\& Polarization}\\[-1pt]
    {\sffamily\fontsize{6.2}{7.0}\selectfont\color{black!58} spatial / polarization\\discrimination}
};

\node[learn] at (5.50,5.55)
{
    {\sffamily\fontsize{7.1}{7.8}\selectfont\bfseries Classical ML}\\[-1pt]
    {\sffamily\fontsize{6.2}{7.0}\selectfont\color{black!58} engineered observable\\features}
};
\node[learn] at (8.55,7.80)
{
    {\sffamily\fontsize{7.1}{7.8}\selectfont\bfseries Temporal / Attention /\\Graph Models}\\[-1pt]
    {\sffamily\fontsize{6.2}{7.0}\selectfont\color{black!58} sequence and satellite-context\\learning}
};
\node[learn] at (11.20,10.45)
{
    {\sffamily\fontsize{7.1}{7.8}\selectfont\bfseries Correlator-Domain\\Deep Learning}\\[-1pt]
    {\sffamily\fontsize{6.2}{7.0}\selectfont\color{black!58} waveform / baseband\\representation learning}
};

\node[mapm] at (13.70,8.05)
{
    {\sffamily\fontsize{7.1}{7.8}\selectfont\bfseries 3D Map-Based\\Visibility}\\[-1pt]
    {\sffamily\fontsize{6.2}{7.0}\selectfont\color{black!58} a priori geometric\\prediction}
};

\node[sensor] at (11.35,2.80)
{
    {\sffamily\fontsize{7.1}{7.8}\selectfont\bfseries IMU-Aided\\Consistency}\\[-1pt]
    {\sffamily\fontsize{6.2}{7.0}\selectfont\color{black!58} kinematic / attitude\\verification}
};
\node[sensor] at (14.50,5.80)
{
    {\sffamily\fontsize{7.1}{7.8}\selectfont\bfseries Camera-Based Visibility\\Screening}\\[-1pt]
    {\sffamily\fontsize{6.2}{7.0}\selectfont\color{black!58} sky / non-sky geometric\\screening}
};
\node[sensor] at (17.30,3.75)
{
    {\sffamily\fontsize{7.1}{7.8}\selectfont\bfseries LiDAR Geometric\\Verification}\\[-1pt]
    {\sffamily\fontsize{6.2}{7.0}\selectfont\color{black!58} 3D blockage /\\reflected-path geometry}
};

\node[hybrid] at (17.45,10.15)
{
    {\sffamily\fontsize{7.1}{7.8}\selectfont\bfseries Multimodal / Hybrid\\Fusion}\\[-1pt]
    {\sffamily\fontsize{6.2}{7.0}\selectfont\color{black!58} complementary cross-modal\\evidence}
};

\end{tikzpicture}

\caption{Qualitative comparison of GNSS reception-state and satellite-visibility methods. The axes reflect implementation burden and dependence on labeled data or environmental priors; positions indicate relative methodological trade-offs rather than quantitative performance rankings.}
\label{fig:tradeoff_map}
\end{figure*}
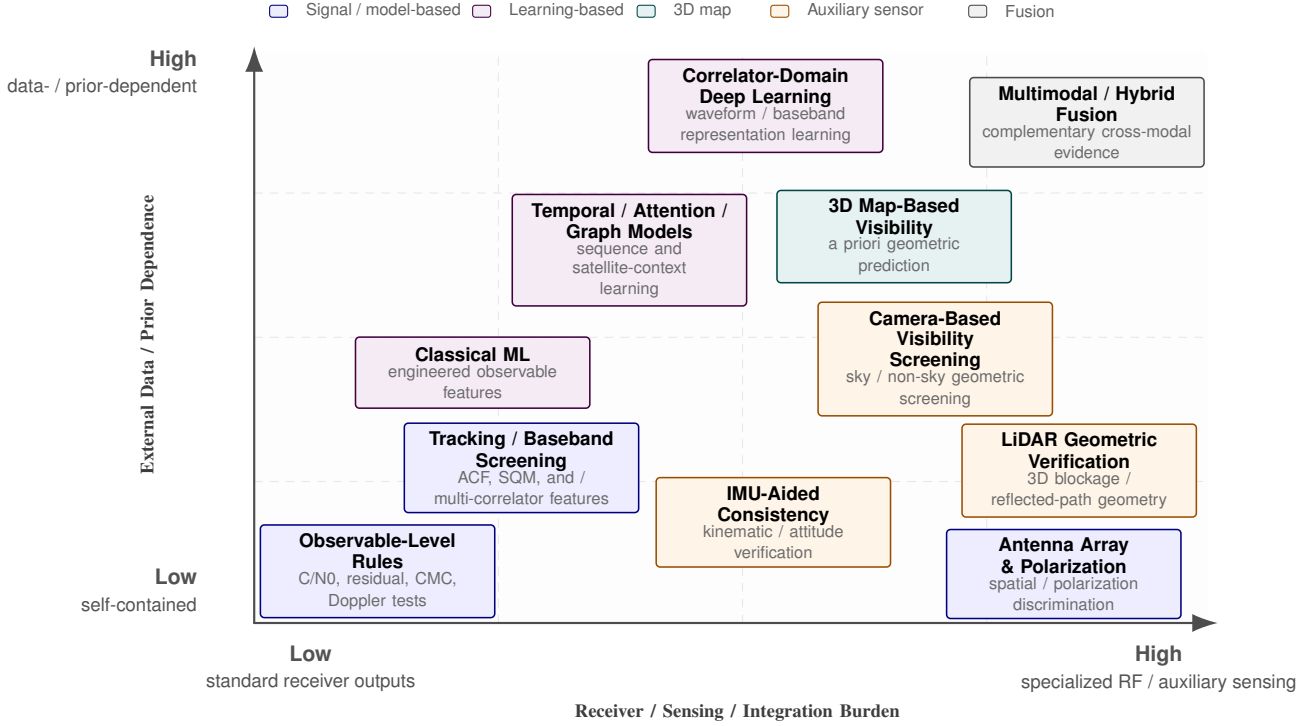

\subsubsection{Visibility and Reflected-Path Inference}
Once a local point-cloud representation is available, the dominant detection mechanism is geometric visibility testing. For each satellite, the line from the receiver antenna toward the satellite direction is evaluated against the LiDAR-derived scene. If this line of sight remains unobstructed within the modeled environment, the measurement is treated as a candidate LOS signal. 
If it intersects a reconstructed surface, object, or occupied cell, the direct path is classified as geometrically blocked; if a GNSS signal from that satellite is nevertheless tracked, it becomes a candidate NLOS reception~\cite{wen2018correcting,pugliese2023nlos}.
This point-cloud visibility test is conceptually similar to 3D-map geometric screening, but with one crucial difference: the occlusion model is sensed onboard and can therefore reflect the current urban state. Subsequent works extended this idea using denser point-cloud representations, local sliding-window maps, improved surface reconstruction, and real-time point-cloud updating to support GNSS real-time kinematic (RTK) and integrated GNSS/LiDAR/INS operation~\cite{wen20223d,zhang2025lidarnlos}. More recent approaches also combine LiDAR-derived visibility evidence with GNSS measurement consistency, rather than relying on geometry alone; for example, LiDAR ray tracing has been coupled with pseudorange-innovation information to improve identification of NLOS measurements~\cite{liu2026lidar}.

While line-of-sight intersection directly tests for geometric blockage, reflected propagation requires a higher-level geometric analysis. A blocked satellite may be received through an NLOS reflection, while an otherwise visible satellite may also contain reflected multipath components from nearby facades, vehicles, or overhangs. 
LiDAR is valuable here because it provides metric 3D surface geometry that camera sky masks cannot recover.
Accordingly, LiDAR-based reflected-path analysis serves as geometry-supported propagation inference.
The reconstructed surfaces provide candidate reflectors, their orientations, and geometrically feasible reflection locations, enabling reflected propagation paths and the associated pseudorange biases to be inferred for NLOS measurements~\cite{wen20223d,pugliese2023nlos,zhang2025lidarnlos}. These approaches primarily determine the reflection geometry from the sensed environment together with the known satellite direction. More recent work further strengthens this inference by incorporating GNSS-side directional evidence, combining LiDAR geometry with Doppler-derived direction information to better constrain the likely NLOS propagation path~\cite{liu20253d}.

LiDAR has also been combined with sky-pointing cameras to exploit their complementary environmental information~\cite{bai2020using}. In this approach, the fisheye camera provides wide-field sky/non-sky visibility information for satellite screening, while LiDAR supplies metric range and surface geometry that can support localization of candidate reflectors and correction of NLOS propagation errors. The modalities therefore address complementary limitations: camera-based screening provides direct angular visibility but limited depth information and remains sensitive to illumination and visual conditions, whereas LiDAR provides metric 3D geometry but is constrained by range, reflectivity, sparsity, and field of view.

\section{Comparative Insights and Design Trade-Offs}
\label{sec:tradeoffs}

The methods reviewed in this survey differ less in objective than in evidence source. All ultimately aim to assess the reception state, visibility, or viability of a GNSS measurement before unreliable observations corrupt the navigation solution, using RF observables, learned signal patterns, environmental geometry, or auxiliary sensing. Since the preceding sections detailed their operating principles, this section synthesizes their deployment characteristics from a systems-engineering perspective. 
The level of experimental evidence also varies substantially across the reviewed literature. Some studies establish feasibility analytically or through simulation under simplified propagation, calibration, or environmental assumptions, whereas others evaluate controlled experiments, real-world static or kinematic data, or real-time implementations. Reported performance should therefore be interpreted together with the validation setting, receiver hardware, data access, and assumptions of each study rather than compared directly across different experimental regimes.
Fig.~\ref{fig:tradeoff_map} provides a qualitative design-space view, positioning the major method families across \emph{hardware/deployment burden} and \emph{data/prior dependence}. Several key observations follow.

First, as shown in the lower-left region of Fig.~\ref{fig:tradeoff_map}, signal-observable rules and classical machine learning on engineered features remain attractive because they require no additional exteroceptive infrastructure and operate on the standard GNSS measurement stream. They are therefore well suited to low-latency measurement screening, exclusion, or reweighting. However, because observable-level indicators are indirect proxies for the underlying propagation state, and their distributions can overlap substantially across LOS, NLOS, and multipath conditions. Their decisions are consequently sensitive to the receiver, environment, and operating scenario.

Second, methods closer to the received waveform or spatial aperture provide evidence more directly linked to propagation structure. Correlator-domain learning, antenna arrays, and polarization-aware methods are more directly tied to propagation mechanisms than scalar observables. Their main limitation is architectural access: they often require receiver-internal data, software-defined radio architectures, calibrated RF hardware, or multi-antenna installations.

Third, map-based priors and auxiliary sensors introduce environmental awareness that GNSS-only observables cannot provide. 3D maps support geometry-driven satellite-visibility determination, while cameras provide real-time sky/occlusion evidence and LiDAR supplies metric 3D structure for blockage testing and reflected-path inference. These spatial sources do not directly measure the RF propagation path; instead, they constrain blockage and reflection hypotheses geometrically. Distinguishing reflected NLOS reception or multipath contamination can therefore benefit from combining this geometric evidence with GNSS-domain observables.

Finally, no method family is universally dominant. Deployment choices depend on platform size, weight, power, cost, receiver accessibility, map availability, and integrity requirements. Low-cost receivers favor signal-observable rules and lightweight learning; mapped urban corridors benefit from 3D visibility determination; and autonomous platforms can exploit existing camera or LiDAR payloads for local geometric screening. For higher-integrity applications, multimodal fusion can provide complementary or redundant evidence across independent information sources, but it also introduces additional requirements for uncertainty propagation, calibration, synchronization, failure detection, and computational resources.

\section{Conclusion and Open Directions}
\label{sec:future}

This survey has synthesized the landscape of GNSS signal-reception classification and satellite-visibility determination in urban canyons across three main methodological categories: 
(i) \emph{signal-domain reception-state inference}, which includes intrinsic RF, tracking-level, and observable-level features together with rule-based, statistical, and learning-based classifiers; 
(ii) \emph{3D map-aided visibility determination}, which uses environmental geometry to predict/detect satellite blockage and measurement viability; and 
(iii) \emph{auxiliary-sensor spatial screening}, where cameras, LiDAR, and IMU provide real-time occlusion, reflection-plausibility, and kinematic-consistency evidence. 
The comparative analysis reveals that no single category is universally dominant. Observable-level rules provide lightweight baseline screening, learning-based models exploit nonlinear, temporal, and cross-satellite structure across observable- and receiver-internal features, map-based priors offer geometry-driven blockage determination, and auxiliary sensors provide complementary geometric and kinematic evidence about local occlusions, dynamic scene changes, and measurement consistency. Taken together, these complementary capabilities motivate layered architectures that combine multiple evidence streams to evaluate measurement viability before corrupted observations enter the navigation engine.

Despite substantial progress in GNSS reception-state and visibility inference, several challenges remain before these methods can generalize reliably beyond specific datasets, receivers, and environments. Key directions concern transferable learning, uncertainty-aware decisions, probabilistic modeling of urban visibility, reliable benchmarking, and practical real-time deployment:

\begin{itemize}

    \item \textbf{Domain Generalization and Label-Efficient Learning:}
    Data-driven detectors remain sensitive to changes in receiver, environment, and operating conditions. Domain adaptation and self-supervised or contrastive learning have begun to reduce this dependence~\cite{zawislak2022gnss,zeng2026environment}, but robust transfer across cities, receivers, constellations, and mobility modes with little or no relabeling remains an open challenge.

    \item \textbf{Uncertainty-Aware Reception-State Inference:}
    Hard LOS/NLOS or multipath decisions provide limited information when propagation states overlap or observations are ambiguous. Recent uncertainty- and distribution-aware approaches provide initial steps~\cite{zhang2023learning,roessl2025quantile,dobre2025uncertainty}; the broader challenge is to produce calibrated measurement-level reliability that can directly support weighting, exclusion, correction, and integrity monitoring.

    \item \textbf{Stochastic Geometry for Urban Visibility:}
    Stochastic geometry has begun to characterize satellite availability and localizability from the spatial distribution of medium Earth orbit (MEO)/low Earth orbit (LEO) constellations~\cite{wang2025localizability}, while complementary models characterize LOS probability from statistical urban geometry~\cite{alhourani2020los,alhourani2024los}. A promising direction is to jointly model satellite distributions and urban occluder/building distributions, enabling analytical prediction of LOS availability, usable satellite geometry, and ultimately positioning localizability without requiring a deterministic 3D map for every environment.

    \item \textbf{Benchmarking and Label Quality:}
    Reception-state labels derived from 3D maps, cameras, LiDAR, ray tracing, or GNSS measurements do not always represent identical physical conditions and can disagree in complex scenes~\cite{beaucamp2025comparison}. Future benchmarks should therefore standardize state definitions and labeling uncertainty and emphasize cross-city, cross-receiver, and cross-platform evaluation rather than only within-dataset accuracy.

    \item \textbf{Real-Time Deployment and Environmental Maintenance:}
    Richer methods increasingly require receiver-internal data, learned models, environmental priors, or auxiliary sensors, creating latency and integration constraints. Practical systems should explore resource-adaptive architectures that invoke expensive inference only when lightweight GNSS screening is insufficient, while map-based approaches additionally require mechanisms for detecting and updating stale environmental information~\cite{fredeluces2025nlos,xu2024urban}.

\end{itemize}

Ultimately, robust urban GNSS should progress from hard anomaly labels toward uncertainty-aware measurement-level reliability assessment. The central objective is to determine, with quantified confidence, whether each observation should be trusted, reweighted, corrected, or excluded before it enters the navigation solution.

\section*{Acknowledgment}

This publication is based upon work supported by King Abdullah University
of Science and Technology (KAUST) under Award No.~ORFS-CRG12-2024-6478.

The authors used AI-assisted writing tools for language refinement and structural editing. All technical content and conclusions were reviewed and verified by the authors, who take full responsibility for the manuscript.

\bibliographystyle{IEEEtran}
\bibliography{references}

\end{document}